\documentclass[11pt]{iopart}
\usepackage{epsfig}
\usepackage{graphicx}
\usepackage{dcolumn}
\usepackage{bm}
\usepackage{amssymb}
\usepackage{cite}
\usepackage{enumerate}
\usepackage{scrtime}
\usepackage{color}
\usepackage[sans]{dsfont} 
\usepackage{colortbl}
\usepackage{subfigure}
\usepackage{paralist}
\usepackage{lscape}
\usepackage{longtable}
\usepackage{ulem}
\usepackage{cancel}

\def\be{\begin{equation}}
\def\ee{\end{equation}}
\def\bea{\begin{eqnarray}}
\def\eea{\end{eqnarray}}
\def\nn{\nonumber}
\def\mz{\mathbb Z}

\def\ms{{M_{\rm string}}}

\def\mbsix{\mathbf{6}}
\def\mbsixb{\mathbf{\overline{6}}}
\def\mbthree{\mathbf{3}}
\def\mbthreeb{\mathbf{\overline{3}}}

\newcommand{\lgp}[2]{{\rm {#1}_{#2}}}

\newcommand{\SU}[1]{\ensuremath{\mathrm{SU}(#1)}}
\newcommand{\BmL}{\ensuremath{B\!-\!L} }
\newcommand{\ov}{\overline}
\newcommand{\mc}{\mathcal}

\newcommand{\fk}{\ensuremath{\mathcal{K}}}

\newcommand{\fw}{\ensuremath{\mathcal{W}}}

\def\ch2{$\chi^2$}

\begin{document}

\title{Searching for the Standard Model in the
String Landscape :  SUSY GUTs}

\author{Stuart Raby\dag\
\footnote[3]{raby@pacific.mps.ohio-state.edu, OHSTPY-HEP-T-10-004}}
\address{\dag\ Department of Physics, The Ohio State University, 191 W. Woodruff Ave., Columbus, OH 43210, USA}

\begin{abstract}
The Standard Model is the theory describing all observational data
from the highest energies to the largest distances.  [There is
however one caveat; additional forms of energy, not part of the
Standard Model, known as dark matter and dark energy must be
included in order to describe the universe at galactic scales and
larger.] High energies refers to physics at the highest energy
particle accelerators, including CERN's LEP II (which ceased
operation in 2000 to begin construction of the LHC or Large Hadron
Collider now operating) and Fermilab's Tevatron, as well as to the
energies obtained in particle jets created in so-called active
gallactic nuclei scattered throughout the visible universe. Some of
these extra-galactic particles bombard our own earth in the form of
cosmic rays, or super-energetic protons which scatter off nucei in
the upper atmosphere.

String theory is, on the other hand, an unfinished theoretical
construct which attempts to describe all matter and their
interactions in terms of the harmonic oscillations of open and/or
closed strings.   It is regarded as unfinished since at the present
it is a collection of ideas, tied together by powerful consistency
conditions, called dualities, with the ultimate goal of finding the
completed String Theory.  At the moment we only have descriptions
which are valid in different mutually exclusive limits with names
like, Type I, IIA, IIB, heterotic, M and F theory.  The string
landscape has been described in the pages of many scholarly and
popular works. It is perhaps best understood as the collection of
possible solutions to the string equations. Albeit these solutions
look totally different in the different limiting descriptions.  What
do we know about the String Landscape?  We know that there are such
a large number of possible solutions that the only way to represent
this number is as $10^{500}$ or a one followed by 500 zeros.  Note
this isn't a precise value since the uncertainty is given by a
number just as large. Moreover, we know that most of these string
states look nothing like the Standard Model.  They have the wrong
matter and wrong forces.   Moreover they are not off by a small
amount, they are totally wrong.

So the question becomes, does string theory really describe our
observable world?   In order to address this question, one must find
at least one string state that resembles it.   One possibility is
that our observable world is in fact a unique string state.   If
this is the case, then the problem becomes one of finding the
proverbial needle in the largest possible haystack!   On the other
hand, there may be many states which are sufficiently close to the
observable world, and we need only to understand why we are in this
finite subspace of the string landscape.  And perhaps there are good
reasons, why this subspace is preferred over 99.999999999...\% of
the myriad of non-Standard-Model-like string states.   Perhaps, just
by confining our attention to this subspace we can learn something
about our observable world which we can not learn otherwise.  Thus
the goal of the present essay is to understand what it takes to find
the Standard Model in the string landscape.

\end{abstract}

\maketitle

\section{Introduction}

\subsection{Prologue}

In this article I will discuss several recent attempts to find the
Standard Model in the string landscape.   I will argue that the most
efficient path is paved with two new proposed symmetries of Nature,
called grand unification and supersymmetry.  Together they make the
journey exciting and inspiring.   In order to bring this adventure
to you, the reader, I should first define what I mean when I talk of
the Standard Model, then describe these two new symmetries and
finally discuss how to find the Standard Model in the string
landscape.   This is the goal of the present article.

\subsection{The Standard Model}

The Standard Model is defined by the collection of four fundamental
forces - strong, weak, electromagnetic and gravitation and the
matter particles which interact via these forces.  The last
ingredient of the Standard Model, yet to be observed directly in an
experiment, is a fifth force, described by the Higgs interaction.
The name {\it the Standard Model} is archaic and no longer
appropriate.  It was coined in the early 1970s when the Standard
Model was the best proposal for a new theory describing all known
phenomena and thus THE THEORY to be tested in every possible way.
After some 30 odd years and millions of experimental tests later,
the {\it Standard Model} is now the accepted {\it Theory of all
observed properties of Nature}.  This is not to say, that the
original proposal was found to be without any unacceptable
blemishes.  In fact, a major discovery in the last ten years has
been the realization that neutrinos, which were believed to be
massless in the simplest version of the SM, are in fact massive.
More about this later.   However with this major correction the SM
is truly an accepted theory.\footnote{That said, the SM has two
major deficiencies with regards to cosmology and astrophysics. It
lacks an explanation for the observed dark matter and dark energy of
the universe.  These topics are typically reserved for a discussion
of the physics beyond the Standard Model.}

The Standard Model includes three families of quarks and leptons.
The quarks come in six different flavors, up, down,  charm, strange,
top, and bottom, with each flavor coming in three colors. Quarks
feel the strong force, exchanging gluons which themselves come in 8
colors. Leptons are however color singlets. The three-fold symmetry
of the strong interactions is described by the mathematical group
called $SU(3)_{color}$. Prior to obtaining mass via the Higgs VEV,
massless quarks separate into left-handed and right-handed
components.   So what is handedness? Quarks and leptons are spin 1/2
particles, called Fermions.  If a fermion's spin points in the
direction opposite to its motion, it is called left-handed and if
its spin points in the same direction as its motion, it is called
right-handed.  This may seem like a trivial distinction, but in fact
the weak doublets are only left-handed. So it is necessary to
distinguish left-handed and right-handed Fermion fields, because
Nature does.    To describe an up quark we then need two fields, $u$
and $\bar u$.  The first up field annihilates a left-handed up quark
and creates a right-handed anti-up quark. The second annihilates a
left-handed anti-up quark (hence the bar) and creates a right-handed
up quark. Then the weak boson $W^+$ takes $d$ into $u$, and $W^-$
does the reverse. We describe this doublet (under the weak $SU(2)_L$
group) as a single field $q$ with two pieces, i.e. $q =
\left(\begin{array}{c} u
\\ d \end{array} \right)$.

Note, in the Standard Model, all particles are described by
relativistic quantum fields.\footnote{Again, one caveat with respect
to gravity and the rest of the Standard Model.  A self-consistent
quantum mechanical treatment of gravity is problematic.   This
concerns the description of black holes and the so-called
``information loss paradox."  It is this paradox which calls for a
better description of gravity and the Standard Model.} It is a
property of relativity that each particle has an anti-particle with
identical mass but opposite charges under all three forces.
Moreover, when a particle and it's own anti-particle meet, they
annihilate into anything which can be produced while conserving
energy.   This may sound like an esoteric, Star Wars like, concept;
it is reality.  In fact, a PET scan that you may have seen in your
local hospital is a manifestation of this fact.   PET scan or
Positron Emission Tomography is a process whereby a radioactive
isotope is administered to the patient.  This radioactive isotope
emits positrons (or anti-electrons).  The isotope travels to a part
of your body (through the blood stream) and sticks to a spot
depending on the chemical properties of the isotope.  For example,
$^{58}$Ga may be used for a PET scan of the brain.   When the
positrons are emitted they quickly annihilate with local electrons.
Two photons are created which travel to the detector moving out in
opposite directions.   Each photon, sometimes called a gamma ray,
$\gamma$, has energy equal to the mass of the electron which
disappeared. The energy of the photon satisfies the famous Einstein
relation, $E_\gamma = m_e c^2$.

So the first family of quarks and leptons is obtained from the
Fermion fields - \be \{ q = \left(\begin{array}{c} u \\ d
\end{array} \right), \ \bar u, \bar d, \ l = \left(\begin{array}{c}
\nu \\ e \end{array} \right), \ \bar e, \bar \nu \}. \ee  Note, the
quark fields $u, \ d$ are color triplets,  $\bar u, \ \bar d$ are
anti-triplets and leptons are color singlets, i.e. they do not have
color charge. The fields $q, \ l$ are weak $SU(2)_L$ doublets and
under weak hypercharge, $U(1)_Y$, these fields have charge - \be Y =
\{ 1/3, \ -4/3, \ 2/3, \ -1, \ 2, \ 0 \}. \ee Note, electric charge,
$Q$, of these fields is given by the simple formula - \be Q = T_3 +
\frac{Y}{2} .\ee   For example, in the lepton doublet we have $T_3 =
+ 1/2$ for the neutrino, $\nu$, and $T_3 = - 1/2$ for the electron,
$e$. Hence the neutrino (electron) have electric charge 0 (-1),
respectively. The anti-electron field, $\bar e$, has $T_3 = 0$,
since it is not part of a weak doublet (we say it is a weak singlet)
and thus has electric charge +1.   It is then a simple exercise to
work out the electric charge of all the fields in one family.  We
find that quarks have fractional charge, 2/3 for $u$ and -1/3 for
$d$. Finally, $\bar \nu$ is neutral under all three SM symmetries,
$SU(3)_{color} \times SU(2)_L \times U(1)_Y$.   It is a so-called
``sterile" neutrino whose sole purpose, as we shall discuss later,
is to give neutrinos mass.   Note, there are inequivalent ways of
introducing neutrino masses in the literature. In the later
discussion of neutrino masses, I will only consider the simplest
possibility known as the Type I See-Saw mechanism.  In this case
only SM singlet states, such as the ``sterile" neutrinos or SM
singlet scalars are added to the original version of the SM, where
neutrinos were, in fact, massless.

Yet the only particles we observe in nature have integral charge.
This is because a proton is a bound state of two up quarks and one
down quark, while a neutron is a bound state of one up quark and two
down quarks.   Protons and neutrons have electric charge +1 and 0,
respectively.   The strong force is extremely strong and is so
strong that an isolated free quark cannot exist.   Only color
singlet combinations of a red up, a blue up and green down is
allowed.  However the residual strong force between color singlet
combinations of quarks in protons and neutrons hold these particles
tightly bound in the nuclei of atoms.  The electromagnetic force is
not as strong but then binds electrons and protons together to form
electrically neutral atoms.

In fact, the electromagnetic force is many times stronger than
gravity.   Gravity holds us to the earth, but the electromagnetic
force between atoms in matter keep us from falling through the
floor!   The electromagnetic force is so very strong, that no person
on earth is strong enough to hold two electrons together, assuming
these were the only two charged particles around.   So if atoms made
of electrons and protons were not electrically neutral, we could not
possibly have structure in the universe.  Tables, chairs, stars and
galaxies would be pulled apart by the strong electromagnetic
repulsion.   So it is a wonderful fact that charge is quantized and
protons and electrons have equal and opposite charges, so when they
are put together they make electrically neutral atoms.   This
property of the Standard Model is known as charge quantization.

There are three families of quarks and leptons. All the states of
the first family, with all their charges under the strong and
electroweak forces, are duplicated with a second family and then a
third family.  The only difference is that the second family
particles are heavier than the first and the third family is even
heavier than the second family.   It is a complete mystery why
Nature would require three {\it almost} identical families of quarks
and leptons.

The Standard Model describes all known phenomena both on experiments
performed on earth as well as a description of cosmology and
astrophysics starting as way back in time as the first 3 minutes of
the universe.   All of this can be described with just 28
fundamental parameters (see Table \ref{tab:SMparameters}).   These
are the three low energy gauge coupling constants, 9 parameters
associated with quark and charged lepton masses,  4 quark weak
mixing angles contained in the CKM matrix and 9 parameters
associated with neutrino masses and mixing.

\begin{table} \begin{center}
\begin{tabular}{|lcc|}
\hline
Sector & \# & Parameters \\
\hline \hline
gauge & 3 & $\alpha_1$, $\alpha_2$, $\alpha_3$, \\
quark masses & 6 & $m_{u}$, $m_{d}$, $m_c$, $m_s$, $m_t$, $m_b$, \\
quark weak mixing angles & 4 & $V_{us}$, $V_{cb}$, $V_{ub}$, $\delta$, \\
charged lepton masses & 3 & $m_e$, $m_\mu$, $m_\tau$,  \\
neutrino masses & 3 & $m_{\nu_1}$, $m_{\nu_2}$, $m_{\nu_3}$, \\
lepton weak mixing angles & 6 &  $\theta_{\rm sol}$, $\theta_{\rm atm}$,  $\theta_{13}$, $\delta_i, \ i=1,2,3$, \\
electroweak scales & 2 & $M_Z$, $m_{\rm Higgs}$ \\
strong CP angle & 1 & $\bar \theta$ \\
\hline \hline
\end{tabular}
\caption{The 28 parameters of the Standard Model.  The values for
all of these parameters are set by experiment.  Note, the charges of
the quarks and leptons under $SU(3)_{color} \times SU(2)_L \times
U(1)_Y$ are typically not included in this list.  They are
nevertheless important ingredients of the theory, whose origin is a
great mystery.} \label{tab:SMparameters}
\end{center}
\end{table}

\subsection{Supersymmetric grand unified theories [SUSY GUTs]}

SUSY GUTs provide a framework for solving many of the problems of
the Standard Model.  But let us first describe these two symmetries
one-by-one.

\subsubsection{Grand unification}

Grand unified theories unify the three fundamental interactions of
the Standard Model. These are the strong forces represented by a
three-fold symmetry called color; the weak forces are responsible
for one form of radioactive nuclear decay which powers the engine
for stellar burning, including our own sun, and the electromagnetic
force, associated with the first relativistic field theory unifying
the theories of electric and magnetic phenomenon into one.   The
latter two forces are themselves intertwined in what is called the
electroweak theory, represented by a two-fold symmetry called weak
isospin and and a phase symmetry called weak hypercharge.   It is an
amazing fact of nature that the electroweak symmetry is a symmetry
of equations of motion of the theory, but not a symmetry of the
vacuum.   In fact, this very symmetry breaking is due to the vacuum
expectation value of the Higgs field.  And this non-zero expectation
value is responsible for the mass of the weak bosons, $W^\pm$ and
$Z^0$, and all quarks and leptons.

Grand unified theories provide an explanation of these amazing
facts. By unifying the three symmetries,  $SU(3)_{color}, \ SU(2)_L,
\ U(1)_Y$, one also unifies the fermions of the Standard Model into
multiplets transforming under the new grand unified symmetry group.
In fact the simplest extension of the Standard Model, the symmetry
group $SO(10)$ \cite{SO(10)}, puts all the fermions of one family
into one single representation, see Table \ref{tab:spinor}.  This is
a phenomenal fact and a significant point in favor of grand
unification.

$SO(10)$ contains the SM as subgroup.  Note, the spinor
representation of $SO(10)$ is represented as a direct product of 5
spin 1/2 states with spin in the 3rd direction labelled $\pm (1/2)$.
$SU(3)$ acts on the first 3 spin states by raising (or lowering) one
spin and lowering (or raising) another.  Hence the states
represented by $ - \, + \, + ; + \, - \, + ; + \, + \, - $ are an
$SU(3)$ triplet and $ + \, + \, + $, $ - \, - \, - $ are singlets.
$SU(2)$ acts on the last two spin states, with $ - \, + ; + \, - $
representing an $SU(2)$ doublet and $ + \, + $, $ - \, - $ are
singlets.   Finally,  electroweak hypercharge $Y$ is given by $Y =
\frac{2}{3} \sum_{i = 1}^3 S_i - \sum_{i = 4}^5 S_i$ with $S_i = \pm
\frac{1}{2}$.

$SO(10)$ also has two very interesting subroups, Pati-Salam
\cite{ps} ($SO(6) \otimes SO(4) \equiv SU(4)_C \otimes SU(2)_L
\otimes SU(2)_R$) and Georgi-Glashow \cite{gg} ($SU(5)$).  $SU(5)$
multiplets can be gotten by raising one spin and lowering another
spin out of the full tensor product state.   As such, one sees that
the spinor representation of $SO(10)$ breaks up into 3 irreducible
representations of $SU(5)$ given by the multiplets with zero, two or
four spin down states, corresponding to an $SU(5)$ ${\bf 1}$, ${\bf
10}$ or ${\bf \bar 5}$. In particular, note that the ${\bf 10} =
\{\bar u, \ q, \ \bar e \}$, ${\bf \bar 5} = \{\bar d, \ l \}$ and
${\bf 1} = \bar \nu$. The full $SO(10)$ operations then allow for
simultaneously raising or lowering two different spins. Similarly,
$SU(4)_C$ is an extension of $SU(3)_C$ with the additional
operations of raising or lowering two different color spins. Hence
the spinor representation of $SO(10)$ decomposes to two irreducible
multiplets of Pati-Salam, with $Q = \{ q , l \} = (4, 2, 1)$ and
$\bar Q = \{ \bar q, \bar l \} = (\bar 4, 1, \bar 2)$ with $SU(2)_R$
represented by the operation of raising or lowering two different
weak indices.

\begin{table}
$$
\begin{array}{|l|c|c|c|}
\hline \multicolumn{4}{|l|}{\rm Grand \;\; Unification \; - \; SO(10)}\\
 \hline
 {\rm State} & { \rm Y}   & {\rm Color} & {\rm Weak} \\
 & =  \frac{2}{3} \Sigma ({\rm C}) - \Sigma ({\rm W}) &
 {\rm C \; spins} & {\rm W \; spins} \\
\hline\hline
{\bf \bar \nu}  & { 0} &  + \, + \, +  &     + \, +  \\
\hline\hline
{\bf \bar e}  & { 2} &  + \, + \, + & - \, -  \\
\hline
{\bf u_r} &  &  - \, + \, + &   + \, - \\
{\bf d_r} &  &  - \, + \, + &   - \, + \\
{\bf u_b} & { \frac{1}{3}} &  + \, - \, + &   + \, -  \\
{\bf d_b} &  &  + \, - \, + &  - \, +  \\
{\bf u_y} &  & + \, + \, - &    + \, - \\
{\bf d_y} &  & + \, + \, - &   - \, + \\
\hline
{\bf \bar u_r} &  & + \, - \, - & + \, + \\
{\bf \bar u_b} & { -\frac{4}{3}} & - \, + \, - & + \, + \\
{\bf \bar u_y} &  & - \, - \, + & + \, + \\
\hline\hline
{\bf \bar d_r} &  & + \, - \, - & - \, - \\
{\bf \bar d_b} & { \frac{2}{3}} & - \, + \, - & - \, - \\
{\bf \bar d_y} &  & - \, - \, + & - \, - \\
\hline
{\bf \nu}  & { -1} &  - \, - \, - & + \, - \\
{\bf e}  &  & - \, - \, - & - \, + \\
\hline
\end{array}
$$
\caption{Spinor representation of $SO(10)$ where this table
explicitly represents the Cartan-Weyl weights for the states of one
family of quarks and leptons.  The double lines separate irreducible
representations of $SU(5)$.  \label{tab:spinor}}
\end{table}

\begin{figure}[t!]
\hspace{-.70in}
\begin{center}
\begin{minipage}{5in}
\epsfig{file=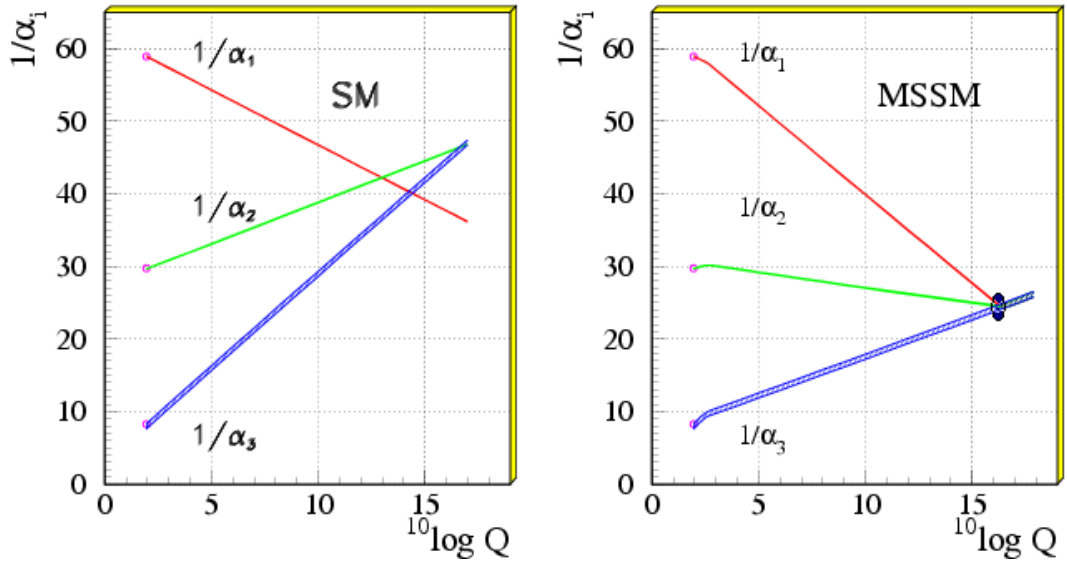,height=3in}
\end{minipage}
\caption{\label{fig:guts} {\small Gauge coupling unification in non-SUSY GUTs on the left vs. SUSY GUTs on the
right using the LEP data as of 1991. Note, the difference in the running for SUSY is the inclusion of
supersymmetric partners of standard model particles at scales of order a TeV.}  Given the present
accurate measurements of the three low energy couplings, in particular
$\alpha_s(M_Z)$, GUT scale threshold corrections are now needed to precisely fit the low energy data. The dark
blob in the plot on the right represents these model dependent corrections.}
\end{center}
\end{figure}

The three fundamental forces, defined by the symmetries
$SU(3)_{color}, \ SU(2)_L, \ U(1)_Y$ couple to matter proportional
to the three fine-structure constants, $\alpha_3(\mu), \
\alpha_2(\mu), \ \alpha_1(\mu)$ defined at some energy scale $\mu$.
A grand unified gauge theory, such as $SO(10)$, on the other hand
has only one coupling constant, $\alpha_{GUT}$.  Note, the values of
the observed coupling constants depend on the energy scale at which
they are measured.  Indeed we are able to calculate the energy
dependence of the fine-structure constants, IF one knows the mass
spectrum of particles which carry strong and electroweak charge. If
one assumes that only particles observed to date and the
theoretically predicted Higgs boson contribute to the so-called
renormalization group running of the Standard Model couplings one
finds the result given in Fig. \ref{fig:guts} (left).  Although the
three fine-structure constants approach to a common value, they do
NOT actually meet.   This is a significant problem with gauge
coupling unification.  However, this calculation explicitly assumes
no new particle states above the weak scale, since only particle
states with mass less than the relevant energy scale affect the
running of the coupling.  This assumption will shortly be
challenged.

Assuming grand unification at a scale, $M_{GUT} \sim 10^{15}$ GeV,
introduces a fundamental problem.  Why is the weak scale,  $M_{weak}
\sim 100$ GeV,  so much smaller than the GUT scale.   This is
particularly problematic since the weak scale is set by the Higgs
boson mass. Note boson masses are sensitive to physics at the
highest scales. This is because quantum mechanical corrections to
any boson mass are proportional to some effective coupling $\alpha$
times the largest scale in the theory.  In this case we would expect
the Higgs mass to be given at leading order by $m_H^2 \approx m_0^2
+ \alpha M_{GUT}^2 + \cdots$ where $m_0$ is the Higgs mass prior to
quantum corrections.   Why the ratio $m_H/M_{GUT} \sim 10^{-13}$ is
known as the ``gauge hierarchy problem."   One possible solution to
the gauge hierarchy problem is to postulate a new symmetry of
nature, so-called ``supersymmetry."  {\bf Supersymmetry} is, as it
sounds, a pretty amazing symmetry.

\subsubsection{Supersymmetry}

Supersymmetry is the largest possible symmetry of space-time, i.e.
it is an extension of Einstein's special theory of relativity and
ordinary space-time translation invariance which adds a new quantum
direction of space. This quantum direction of space, $\theta$, is,
in fact, infinitesimal since $\theta^2 = 0$. Moreover, there is at
least two such directions,  $\theta_1, \ \theta_2$ with the property
that $\theta_1 \theta_2 = - \theta_2 \theta_1$. Ordinary coordinates
of space and time are pure numbers, where given two different
directions $x$ and $y$, we have $x y = y x$.  One obtains the same
number no matter which order these two numbers are multiplied.   The
difference in these multiplication rules is identical to the
difference between the quantum statistics of identical fermions and
identical bosons.   Supersymmetry is a rotation of ordinary space
into superspace where the $\theta$-like coordinates rotate into the
ordinary space-time coordinates.

\begin{figure}[t!]
\hspace{-.70in}
\begin{center}
\epsfig{file=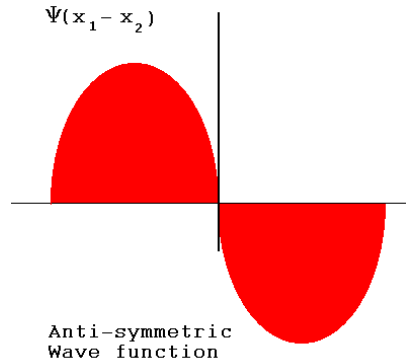,height=2in}
\end{center}
\caption{\label{fig:antisym} {\small This figure represents the anti-symmetric wave function for two identical fermions at
positions $x_1, \ x_2$.  Note, due to the anti-symmetry of the wave function, it is NOT possible to have two fermions at
the same position in space, at the same time.}}

\end{figure}

Let's take a small digression which we will quickly relate to the
above discussion. Matter particles, such as electrons, protons and
neutrons, are spin 1/2 particles, known as Fermions. They satisfy
the Pauli exclusion principle. This is the property postulated by
Wolfgang Pauli which says that no two identical fermions can be put
in the same position at the same time. It is mathematically
expressed by the form of the quantum mechanical wave function for
two identical fermions. The square of the wave function gives the
probability for finding the fermions at points $x_1$ and $x_2$ in
space, at the same time (see Fig. \ref{fig:antisym}). According to
the Pauli exclusion principle, this wave function must be
anti-symmetric under interchanging the two fermion positions.  As a
result the probability for finding two identical fermions at the
same point in space, at the same time, vanishes. This property is in
agreement with the intuitive property of matter.

\begin{figure}[t!]
\hspace{-.70in}
\begin{center}
\epsfig{file=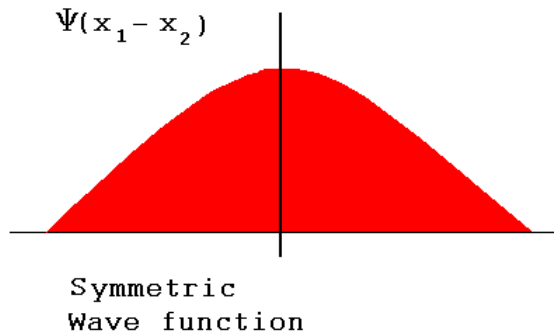,height=2in}
\end{center}
\caption{\label{fig:sym} {\small This figure represents the symmetric wave function for two identical bosons at
positions $x_1, \ x_2$.  Note, due to the symmetry of the wave function, two identical bosons prefer to be at
the same position in space, at the same time.}}

\end{figure}

All the force particles of nature have integral spin.  Photons, the
weak particles,  $W^\pm, \ Z^0$, and strong interaction gluons are
all spin 1. Gravitons are spin 2 and the postulated Higgs particle
has spin 0. According to the physicists Bose and Einstein, all
identical integral spin particles, called Bosons, satisfy
Bose-Einstein statistics.  The quantum mechanical wave function
describing two identical Bosons must be symmetric under interchange
(see Fig. \ref{fig:sym}). Hence it is most probable to find two
identical Bosons at the same position in space at the same time.
Moreover, it is more probable to have any number of Bosons sitting
right on top of each other. In fact, with millions and millions of
photons sitting in the same quantum state one obtains a macroscopic
electric and magnetic field. It is no wonder that all force
particles are Bosons.

The bosonic wave function behaves as ordinary spatial coordinates.
We say that they commute with each other.  On the other hand,
interchanging two fermions changes the sign of the wave function. We
say that they anti-commute,  {\em just like the new anti-commuting
coordinates of supersymmetry}.   In fact,  supersymmetry is a
symmetry which rotates bosonic particles into fermionic ones and
vice versa.   Supersymmetric rotations change the spin and
statistics of the particles,  BUT NOT their other properties.   For
example, if supersymmetry were an exact symmetry, then for every
spin 1/2 electron there would necessarily exist a spin 0 bosonic
electron, a scalar electron or ({\em selectron}).   For every spin 1
photon, there would necessarily exist a spin 1/2 fermionic photon or
({\em photino}).   In the minimal supersymmetric extension of the
Standard Model or [MSSM], every particle has its supersymmetric
partner.   So why would we want to postulate doubling the entire
particle spectrum?   Besides the fact that supersymmetry is the
unique extension of special relativity, i.e. the largest possible
symmetry of nature in four dimensions; supersymmetric theories, as
developed by Wess and Zumino, and Salam and Strathdee, are much
better behaved than ordinary relativistic field theories. As a
result {\em supersymmetry can solve the gauge hierarchy problem}
\cite{witten}. Although an ordinary boson gets quantum mass
corrections which scale like the largest mass in the theory,
fermions on the other hand are protected from such large
corrections.   In fact it was shown long ago that due to the
handedness (or chiral) symmetry of fermions their mass terms receive
quantum corrections which are only logarithmically sensitive to the
highest mass scale in the theory. For example the electron mass at
leading order receives a quantum correction of the form, $m_e
\approx m_0 ( 1 + \log M_{GUT}/m_0 )$. In supersymmetric theories
bosons are protected by the chiral symmetry of their fermionic
partners.  In the MSSM all quarks and leptons have scalar partners,
called squarks and sleptons and all gauge bosons have fermionic
partners, called gauginos. Moreover the Higgs doublets have
fermionic partners, called Higgsinos.   As long as supersymmetry is
unbroken, then the SM particles and their superpartners are
degenerate. Hence the Higgs scalars are protected from large
radiative corrections.  However, once SUSY is spontaneously broken
then scalars will naturally obtain mass of order the SUSY breaking
scale, due to radiative corrections. However, for gauginos to obtain
mass, SUSY as well as the chiral symmetry for gauginos, known as R
symmetry, must be broken.

Including the supersymmetric particle spectrum (at a scale of order
1 TeV, as required to solve the gauge hierarchy problem) in the
renormalization group running of the three low energy fine structure
constants results in the graph of Fig. \ref{fig:guts} (right).
Miraculously the three gauge couplings constants now unify at a GUT
scale of order $3 \times 10^{16}$ GeV \cite{drw}.   As we shall
argue there are three experimental pillars of SUSY GUTs:
\begin{enumerate}
\item Gauge coupling unification,
\item Low energy supersymmetry,
\item Proton decay.
\end{enumerate}
Hence the first experimental pillar of SUSY GUTs stands firm due to
the LEP data from 1991.   The CDF and DZero experiments at the
Tevatron accelerator at Fermi National Accelerator Laboratory in
Batavia, Illinois are searching for the SUSY particles.    AND NOW
the LHC at CERN will begin the search for supersymmetry.   The
discovery of super particles at the Fermilab Tevatron or by the CMS
or ATLAS detectors at the LHC will verify the second pillar of SUSY
GUTs. Supersymmetry is clearly the New Standard Model of the new
millenium.  It is a new symmetry of Nature which most elementary
particle theorists would expect to show up at the TeV energies
accessible to the LHC.  The third pillar, which may take some time,
is the observation of proton decay!   Experiments searching for
proton decay are on-going at SUPER-KAMIOKANDE in Kamioka, Japan.
However, we may have to wait for the next generation of proton decay
experiments, such as the large water \v{c}erenkov detector at the
planned Deep Underground Science and Engineering Laboratory [DUSEL]
in South Dakota.

\subsection{Virtues of SUSY GUTs}

The framework provided by SUSY GUTs can be used to address the
following open problems of the Standard Model. In particular,
\begin{enumerate}
\item it can ``naturally" explain why the weak scale $M_Z << M_{GUT}$;
\item it explains charge quantization, and
\item predicts gauge coupling unification!
\item It predicts SUSY particles at the LHC, and
\item it predicts proton decay.
\end{enumerate}
In addition, it is a natural framework for understanding quark and
lepton masses and it has non-trivial consequences for astrophysics
and cosmology.
\begin{enumerate}
\item It predicts Yukawa coupling unification,
\item and with Family symmetry can explain the family mass
hierarchy.
\item It can accomodate neutrino masses and mixing via the See-Saw
mechanism, and
\item  it can generate a cosmological asymmetry in the number of baryons minus
anti-baryons, i.e. baryogenesis via leptogenesis, via the decay of
the heavy right-handed neutrinos.
\item In the minimal supersymmetric standard model with R
parity, the lightest supersymmetric particle [LSP] is a Dark Matter
candidate.
\end{enumerate}
For all of these reasons we might suspect that SUSY GUTs are a
fundamental component of any realistic string vacuum.   And thus if
one is searching for the MSSM in the mostly barren string landscape,
one should incorporate SUSY GUTs at the first step.   So what do
string theories have to offer?

\subsection{String Theory}

As I explained earlier, the term the {\it Standard Model} is a
severe understatement and misnomer, since the {\bf Standard Model}
has now been verified in millions of ways at multiple laboratories
on earth and in the heavens.   Similarly the term {\it string
theory} is a misnomer and a major overstatement, since there is no
overall theoretical construct which describes in one formalism all
the properties attributed to the many faces of {\bf string theory}.
String theory evolves by studying the several different limiting
forms of the theory, called {\it Type I, Type IIA, Type IIB,  $E_8
\otimes E_8$ and $SO(32)$ heterotic,  M and F theory}. These
limiting forms can all be studied in their respective perturbative
limits.  Then they are all tied together via so-called duality
transformations which relate the solutions of the different
theories.

That said, perturbative string theory starts by studying the
oscillations of a one dimensional vibrating string moving along some
trajectory in space with time.   Type I describes open and closed
strings, while Type II/F and heterotic theories describe the motion
of closed strings.   The space time in which they move is 10
dimensional. Finally,  M theory is an 11 dimensional supersymmetric
field theory whose topological excitations can be related to all the
other pure string theories via specific duality transformations.

The normal oscillation modes of the string are set by the string
tension (or string scale - $M_S$) which is typically taken to be a
scale of order the four dimensional Planck scale - $M_{Pl}$. At
energies below $M_S$ only massless string modes can be excited.
String theories also have the amazing property that they are finite
theories, i.e. all field theoretic divergences are naturally cut-off
at $M_S$.  The effective theory of the massless excitations of the
string is typically described by supergravity field theory
(supersymmetric field theory, plus the supersymmetric version of
Einstein gravity) including all the gauge interactions and matter
multiplets described by the massless modes of the string. In fact,
the amazing property of strings is that a massless spin 2 graviton
(the particle of Einstein gravity) naturally appears in the
spectrum.  This is why Scherk and Schwarz first proposed that the
string scale should satisfy $M_S \sim M_{Pl}$, since $M_S$ sets the
scale for the graviton to couple to matter (recall Newton's constant
of gravity $G_N = M_{Pl}^{-2}$). String theory excitations also
naturally contain massless spin one excitations with properties like
the Standard Model gauge particles, such as the photon, and even
matter multiplets. However the particular massless spectrum of the
string depends in detail on the choice of a {\it string vacuum}. The
problem is that there does not appear to be a unique choice of
string vacuum.  Moreover, in the supersymmetric limit, there are a
continuous infinity of possible string vacua.   And even after
supersymmetry is spontaneously broken,  there are still estimated to
be of order $10^{500}$ string vacua.

So how would one ever expect to be able to find the Standard Model
in this huge landscape of string vacua.   The analog of finding the
Standard Model in the string landscape may be like finding a golf
ball on the surface of the earth.  The surface of the earth is huge
with many mountains, valleys, deserts and oceans and any random
search over this huge landscape will have almost zero probability of
finding a golf ball.   However if the search was not random and one
first found all the golf courses on the earth,  then the probability
of finding a golf ball would increase dramatically!  I propose that
the analog of the golf course for our problem is SUSY GUTs. When one
first looks for SUSY GUTs in the string landscape,  the probability
of finding the Standard Model jumps by many, many orders of
magnitude.

\subsection{Preview}

In the introduction I have tried to provide the basic motivation and
phenomena associated with supersymmetric grand unified theories and
super strings. In the rest of this article I will discuss more
details associated with SUSY GUTs in four dimensions.  I will then
introduce the concept of GUTs in extra dimensions, so-called
orbifold GUTs, which will then take us to the goal of String GUTs.
Or finding the MSSM in the string landscape. The bottom line of this
discussion is the assertion that in order to find the MSSM one must
first find regions of the string landscape containing SUSY GUTs. In
these regions the MSSM becomes natural.

\section{Four dimensional SUSY GUTs and Gauge coupling unification}

 Now that we have identified the states of one family in $SU(5)$, let
us exhibit the fermion Lagrangian (with gauge interactions).  We
have \be {\cal L}_{fermion} = \bar 5_\alpha^\dagger i (\bar
\sigma_\mu \ {D^\mu )_\alpha}^\beta \ \bar 5_\beta  + {10^{\alpha
\beta}}^\dagger i (\bar \sigma_\mu \ D^\mu )^{\alpha \beta}_{\gamma
\delta} \ 10^{\gamma \delta} \ee where \be D^\mu = \partial^\mu + i
g_5 T_A \ A^\mu_A  \ee  and $T_A$ is in the $\bar 5$ or $10$
representation.   We see that since there is only one gauge coupling
constant at the GUT scale we have  \be g_3 = g_2 = g_1 \equiv g_5
\ee  where, after weak scale threshold corrections are included, we
have \be g_3 \rightarrow g_s, \; g_2 \rightarrow  g, \; g_1
\rightarrow \sqrt{\frac{5}{3}}  \ g^\prime . \ee  At the GUT scale
we have the relation \be \sin^2\theta_W = \frac{(g^\prime)^2}{g^2 +
(g^\prime)^2} = 3/8 . \ee

But these are tree level relations which do not take into account
threshold corrections at either the GUT or the weak scales nor
renormalization group [RG] running from the GUT scale to the weak
scale.  Consider first RG running.   The one loop RG equations are
given by \be \frac{d \alpha_i}{d t} = - \frac{b_i}{2 \pi} \alpha_i^2
\ee where $\alpha_i = \frac{g_i^2}{4 \pi},  \; i = 1, 2, 3$ and \be
\label{eq:RGgeneral} b_i = \frac{11}{3} C_2(G_i) - \frac{2}{3} T_R \
N_F - \frac{1}{3} T_R \ N_S . \ee  Note, $t = - \ln
(\frac{M_G}{\mu})$,  $\sum_A (T_A^2) = C_2(G_i) \mathbb{I}$ with
$T_A$ in the adjoint representation defines the quadratic Casimir
for the group $G_i$ with $C_2(SU(N)) = N$ and $C_2(U(1)) = 0$.
$Tr(T_A T_B) = T_R \ \delta_{A B}$ for $T_A$ in the representation
$R$ (for $U(1)_Y$, $T_R \equiv \frac{3}{5} \ Tr( \frac{Y^2}{4} )$)
and $N_F (N_S)$ is the number of Weyl fermions (complex scalars) in
representation $R$. For N = 1 supersymmetric theories, Equation
\ref{eq:RGgeneral} can be made more compact.   We have \be
\label{eq:RGsusy} b_i = 3 C_2(G_i) - T_R \ N_\chi  \ee where the
first term takes into account the vector multiplets and $N_\chi$ is
the number of left-handed chiral multiplets in the representation
$R$ \cite{drw,2loop}.  The solution to the one loop RG equations is
given by \be \alpha_i(M_Z)^{-1} = \alpha_G^{-1} - \frac{b_i}{2 \pi}
\ln(\frac{M_G}{M_Z}) . \ee

For the SM we find \be  {\bf b}_{SM} \equiv (b_1, b_2, b_3) =
(-\frac{4}{3} N_{fam} - \frac{1}{10} N_H, \frac{22}{3} - \frac{4}{3}
N_{fam} - \frac{1}{6} N_H, 11 - \frac{4}{3} N_{fam})  \ee where
$N_{fam} (N_H)$ is the number of families (Higgs doublets). For SUSY
we have \be {\bf b}_{SUSY} = (-2 N_{fam} - \frac{3}{5}
N_{(H_u+H_d)}, 6 - 2 N_{fam} - N_{(H_u+H_d)}, 9 - 2 N_{fam} ) \ee
where $N_{(H_u+H_d)}$ is the number of pairs of Higgs doublets. Thus
for the MSSM we have  \be {\bf b}^{MSSM} = (- 33/5, -1, 3) . \ee

The one loop equations can be solved for the value of the GUT scale
$M_G$ and $\alpha_G$ in terms of the values of $\alpha_{EM}(M_Z)$
and $\sin^2\theta_W(M_Z)$.  We have (without including weak scale
threshold corrections)  \be \alpha_2(M_Z) =
\frac{\alpha_{EM}(M_Z)}{\sin^2\theta_W(M_Z)}, \; \alpha_1(M_Z) =
\frac{\frac{5}{3} \alpha_{EM}(M_Z)}{\cos^2\theta_W(M_Z)}  \ee and we
find  \be (\frac{3}{5} -\frac{8}{5} \ \sin^2\theta_W(M_Z))
\alpha_{EM}(M_Z)^{-1} = (\frac{b^{MSSM}_2 - b^{MSSM}_1}{2 \pi}) \ln
(\frac{M_G}{M_Z}) \ee which we use to solve for $M_G$.  Then we use
\be  \alpha_G^{-1} = \sin^2\theta_W(M_Z) \ \alpha_{EM}(M_Z)^{-1} +
\frac{b^{MSSM}_2}{2 \pi} \ln (\frac{M_G}{M_Z}) \ee to solve for
$\alpha_G$. We can then predict the value for the strong coupling
using \be \alpha_3(M_Z)^{-1} = \alpha_G^{-1} - \frac{b^{MSSM}_3}{2
\pi} \ln (\frac{M_G}{M_Z}). \ee

Given the experimental values  $\sin^2\theta_W(M_Z) \approx .23$ and
$\alpha_{EM}(M_Z)^{-1} \approx 128$ we find $M_G \approx 1.3 \times
10^{13}$ GeV with $N_H =1$ and $\alpha_G^{-1} \approx 42$ for the SM
with the one loop prediction for $\alpha_3(M_Z) \approx 0.07$. On
the other hand, for SUSY we find $M_G \approx 2.7 \times 10^{16}$
GeV,  $\alpha_G^{-1} \approx 24$ and the predicted strong coupling
$\alpha_3(M_Z) \approx 0.12$.  How well does this agree with the
data?  According to the PDG the average value of $\alpha_s(M_Z) =
0.1176 \pm 0.002$ \cite{pdg}.  So at one loop the MSSM is quite
good, while non-SUSY GUTs are clearly excluded.

At the present date,  the MSSM is compared to the data using 2 loop
RG running from the weak to the GUT scale with one loop threshold
corrections included at the weak scale.  These latter corrections
have small contributions from integrating out the W, Z, and top
quark. But the major contribution comes from integrating out the
presumed SUSY spectrum.  With a ``typical" SUSY spectrum and
assuming no threshold corrections at the GUT scale, one finds a
value for $\alpha_s(M_Z) \ge 0.127$ which is too large
\cite{Langacker:1995fk}.   It is easy to see where this comes from
using the approximate analytic formula \be \alpha_i^{-1}(M_Z) =
\alpha_G^{-1} - \frac{b^{MSSM}_i}{2 \pi} \ln(\frac{M_G}{M_Z}) +
\delta_i \ee where \be \delta_i = \delta_i^h + \delta_i^2 +
\delta_i^l . \ee The constants $\delta_i^2, \ \delta_i^l, \
\delta_i^h$ represent the 2 loop running effects \cite{2loop}, the
weak scale threshold corrections and the GUT scale threshold
corrections, respectively.  We have \be \delta_i^2 \approx -
\frac{1}{\pi} \sum_{j=1}^3  \frac{b^{MSSM}_{i j}}{b^{MSSM}_j} \log
\left[ 1 - b^{MSSM}_j \left( \frac{3 - 8 \sin^2\theta_W}{36
\sin^2\theta_W - 3} \right) \right] \ee where the matrix $
b^{MSSM}_{i j}$ is given by \cite{2loop} \be b^{MSSM}_{i j} = \left(
\begin{array}{ccc} \frac{199}{100} & \frac{27}{20} & \frac{22}{5} \\
 \frac{9}{20} & \frac{25}{4} & 6 \\ \frac{11}{20} & \frac{9}{4} &
 \frac{7}{2} \end{array} \right).  \ee   The light thresholds are
 given by \be \delta_i^l = \frac{1}{\pi} \sum_j {b^{l}_i}(j) \log
 (\frac{m_j}{M_Z} ) \ee where the sum runs over all states at the
 weak scale including the top, $W$, Higgs and the supersymmetric
 spectrum.  Finally the GUT scale threshold correction is given by
 \be \delta_i^h = - \frac{1}{2 \pi} \sum_\zeta b_i^\zeta \log (\frac{M_\zeta}{M_G}) . \ee

 In general the prediction for $\alpha_3(M_Z)$ is given by \bea
 \alpha_3^{-1}(M_Z) & = (\frac{b_3-b_1}{b_2-b_1}) \alpha_2^{-1}(M_Z) -
 (\frac{b_3-b_2}{b_2-b_1}) \alpha_1^{-1}(M_Z)  & \nn \\
 & + (\frac{b_3-b_2}{b_2-b_1}) \delta_1 - (\frac{b_3-b_1}{b_2-b_1})
 \delta_2 + \delta_3 & \nn \\ & = \frac{12}{7} \alpha_2^{-1}(M_Z) -
 \frac{5}{7} \alpha_1^{-1}(M_Z) + \frac{1}{7} (5 \delta_1 - 12
 \delta_2 + 7 \delta_3) & \nn \\ &  \equiv (\alpha_3^{LO})^{-1} + \delta_s  &\eea
 where  $b_i \equiv b^{MSSM}_i$,  $\alpha_3^{LO}(M_Z)$ is the leading
 order one-loop result and $\delta_s \equiv  \frac{1}{7} (5 \delta_1 - 12
 \delta_2 + 7 \delta_3)$.  We find $\delta_s^2 \approx -0.82$ (Ref. \cite{Alciati:2005ur}) and
 $\delta_s^l = -0.04 + \frac{19}{28 \pi} \ln( \frac{T_{SUSY}}{M_Z})$
 where the first term takes into account the contribution of the
 $W$, top and the correction from switching from the $\overline{MS}$
 to $\overline{DR}$ RG schemes and
 (following Ref.  \cite{Carena:1993ag})
 \be \label{eq:Tsusy} T_{SUSY} = m_{\tilde H} (\frac{m_{\tilde W}}{m_{\tilde g}})^{28/19}
 \left[ (\frac{m_{\tilde l}}{m_{\tilde q}})^{3/19}
 (\frac{m_{H}}{m_{\tilde H}})^{3/19} (\frac{m_{\tilde W}}{m_{\tilde H}})^{4/19} \right]
 .\ee  For a Higgsino mass $m_{\tilde H} = 400$ GeV, a Wino mass $m_{\tilde W} = 300$ GeV, a gluino mass $m_{\tilde g} = 900$
 GeV and all other mass ratios of order one, we find $\delta_s^l
 \approx -0.12$.  If we assume $\delta_s^h = 0$, we find the
 predicted value of $\alpha_3(M_Z) = 0.135$.  In order to obtain a
 reasonable value of $\alpha_3(M_Z)$ with only weak scale threshold
 corrections, we need
$\delta_s^2 + \delta_s^l \approx 0$ corresponding to a value of
$T_{SUSY} \sim 5$ TeV.   But this is very difficult considering the
weak dependence $T_{SUSY}$ (Eqn. \ref{eq:Tsusy}) has on squark and
slepton masses. Thus in order to have $\delta_s \approx 0$ we need a
GUT scale threshold correction \be \label{eq:deltah} \delta_s^h
\approx + 0.94 . \ee

At the GUT scale we have \be \alpha_i^{-1}(M_G) = \alpha_G^{-1} +
\delta_i^h . \ee  Define \be  \tilde \alpha_G^{-1} = \frac{1}{7} [12
\alpha_2^{-1}(M_G) - 5 \alpha_1^{-1}(M_G) ] \ee (or if the GUT scale
is defined at the point where $\alpha_1$ and $\alpha_2$ intersect,
then $\tilde \alpha_G \equiv \alpha_1(M_G) = \alpha_2(M_G)$.  Hence,
in order to fit the data, we need a GUT threshold correction \be
\epsilon_3 \equiv \frac{\alpha_3(M_G) - \tilde \alpha_G}{\tilde
\alpha_G} = - \tilde \alpha_G \ \delta_s^h \approx  - 4 \%. \ee
Note, this result depends implicitly on the assumption of universal
soft SUSY breaking masses at the GUT scale, which directly affect
the spectrum of SUSY particles at the weak scale.  For example, if
gaugino masses were not unified at $M_{G}$ and, in particular,
gluinos were lighter than winos at the weak scale, then it is
possible that, due to weak scale threshold corrections,  a much
smaller or even slightly positive threshold correction at the GUT
scale would be consistent with gauge coupling unification
\cite{Raby:2009sf}.

\subsection{Nucleon Decay}

Baryon number is necessarily violated in any GUT~\cite{grs}.  In
$SU(5)$ nucleons decay via the exchange of gauge bosons with GUT
scale masses, resulting in dimension 6 baryon number violating
operators suppressed by $(1/M_G^2)$.  The nucleon lifetime is
calculable and given by $\tau_N \propto M_G^4/(\alpha_G^2 \;
m_p^5)$.  The dominant decay mode of the proton (and the baryon
violating decay mode of the neutron), via gauge exchange, is $p
\rightarrow e^+ \; \pi^0$ ($n \rightarrow e^+ \; \pi^-$).  In any
simple gauge symmetry, with one universal GUT coupling and scale
($\alpha_G, \; M_G$), the nucleon lifetime from gauge exchange is
calculable.  Hence, the GUT scale may be directly observed via the
extremely rare decay of the nucleon.   In SUSY GUTs, the GUT scale
is of order $3\times 10^{16}$ GeV, as compared to the GUT scale in
non-SUSY GUTs which is of order $10^{15}$ GeV. Hence the dimension 6
baryon violating operators are significantly suppressed in SUSY
GUTs~\cite{drw} with $\tau_p \sim 10^{34 - 38}$ yrs.

However, in SUSY GUTs there are additional sources for baryon number
violation -- dimension 4 and 5 operators~\cite{bviol}.   Although
our notation does not change,  when discussing SUSY GUTs all fields
are implicitly bosonic superfields and the operators considered are
the so-called F terms which contain two fermionic components and the
rest scalars or products of scalars. Within the context of $SU(5)$
the dimension 4 and 5 operators have the form $({\bf 10 \; \bar 5 \;
\bar 5})$ $\supset  (\bar U \; \bar D \; \bar D)  +  (Q \; L \; \bar
D) + (\bar E \; L \; L)$ and $({\bf 10 \; 10 \; 10 \; \bar 5})$
$\supset (Q \; Q\; Q\; L)  +  (\bar U \; \bar U \;  \bar D \; \bar
E) \; + $ $B$ and $L$ conserving terms, respectively.   The
dimension 4 operators are renormalizable with dimensionless
couplings; similar to Yukawa couplings.   On the other hand, the
dimension 5 operators have a dimensionful coupling of order
($1/M_G$).

The dimension 4 operators violate baryon number or lepton number,
respectively, but not both.  The nucleon lifetime is extremely short
if both types of dimension 4 operators are present in the low energy
theory.  However both types can be eliminated by requiring R parity.
In $SU(5)$ the Higgs doublets reside in a ${\bf 5_H,\; \bar 5_H}$
and R parity distinguishes the ${\bf \bar 5}$ (quarks and leptons)
from ${\bf \bar 5_H}$ (Higgs).  R parity~\cite{rparity} (or its
cousin, family reflection symmetry (see Dimopoulos and
Georgi~\cite{drw}and DRW ~\cite{drw2}) takes  $F \rightarrow -F, \;
H \rightarrow H$ with $F = \{ {\bf 10,\;  \bar 5} \}, \; H = \{ {\bf
\bar 5_H,\; 5_H} \}$. This forbids the dimension 4 operator $({\bf
10 \; \bar 5 \; \bar 5})$, but allows the Yukawa couplings of the
form $({\bf 10 \; \bar 5 \; \bar 5_H})$ and $({\bf 10 \; 10 \;
5_H})$.   It also forbids the dimension 3, lepton number violating,
operator $({\bf \bar 5 \; 5_H})$ $\supset (L \; H_u)$ with a
coefficient with dimensions of mass which, like the $\mu$ parameter,
could be of order the weak scale and the dimension 5, baryon number
violating, operator $({\bf 10 \; 10 \; 10 \; \bar 5_H})$ $\supset (Q
\; Q\; Q\; H_d) + \cdots$.

Note, in the MSSM it is possible to retain R parity violating
operators at low energy as long as they violate either baryon number
or lepton number only but not both.  Such schemes are natural if one
assumes a low energy symmetry, such as lepton number, baryon number
or a baryon parity~\cite{ir}.    However these symmetries cannot be
embedded in a GUT. Thus, in a SUSY GUT, only R parity can prevent
unwanted dimension four operators.  Hence, by naturalness arguments,
R parity must be a symmetry in the effective low energy theory of
any SUSY GUT. This does not mean to say that R parity is guaranteed
to be satisfied in any GUT.  A possible exception to this rule using
constrained matter content which generates the effective R parity
violating operators in a GUT can be found in
\cite{Barbieri:1997zn,Giudice:1997wb} or for a review on R parity
violating interactions, see \cite{Barbier:2004ez}.   For example, in
Ref. \cite{Giudice:1997wb}, the authors show how to obtain the
effective R parity violating operator $O^{ijk} = (\bar 5^j \cdot
\bar 5^k)_{\overline{15}} \cdot (10^i \cdot \Sigma)_{15}$ where
$\Sigma$ is an $SU(5)$ adjoint field and the subscripts
$\overline{15}, 15$ indicate that the product of fields in
parentheses have been projected into these $SU(5)$ directions.  As a
consequence the operator $O^{ijk}$ is symmetric under interchange of
the two $\bar 5$ states, $O^{ijk} = O^{ikj}$, and out of ${\bf 10 \
\bar 5 \ \bar 5}$ only the lepton number/R parity violating operator
$Q L \bar D$ survives.

R parity also distinguishes Higgs multiplets from ordinary families.
In $SU(5)$, Higgs and quark/lepton multiplets have identical quantum
numbers; while in $E(6)$, Higgs and families are unified within the
fundamental ${\bf 27}$ representation. Only in SO(10) are Higgs and
ordinary families distinguished by their gauge quantum numbers.
Moreover the $\mathbb{Z}_4$ center of $SO(10)$ distinguishes ${\bf
10}$s from ${\bf 16}$s and can be associated with R
parity~\cite{senjanovic}.

\begin{figure} \begin{center}
\includegraphics[height=.1\textheight]{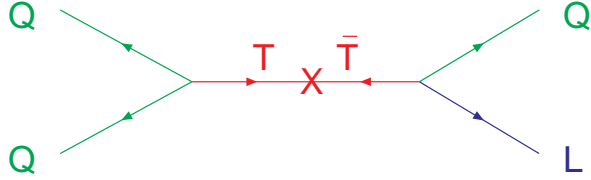} \end{center}
\caption{\label{fig:dim5op} The effective dimension 5 operator for
proton decay.}
\end{figure}

Dimension 5 baryon number violating operators may be forbidden at
tree level by symmetries in an $SU(5)$ model, etc.  These symmetries
are typically broken however by the VEVs responsible for the color
triplet Higgs masses. Consequently these dimension 5 operators are
generically generated via color triplet Higgsino exchange (Fig.
\ref{fig:dim5op}). Hence, the color triplet partners of Higgs
doublets must necessarily obtain mass of order the GUT scale.  [It
is also important to note that Planck or string scale physics may
independently generate dimension 5 operators, even without a GUT.
These contributions must be suppressed by some underlying symmetry.
See the discussion in Section \ref{sec:protondecay}.]

The dominant decay modes from dimension 5 operators are $p
\rightarrow K^+ \; \bar \nu \;\; (n \rightarrow K^0 \; \bar \nu)$.
This is due to a simple symmetry argument;  the operators $(Q_i \;
Q_j\; Q_k\; L_l), \;\; (\bar U_i \; \bar U_j\; \bar D_k\; \bar E_l)$
(where $i,\; j,\; k,\; l = 1,2,3$ are family indices and color and
weak indices are implicit) must be invariant under $SU(3)_C$ and
$SU(2)_L$.  As a result their color and weak doublet indices must be
anti-symmetrized.  However since these operators are given by
bosonic superfields, they must be totally symmetric under
interchange of all indices. Thus the first operator vanishes for $i
= j = k$ and the second vanishes for $i = j$.  Hence a second or
third generation particle must appear in the final
state~\cite{drw2}.

\begin{figure} \begin{center}
\includegraphics[height=.13\textheight]{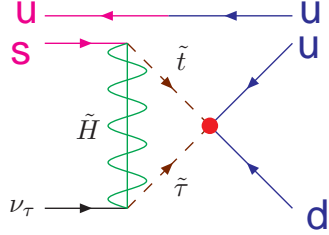} \end{center}
\caption{\label{fig:pdecay} The effective four fermi operator for
proton decay obtained by integrating out sparticles at the weak
scale.}
\end{figure}

The dimension 5 operator contribution to proton decay requires a
sparticle loop at the SUSY scale to reproduce an effective dimension
6 four fermi operator for proton decay (see Fig. \ref{fig:pdecay}).
The loop factor is of the form  \be (LF) \propto \frac{\lambda_t \;
\lambda_\tau}{16 \pi^2} \frac{\sqrt{\mu^2 + M_{1/2}^2}}{m_{16}^2}
\ee leading to a decay amplitude  \be A(p \rightarrow K^+ \bar \nu)
\propto  \frac{c \ c}{M_T^{eff}} \ (\rm LF),  \ee where $M_T^{eff}$
is the effective Higgs color triplet mass (see Fig.
\ref{fig:dim5op}), $M_{1/2}$ is a universal gaugino mass and
$m_{16}$ is a universal mass for squarks and sleptons. In any
predictive SUSY GUT, the coefficients $c$ are 3 $\times$ 3 matrices
related to (but not identical to) Yukawa matrices.  Thus these tend
to suppress the proton decay amplitude. However this is typically
not sufficient to be consistent with the experimental bounds on the
proton lifetime. Thus it is also necessary to minimize the loop
factor, (LF).   This can be accomplished by taking $\mu, M_{1/2}$
{\em small} and $m_{16}$ {\em large}.   Finally the effective Higgs
color triplet mass $M_T^{eff}$ must be MAXIMIZED. With these
caveats, it is possible to obtain rough theoretical bounds on the
proton lifetime given by \cite{Lucas:1995ic,Altarelli:2000fu,dmr}
\be \tau_{p \rightarrow K^+ \bar \nu} \leq ( \frac{1}{3} - 3 )
\times 10^{34} \; {\rm yrs.} . \ee

\subsubsection{Gauge Coupling Unification and Proton Decay}

The dimension 5 operator (see Fig. \ref{fig:dim5op}) is given in
terms of the matrices $c$ and an effective Higgs triplet mass by \be
\frac{1}{M^{eff}_T} \left[ \ Q \ \frac{1}{2} \ c_{qq} Q \ Q \ c_{ql}
L + \overline U \ c_{ud} \overline D \ \overline U \ c_{ue}
\overline E \ \right] . \ee Note, $M^{eff}_T$ can be much greater
than $M_G$ without fine-tuning and without having any particle with
mass greater than the GUT scale. Consider a theory with two pairs of
Higgs $5_i$ and $\bar 5_i$ with $i = 1,2$ at the GUT scale with only
$5_1, \ \bar 5_1$ coupling to quarks and leptons. Then we have \be
\frac{1}{M^{eff}_T} = ( M_T^{-1} )_{11} . \ee

If the Higgs color triplet mass matrix is given by  \be M_T = \left(
\begin{array}{cc} 0 &  M_G \\  M_G & X
\end{array} \right) \ee then we have \be \frac{1}{M^{eff}_T} \equiv
\frac{X}{M_G^2}. \ee  Thus for $X << M_G$ we obtain $M^{eff}_T
>> M_G$.

We assume that the Higgs doublet mass matrix, on the other hand, is
of the form \be M_D = \left(
\begin{array}{cc} 0 &  0 \\  0 & X
\end{array} \right) \ee  with two light Higgs doublets.   Note this
mechanism is natural in $S0(10)$
\cite{Dimopoulos:1981xm,Babu:1993we} with a superpotential of the
form   \be W \supset 10 \ 45 \ 10^\prime + X \ (10^\prime)^2 \ee
with only $10$ coupling to quarks and leptons, $X$ is a gauge
singlet and $\langle 45 \rangle = (B-L) \ M_G$.

Recall $\epsilon_3 \equiv \frac{(\alpha_3(M_G) - \tilde
\alpha_G)}{\tilde \alpha_G} \sim - 4\%$.   At one loop we find \be
\epsilon_3 = \epsilon_3^{\rm Higgs} + \epsilon_3^{\rm GUT \;
breaking} + \cdots . \ee  Moreover \be \epsilon_3^{\rm Higgs} =
\frac{3 \alpha_G}{5 \pi} \ln(\frac{M^{eff}_T}{M_G}) . \ee  See Table
\ref{tab:ep3} for the contribution to $\epsilon_3$ in Minimal SUSY
$SU(5)$, and in an $SU(5)$ and $SO(10)$ model with natural Higgs
doublet-triplet splitting.

\begin{table} \caption{Contribution to $\epsilon_3$ in three different GUT
models.} \label{tab:ep3}
$$
\begin{array}{|l|c|c|c|} \hline
{\rm Model}  &   {\rm Minimal}   & {SU_5}  & {\rm Minimal}  \\
& SU_5 & {\rm ``Natural" D/T} $\cite{Altarelli:2000fu}$ & SO(10) $\cite{dmr}$ \\
\hline
\epsilon_3^{\rm GUT breaking} &  0  &  - 7.7 \%  &  -10 \% \\
\epsilon_3^{\rm Higgs} &  - 4 \% & + 3.7 \% &  + 6 \% \\
\hline M^{eff}_T {\rm [GeV]} &  2 \times 10^{14} & 3 \times 10^{18}
& 6 \times
10^{19} \\
\hline
\end{array}$$
\end{table}

Recent Super-Kamiokande bounds on the proton lifetime severely
constrain these dimension 6 and 5 operators with $\tau_{(p
\rightarrow e^+ \pi^0)} > 1.0 \times 10^{34}$ yrs (172.8 ktyr) at
(90\% CL) \cite{ichep2010}, and $\tau_{(p \rightarrow  K^+ \bar
\nu)}
> 2.3 \times 10^{33}$ yrs (92 ktyr) at (90\% CL) based on the listed
exposures~\cite{superk}. These constraints are now sufficient to
rule out minimal SUSY $SU(5)$~\cite{murayama}.\footnote{Note,  I
have implicitly assumed a hierarchical structure for Yukawa matrices
in this analysis.  It is however possible to fine-tune a
hierarchical structure for quarks and leptons which baffles the
family structure.  In this case it is possible to avoid the present
constraints on minimal SUSY SU(5), for example see
\cite{Choi:2008zw}.  It is also possible to ameliorate the conflict
if one forgoes universal gaugino masses at the GUT scale and gluinos
are lighter than winos at the weak scale \cite{Raby:2009sf}.} The
upper bound on the proton lifetime from these theories (particularly
from dimension 5 operators) is approximately a factor of 5 above the
experimental bounds. These theories are also being pushed to their
theoretical limits.  Hence if SUSY GUTs are correct, then nucleon
decay must be seen soon.

\subsection{Yukawa coupling unification}
\subsubsection{3rd generation, $b - \tau$  or $t - b - \tau$ unification}

In $SU(5)$, there are two independent renormalizable Yukawa
interactions given by  $\lambda_t \; ({\bf 10 \; 10 \;  5_H}) $ $+
\; \lambda \; ({\bf 10 \; \bar 5 \; \bar 5_H})$.   These contain the
SM interactions $\lambda_t \; ({\bf Q \; \bar u \; H_u})$  $ + \;
\lambda \; ({\bf Q \; \bar d \;  H_d} \; + \; {\bf \bar e \; L \;
H_d})$. Hence, at the GUT scale we have the tree level relation,
$\lambda_b = \lambda_\tau \equiv \lambda$~\cite{chanowitz}.   In
$SO(10)$ (or Pati-Salam) there is only one independent
renormalizable Yukawa interaction given by $\lambda \; ({\bf 16 \;
16 \;  10_H})$ which gives the tree level relation, $\lambda_t =
\lambda_b = \lambda_\tau \equiv
\lambda$~\cite{so10yuk,hrr,so10yuksusy}. Note, in the discussion
above we assume the minimal Higgs content with Higgs in ${\bf 5,\;
\bar 5}$ for $SU(5)$ and ${\bf 10}$ for $SO(10)$.  With Higgs in
higher dimensional representations there are more possible Yukawa
couplings~\cite{Lazarides:1980nt,Bajc:2002iw,Goh:2003sy}.

In order to make contact with the data, one now renormalizes the
top, bottom and $\tau$ Yukawa couplings, using two loop RG
equations, from $M_G$ to $M_Z$.   One then obtains the running quark
masses  $m_t(M_Z)\; = \;\lambda_t(M_Z)\; v_u$, $\;\; m_b(M_Z)\;  =\;
\lambda_b(M_Z)\; v_d$ and   $\; m_\tau(M_Z)\; =\;
\lambda_\tau(M_Z)\; v_d$ where $<H_u^0> \equiv v_u = \sin\beta \;
v/\sqrt{2}$, $<H_d^0> \equiv v_d = \cos\beta\; v/\sqrt{2}$, $v_u/v_d
\equiv \tan\beta$ and  $v \sim 246$ GeV is fixed by the Fermi
constant, $G_{\mu}$.

Including one loop threshold corrections at $M_Z$ and additional RG
running, one finds the top, bottom and $\tau$ pole masses. In SUSY,
$b - \tau$ unification has two possible solutions with $\tan\beta
\sim 1$ or $40 - 50$. The small $\tan\beta$ solution is now
disfavored by the LEP limit, $\tan\beta > 2.4$~\cite{lep}. However,
this bound disappears if one takes $M_{SUSY} = 2$ TeV and $m_t =
180$ GeV~\cite{Carena:2002es}. The large $\tan\beta$ limit overlaps
the $SO(10)$ symmetry relation.

When  $\tan\beta$ is large there are significant weak scale
threshold corrections to down quark and charged lepton masses from
either gluino and/or chargino loops~\cite{yukawacorr}. Yukawa
unification (consistent with low energy data) is only possible in a
restricted region of SUSY parameter space with important
consequences for SUSY searches~\cite{bdr}.

Consider a minimal $SO(10)$ SUSY model [MSO$_{10}$SM]~\cite{bdr}.
Quarks and leptons of one family reside in the $\bf 16$ dimensional
representation, while the two Higgs doublets of the MSSM reside in
one $\bf 10$ dimensional representation.  For the third generation
we assume the minimal Yukawa coupling term given by $ {\bf \lambda \
16 \ 10 \ 16 }. $ On the other hand, for the first two generations
and for their mixing with the third, we assume a hierarchical mass
matrix structure due to effective higher dimensional operators.
Hence the third generation Yukawa couplings satisfy $\lambda_t =
\lambda_b = \lambda_\tau = \lambda_{\nu_\tau} = {\bf \lambda}$.
Note, that these relations are only approximate once the
off-diagonal elements of the Yukawa matrices are taken into account.
If the Yukawa matrices are hierarchical then the corrections are
typically $\leq$ 10\%.

Soft SUSY breaking parameters are also consistent with $SO(10)$ with
(1) a universal gaugino mass $M_{1/2}$, (2) a universal squark and
slepton mass $m_{16}$, [$SO(10)$ does not require all sfermions to
have the same mass. This however may be enforced by non--abelian
family symmetries or possibly by the SUSY breaking mechanism.] (3) a
universal scalar Higgs mass $m_{10}$, and (4) a universal A
parameter $A_0$. In addition we have the supersymmetric (soft SUSY
breaking) Higgs mass parameters $\mu$ ($B \mu$).  $B \mu$ may, as in
the CMSSM, be exchanged for $\tan\beta$. Note, not all of these
parameters are independent. Indeed, in order to fit the low energy
electroweak data, including the third generation fermion masses, it
has been shown that $A_0, \ m_{10}, \ m_{16}$ must satisfy the
constraints~\cite{bdr}
\bea A_0 \approx - 2 \ m_{16}; & m_{10} \approx \sqrt{2} \ m_{16}  
\label{eq:constraint1}
\\
m_{16} > 1.2 \; {\rm TeV}; & \mu, \ M_{1/2} \ll m_{16} 
\label{eq:constraint2} \eea with \be \tan\beta \approx 50.
\label{eq:tanbeta} \ee
This result has been confirmed by several independent
analyses~\cite{Tobe:2003bc,Auto:2003ys,Baer:2008jn}.  Note,
different regions of parameter space consistent with Yukawa
unification have also been discussed
in~\cite{Tobe:2003bc,Auto:2003ys,Balazs:2003mm}. Although the
conditions (Eqns.~\ref{eq:constraint1}, \ref{eq:constraint2}) are
not obvious, it is however easy to see that
(Eqn.~(\ref{eq:tanbeta})) is simply a consequence of third
generation Yukawa unification, since $m_t(m_t)/m_b(m_t) \sim
\tan\beta$.

Finally, as a bonus, these same values of soft SUSY breaking
parameters, with $m_{16} \gg$ TeV, result in two very interesting
consequences.  Firstly, it ``naturally" produces an inverted scalar
mass hierarchy [ISMH]~\cite{scrunching}. With an ISMH, squarks and
sleptons of the first two generations obtain mass of order $m_{16}$
at $M_Z$. The stop, sbottom, and stau, on the other hand, have mass
less than (or of order) a TeV. An ISMH has two virtues. (1) It
preserves ``naturalness" (for values of $m_{16}$ which are not too
large), since only the third generation squarks and sleptons couple
strongly to the Higgs. (2) It ameliorates the SUSY CP and flavor
problems, since these constraints on CP violating angles or flavor
violating squark and slepton masses are strongest for the first two
generations, yet they are suppressed as $1/m_{16}^{2}$.  For $m_{16}
> $ a few TeV, these constraints are
weakened~\cite{masieroetal}.  Secondly, Super--Kamiokande bounds on
$\tau(p \rightarrow K^+ \bar \nu) \ > 2.3 \times 10^{33}$
yrs~\cite{superk} constrain the contribution of dimension 5 baryon
and lepton number violating operators. These are however minimized
with $\mu, \ M_{1/2} \ll m_{16}$~\cite{dmr}.

\subsubsection{Three families}

Simple Yukawa unification is not possible for the first two
generations of quarks and leptons.  Consider the $SU(5)$ GUT scale
relation $\lambda_b = \lambda_\tau$.  If extended to the first two
generations one would have $\lambda_s = \lambda_\mu$, $\lambda_d =
\lambda_e$ which gives $\lambda_s/\lambda_d =
\lambda_\mu/\lambda_e$. The last relation is a renormalization group
invariant and is thus satisfied at any scale. In particular, at the
weak scale one obtains $m_s/m_d = m_\mu/m_e$ which is in serious
disagreement with the data with $m_s/m_d \sim 20$ and $m_\mu/m_e
\sim 200$.  An elegant solution to this problem was given by Georgi
and Jarlskog~\cite{gj}. Of course, a three family model must also
give the observed CKM mixing in the quark sector. Note, although
there are typically many more parameters in the GUT theory above
$M_G$, it is possible to obtain effective low energy theories with
many fewer parameters making strong predictions for quark and lepton
masses.

It is important to note that grand unification alone is not
sufficient to obtain predictive theories of fermion masses and
mixing angles.  Other ingredients are needed.  In one approach
additional global family symmetries are introduced (non-abelian
family symmetries can significantly reduce the number of arbitrary
parameters in the Yukawa matrices).  These family symmetries
constrain the set of effective higher dimensional fermion mass
operators.  In addition, sequential breaking of the family symmetry
is correlated with the hierarchy of fermion masses. Three-family
models exist which fit all the data, including neutrino masses and
mixing~\cite{guts}.  In a completely separate approach for $SO(10)$
models, the Standard Model Higgs bosons are contained in the higher
dimensional Higgs representations including the {\bf 10},
$\overline{\bf 126}$ and/or {\bf 120}.   Such theories have been
shown to make predictions for neutrino masses and mixing
angles~\cite{Lazarides:1980nt,Bajc:2002iw,Goh:2003sy}.   Some simple
patterns of fermion masses (see Table \ref{tab:patterns}) must be
incorporated into any successful model.

\begin{table} \begin{center} \caption{Patterns of Masses and Mixing}
\begin{tabular}{|ll|}  \hline
$\lambda_t = \lambda_b = \lambda_\tau = \lambda_{\nu_\tau}$ &
$SO(10) @ M_G$  \\
\hline $ \lambda_s \sim \frac{1}{3} \lambda_\mu , \;\; \lambda_d
\sim 3 \lambda_e $ & $@ M_G $ \cite{gj,hrr} \\
$ m_s \approx 4 \cdot \frac{1}{3} m_\mu ,  \;\; m_d \approx 4 \cdot
3 m_e $ & $ @ M_Z $ \\ \hline $ \lambda_d \lambda_s \lambda_b
\approx
\lambda_e \lambda_\mu \lambda_\tau$ & $SU(5) @ M_G$  \\
$ Det(m_d) \approx  Det(m_e)$ & $ @ M_G $ \\ \hline $ V_{us} \approx
(\sqrt{m_d/m_s} - i \sqrt{m_u/m_c})$ & \cite{fritzsch,Kim:2004ki} \\
$V_{ub}/V_{cb} \approx \sqrt{m_u/m_c}$ & \cite{Hall:1993ni} \\
$ V_{cb}  \sim  m_s/m_b \sim \sqrt{m_c/m_t}$ & \cite{hrr}
\\ \hline
\end{tabular} \label{tab:patterns}   \end{center}
\end{table}

\subsection{Neutrino Masses}

Atmospheric and solar neutrino oscillations require neutrino masses.
Using the three ``sterile" neutrinos $\bar \nu$ with the Yukawa
coupling $\lambda_\nu \; ({\bf \bar \nu \; L \; H_u})$, one easily
obtains three massive Dirac neutrinos with mass $m_\nu = \lambda_\nu
\; v_u$. Note, these ``sterile" neutrinos are quite naturally
identified with the right-handed neutrinos necessarily contained in
complete families of $SO(10)$ or Pati-Salam. However in order to
obtain a tau neutrino with mass of order $ 0.1\; {\rm eV} $, one
needs $\lambda_{\nu_\tau}/\lambda_\tau \leq 10^{-10}$. The see-saw
mechanism, on the other hand, can naturally explain such small
neutrino masses~\cite{minkowski,yanagida}. Since $\bar \nu$ has no
$SU(3)_{color} \times SU(2)_L \times U(1)_Y$ quantum numbers, there
is no symmetry (other than global lepton number) which prevents the
mass term $\frac{1}{2} \; \bar \nu \; M \; \bar \nu$. Moreover one
might expect $M \sim M_G$. Heavy ``sterile" neutrinos can be
integrated out of the theory, defining an effective low energy
theory with only light active Majorana neutrinos with the effective
dimension 5 operator $\frac{1}{2} \; ({\bf L\;  H_u}) \;
\lambda_\nu^T \; M^{-1} \; \lambda_\nu \; ({\bf L\; H_u})$.  This
then leads to a $3 \times 3$ Majorana neutrino mass matrix ${\bf m}
= m_\nu^T \; M^{-1} \; m_\nu$.

Atmospheric neutrino oscillations require neutrino masses with
$\Delta m_\nu^2 \sim 3 \times 10^{-3}$ eV$^2$ with maximal mixing,
in the simplest two neutrino scenario.  With hierarchical neutrino
masses $m_{\nu_\tau} = \sqrt{\Delta m_\nu^2} \sim 0.055$ eV.
Moreover via the ``see-saw" mechanism $m_{\nu_\tau} = m_t(m_t)^2/(3
M)$. Hence one finds $M \sim 2 \times 10^{14}$ GeV;  remarkably
close to the GUT scale.  Note we have related the neutrino Yukawa
coupling to the top quark Yukawa coupling  $\lambda_{\nu_\tau} =
\lambda_t$ at $M_G$ as given in $SO(10)$ or $SU(4)\times SU(2)_L
\times SU(2)_R$.  However at low energies they are no longer equal
and we have estimated this RG effect by $\lambda_{\nu_\tau}(M_Z)
\approx \lambda_t(M_Z)/\sqrt{3}$.

\subsection{$SO(10)$ GUT with $[D_3 \times U(1)]$ Family Symmetry}

A complete model for fermion masses was given in Refs.
\cite{Dermisek:2005ij,Dermisek:2006dc}.    Using a global $\chi^2$
analysis, it has been shown that the model fits all fermion masses
and mixing angles, including neutrinos, and  a minimal set of
precision electroweak observables.  The model is consistent with
lepton flavor violation and lepton electric dipole moment bounds. In
two papers, Ref. \cite{Albrecht:2007ii,Altmannshofer:2008vr}, the
model was also tested by flavor violating processes in the B system.

The model is an $SO(10)$ SUSY GUT with an additional $D_3 \times
[U(1) \times Z_2 \times Z_3]$ family symmetry. The symmetry group
fixes the following structure for the superpotential \be W = W_f +
W_\nu ~,  \ee with \bea \label{Wcf}
W_f &=& \textbf{16}_3 \, \textbf{10}\, \textbf{16}_3+\textbf{16}_a \, \textbf{10}\, \chi_a \nn \\
&+&\bar{\chi}_a (M_{\chi}\, \chi_a+\textbf{45}
\,\frac{\phi_a}{\hat{M}}\, \textbf{16}_3
+\textbf{45} \,\frac{\tilde{\phi}_a}{\hat{M}} \, \textbf{16}_a+A\, \textbf{16}_a)~,\\
W_\nu &=&\ov{\textbf{16}} (\lambda_2\,
N_a\,\textbf{16}_a+\lambda_3\,N_3\,\textbf{16}_3) +\frac{1}{2}
(S_a\,N_a\,N_a+S_3\,N_3\, N_3)~. \label{eq:wnu} \eea
The first two families of quarks and leptons are contained in the
superfield $\textbf{16}_a,\:a=1,2$, which transforms under
SO(10)$\times D_3$ as $(\textbf{16}, \textbf{$2_A$})$, whereas the
third family in $\textbf{16}_3$ transforms as $
(\textbf{16},\textbf{$1_B$})$. The two MSSM Higgs doublets $H_{u}$
and $H_d$ are contained in a $\textbf{10}$. As can be seen from the
first term on the right-hand side of (\ref{Wcf}), Yukawa unification
$\lambda_t=\lambda_b=\lambda_\tau=\lambda_{\nu_\tau}$ at $M_G$ is
obtained {\em only} for the third generation, which is directly
coupled to the Higgs $\textbf{10}$ representation. This immediately
implies large $\tan\beta \approx 50$ at low energies and constrains
soft SUSY breaking parameters.

The effective Yukawa couplings of the first and second generation
fermions are generated hierarchically via the Froggatt-Nielsen
mechanism \cite{froggatt} as follows. Additional fields are
introduced, i.e. the $\textbf{45}$ which is an adjoint of SO(10),
the SO(10) singlet flavon fields $\phi^a, \tilde {\phi^a}, A$ and
the Froggatt-Nielsen states $\chi_a, \bar{\chi}_a$. The latter
transform as a $(\textbf{16},\textbf{$2_A$})$ and a
$(\ov{\textbf{16}},\textbf{$2_A$})$, respectively, and receive
masses of O$(M_{G})$ as $M_\chi$ acquires an SO(10) breaking VEV.
Once they are integrated out, they give rise to effective mass
operators which, together with the VEVs of the flavon fields, create
the Yukawa couplings for the first two generations. This mechanism
breaks systematically the full flavor symmetry and produces the
right mass hierarchies among the fermions.

\begin{table} \begin{center}
\begin{tabular}{|lcc|}
\hline
Sector & \# & Parameters \\
\hline \hline
gauge & 3 & $\alpha_G$, $M_G$, $\epsilon_3$, \\
SUSY (GUT scale) & 5 & $m_{16}$, $M_{1/2}$, $A_0$, $m_{H_u}$, $m_{H_d}$, \\
textures & 11 & $\epsilon$, $\epsilon^\prime$, $\lambda$, $\rho$, $\sigma$, $\tilde \epsilon$, $\xi$, \\
neutrino & 3 & $M_{R_1}$, $M_{R_2}$, $M_{R_3}$, \\
SUSY (EW scale) & 2 & $\tan\beta$, $\mu$ \\
\hline \hline
\end{tabular}
\caption{The 24 parameters defined at the GUT scale which are used
to minimize $\chi^2$.} \label{tab:parameters} \end{center}
\end{table}

Upon integrating out the FN states one obtains Yukawa matrices for
up-quarks, down-quarks, charged leptons and neutrinos given by \bea
&Y_u=\left(
\begin{array}{ccc}
0 & \varepsilon' \,\rho & -\varepsilon\,\xi \\
-\varepsilon' \, \rho & \tilde{\varepsilon} \,\rho & -\varepsilon \\
\varepsilon\,\xi & \varepsilon & 1
\end{array}
\right)\,\lambda~,&~~ Y_d=\left(
\begin{array}{ccc}
0 & \varepsilon'  & -\varepsilon\,\xi\,\sigma \\
-\varepsilon'  & \tilde{\varepsilon}  & -\varepsilon\,\sigma \\
\varepsilon\,\xi & \varepsilon & 1
\end{array}
\right)\,\lambda~,\nn \\
&Y_e=\left(
\begin{array}{ccc}
0 & -\varepsilon'  & 3\,\varepsilon\,\xi \\
\varepsilon'  & 3\,\tilde{\varepsilon}  & 3\,\varepsilon \\
-3\,\varepsilon\,\xi\,\sigma & -3\,\varepsilon\,\sigma & 1
\end{array}
\right)\,\lambda~,&~~ Y_\nu=\left(
\begin{array}{ccc}
0 & -\varepsilon' \,\omega & \frac{3}{2}\,\varepsilon\,\xi \,\omega\\
\varepsilon'  \,\omega& 3\,\tilde{\varepsilon}\,\omega &
\frac{3}{2}\,\varepsilon\,\omega \\ -3\,\varepsilon\,\xi\,\sigma &
-3\,\varepsilon\,\sigma & 1
\end{array}
\right)\,\lambda~. \label{Y-textures} \eea From eqs.
(\ref{Y-textures}) one can see that the flavor hierarchies in the
Yukawa couplings are encoded in terms of the four complex parameters
$\rho, \sigma, \tilde \varepsilon, \xi$ and the additional real ones
$\varepsilon, \varepsilon', \lambda$.

For neutrino masses one invokes the See-Saw
mechanism~\cite{minkowski,yanagida}. In particular, three SO(10)
singlet Majorana fermion fields $N_a, N_3$ $(a=1,2)$ are introduced
via the contribution of $\frac{1}{2}\, (S_a\,N_a \, N_a+S_3
\,N_3\,N_3)$ to the superpotential (Eqn. \ref{eq:wnu}). The mass
term $\frac{1}{2}\,N\,M_N\,N$ is produced when the flavon fields
acquire VEVs $\langle S_a\rangle=M_{N_a}$ and $\langle
S_3\rangle=M_{N_3}$. Together with a $\ov{\textbf{16}}$ Higgs one is
allowed to introduce the interaction terms $\ov{\textbf{16}}
\,(\lambda_2 \, N_a\, \textbf{16}_a+\lambda_3 \, N_3\,
\textbf{16}_3)$ (Eqn. \ref{eq:wnu}).  This then generates a mixing
matrix $V$ between the right-handed neutrinos and the additional
singlets ($\nu^c \,  V \, N$), when the $\ov{\textbf{16}}$ acquires
an SO(10) breaking VEV $\langle \ov{\textbf{16}} \rangle_{\nu^c} =
v_{16}$. The resulting effective right-handed neutrino mass terms
are given by \bea
 W_N= \bar \nu \, V\,N+\frac{1}{2}\,N\,M_N\,N~,
\eea \bea V=v_{16}\left(\begin{array}{ccc}
0 & \lambda_2 & 0 \\
\lambda_2 & 0 & 0 \\
0 & 0 & \lambda_3
\end{array}\right)~,~~~~
M_N={\rm diag}(M_{N_1},M_{N_2},M_{N_3})~. \eea Diagonalization leads
to the effective right-handed neutrino Majorana mass \bea M_R = - V
\, M_N^{-1}\, V^T \equiv - {\rm diag}(M_{R_1},M_{R_2},M_{R_3}) ~.
\eea By integrating out the EW singlets $\nu^c$ and $N$, which both
receive GUT scale masses, one ends up with the light neutrino mass
matrix at the EW scale given by the usual see-saw formula \bea \mc M
= m_\nu\, M_R^{-1}\,m_\nu^T~. \eea

\begin{table} \begin{center}
\begin{tabular}{|lc||lc|}
\hline
Observable & Value($\sigma_{\rm exp}$)  & Observable & Value($\sigma_{\rm exp}$)  \\
\hline \hline
$M_W$ & $80.403(29)$ &  $M_\tau$ & $1.777(0)$  \\
$M_Z$ & $91.1876(21)$ &   $M_\mu$ & $0.10566(0)$ \\
$10^{5} G_\mu$ & $1.16637(1)$ &  $10^3 M_e$ & $0.511(0)$  \\
$1/\alpha_{\rm em}$ & $137.036$  & $|V_{us}|$ & $0.2258(14)$ \\
$\alpha_s(M_Z)$ & $0.1176(20)$  & $10^3 |V_{ub}|$ & $4.1(0.4)$  \\
$M_t$ & $170.9(1.8)$  & $10^2 |V_{cb}|$ & $4.16(7)$ \\
$m_b(m_b)$ & $4.20(7)$  & $\sin 2 \beta$ & $0.675(26)$  \\
$m_c(m_c)$ & $1.25(9)$ & $10^3 \Delta m_{31}^2$ [eV$^2$]& $2.6(0.2)$  \\
$m_s(2~{\rm GeV})$ & $0.095(25)$  & $10^5 \Delta m_{21}^2$ [eV$^2$]& $7.90(0.28)$  \\
$m_d(2~{\rm GeV})$ & $0.005(2)$ & $\sin^2 2 \theta_{12}$ & $0.852(32)$  \\
$m_u(2~{\rm GeV})$ & $0.00225(75)$  & $\sin^2 2 \theta_{23}$ & $0.996(18)$ \\
\hline \hline
\end{tabular}
\caption{Flavor conserving observables used in the fit. Dimensionful
quantities are expressed in GeV, unless otherwise specified
\cite{Albrecht:2007ii}.} \label{tab:obs} \end{center}
\end{table}

\begin{table}[ht]  \begin{center}
\begin{tabular}{|lc|}
\hline
Observable & Value($\sigma_{\rm exp}$)($\sigma_{\rm theo}$) \\
\hline \hline
$10^3 \epsilon_K$ & 2.229(10)(252) \\
$\Delta M_s / \Delta M_d$ & 35.0(0.4)(3.6)  \\
$10^4$ BR$(B \to X_s \gamma)$ & 3.55(26)(46)  \\
$10^6$ BR$(B \to X_s \ell^+ \ell^-)$ & 1.60(51)(40)  \\
$10^4$ BR$(B^+ \to \tau^+ \nu)$ & 1.31(48)(9)  \\
BR$(B_s \to \mu^+ \mu^-)$ & $< 1.0 \times 10^{-7}$ \\
\hline \hline
\end{tabular}
\caption{FC observables used in the fit \cite{Albrecht:2007ii}.}
\label{tab:FCobs} \end{center}
\end{table}

\begin{table}[r] \begin{center}
\begin{tabular}{|lc|}
\hline
Observable & Lower Bound  \\
\hline \hline
$M_{h_0}$ & $114.4$ GeV  \\
$m_{\tilde t}$ & $60$ GeV \\
$m_{\tilde\chi^+}$ & $104$ GeV  \\
$m_{\tilde g}$ & $195$ GeV \\
\hline \hline
\end{tabular}
\caption{Mass bounds used in the fit \cite{Albrecht:2007ii}.}
\label{tab:bounds} \end{center}
\end{table}

The model has a total of 24 arbitrary parameters, with all except
$\tan\beta$ defined at the GUT scale (see Table
\ref{tab:parameters}). Using a two loop RG analysis the theory is
redefined at the weak scale. Then a $\chi^2$ function is constructed
with low energy observables.  In Ref. \cite{Dermisek:2006dc} fermion
masses and mixing angles, a minimal set of precision electroweak
observables and the branching ratio BR($b \rightarrow s \gamma$)
were included in the $\chi^2$ function. Then predictions for lepton
flavor violation, lepton electric dipole moments, Higgs mass and
sparticle masses were obtained.  The $\chi^2$ fit was quite good.
The light Higgs mass was always around 120 GeV.  In the recent
paper, Ref. \cite{Albrecht:2007ii}, precision B physics observables
were added. See Tables \ref{tab:obs}, \ref{tab:FCobs} for the 28 low
energy observables and Table \ref{tab:bounds} for the 4 experimental
bounds included in their analysis.    The fits were not as good as
before with a minimum $\chi^2 \sim 25$ obtained for large values of
$m_{16} = 10$ TeV.

The dominant problem was due to constraints from the processes $B
\rightarrow X_s \gamma, \; B \rightarrow X_s \ell^+ \ell^-$.  The
former process constrains the magnitude of the Wilson coefficient
$C_7$ for the operator \be O_7 = m_b \ \bar s_L \Sigma_{\mu \nu} b_R
\ F^{\mu \nu} \ee with $C_7 \sim |C_7^{\rm SM}|$,  while the latter
process is also sensitive to the sign of $C_7$. Note, the charged
and neutral Higgs contributions to ${\rm BR} (B\to X_s \gamma)$ are
strictly positive. While the sign of the chargino contribution,
relative to the SM, is ruled by the following relation \bea
C_7^{\tilde \chi^+} \propto + \mu A_t \tan\beta \times {\rm
sign}(C_7^{\rm SM})~, \label{C7chi} \eea with a positive
proportionality factor, so it is opposite to that of the SM one for
$\mu > 0$ and $A_t < 0$.  Hence it is possible for $C_7 \approx \pm
|C_7^{\rm SM}|$. Note the experimental result for $B \rightarrow X_s
\ell^+ \ell^-$ seems to favor the sign of $C_7$ to be the same as in
the Standard Model, however the results are inconclusive.  Another
problem was $V_{ub}$ which was significantly smaller than present
CKM fits.

In the recent analysis, Ref. \cite{Altmannshofer:2008vr},  it was
shown that better $\chi^2$ can be obtained by allowing for a 20\%
correction to Yukawa unification.  Note, this analysis only included
Yukawa couplings for the third family.  For a good fit, see Tables
\ref{tab:fit-example} and \ref{tab:fit-example2}. We find
$\tan\beta$ still large, $\tan\beta = 46$ and a light Higgs mass
$m_h = 121$ GeV.   See Table \ref{tab:fit-example} for the sparticle
spectrum which should be observable at the LHC.
\begin{table} \begin{center}
\begin{tabular}{|lccc|}
\hline
Observable  &  Exp.  &  Fit  &  Pull  \\
\hline\hline
$M_W$  &  80.403  &  80.56  &  0.4  \\
$M_Z$  &  91.1876  &  90.73  &  \textbf{1.0}  \\
$10^{5}\; G_\mu$  &  1.16637  &  1.164  &  0.3  \\
$1/\alpha_{\rm em}$  &  137.036  &  136.5  &  0.8  \\
$\alpha_s(M_Z)$  &  0.1176  &  0.1159  &  0.8  \\
$M_t$  &  170.9  &  171.3  &  0.2  \\
$m_b(m_b)$  &  4.20  &  4.28  &  \textbf{1.1}  \\
$M_\tau$  &  1.777  &  1.77  &  0.4  \\
$10^{4}\; {\rm BR} (B \to X_s \gamma)$  &  3.55  &  2.72  &  \textbf{1.6}  \\
$10^{6}\; {\rm BR} (B \to X_s \ell^+\ell^-)$  &  1.60  &  1.62  &  0.0  \\
$\Delta M_s / \Delta M_d$  &  35.05  &  32.4  &  0.7  \\
$10^{4}\; {\rm BR} (B^+ \to \tau^+\nu)$  &  1.41  &  0.726  &  \textbf{1.4}  \\
$10^{8}\; {\rm BR} (B_s \to \mu^+\mu^-)$  &  $<5.8 $ &  3.35  &  --  \\
\hline
\multicolumn{3}{|r}{total $\chi^2$:}  &  \textbf{8.78} \\
\hline
\end{tabular}
\caption{Example of successful fit in the region with $b -\tau$
unification. Dimensionful quantities are expressed in powers of GeV.
Higgs, lightest stop and gluino masses are pole masses, while the
rest are running masses evaluated at $M_Z$
\cite{Altmannshofer:2008vr}.} \label{tab:fit-example} \end{center}
\end{table}

\begin{table} \begin{center}
\begin{tabular}{|lc|lc|}
\hline
\multicolumn{2}{|l}{Input parameters} & \multicolumn{2}{|l|}{Mass predictions} \\
\hline\hline
$m_{16}$  &  $7000$  &  $M_{h^0}$  &  121.5  \\
$\mu$  &  $1369$  &  $M_{H^0}$  &  585  \\
$M_{1/2}$  &  $143$  &  $M_{A}$  &  586  \\
$A_0$  &  $-14301$  &  $M_{H^+}$  &  599  \\
$\tan\beta$  &  $46.1$  &  $m_{\tilde t_1}$  &  783  \\
$1/\alpha_G$  &  $24.7$  &  $m_{\tilde t_2}$  &  1728  \\
$M_G / 10^{16}$  &  $3.67$  &  $m_{\tilde b_1}$  &  1695  \\
$\epsilon_3 / \%$  &  $-4.91$  &  $m_{\tilde b_2}$  &  2378  \\
$(m_{H_u}/m_{16})^2$  &  $1.616$  &  $m_{\tilde \tau_1}$  &  3297  \\
$(m_{H_d}/m_{16})^2$  &  $1.638$  &  $m_{\tilde\chi^0_1}$  &  58.8  \\
$M_{R} / 10^{13}$  &  $8.27$  &  $m_{\tilde\chi^0_2}$  &  117.0  \\
$\lambda_u$  &  $0.608$  &  $m_{\tilde\chi^+_1}$  &  117.0  \\
$\lambda_d$  &  $0.515$  &  $M_{\tilde g}$  &  470  \\
\hline
\end{tabular}
\caption{Example of successful fit in the region with $b -\tau$
unification. Dimensionful quantities are expressed in powers of GeV.
Higgs, lightest stop and gluino masses are pole masses, while the
rest are running masses evaluated at $M_Z$
\cite{Altmannshofer:2008vr}.} \label{tab:fit-example2} \end{center}
\end{table} Finally, an analysis of dark matter for this model has been performed with
good fits to WMAP data \cite{Dermisek:2003vn}. The authors of Ref.
\cite{Baer:2008jn} also analyze dark matter in the context of the
minimal $SO(10)$ model with Yukawa unification.   They have
difficulty fitting WMAP data.  We believe this is because they do
not adjust the CP odd Higgs mass to allow for dark matter
annihilation on the resonance.

\subsection{Problems of 4D GUTs}

There are two aesthetic (perhaps more fundamental) problems
concerning 4d GUTs. They have to do with the complicated sectors
necessary for GUT symmetry breaking and Higgs doublet-triplet
splitting.   These sectors are sufficiently complicated that it is
difficult to imagine that they may be derived from a more
fundamental theory, such as string theory.  In order to resolve
these difficulties, it becomes natural to discuss grand unified
theories in higher spatial dimensions.  These are the so-called
orbifold GUT theories discussed in the next section.

Consider, for example, one of the simplest constructions in $SO(10)$
which accomplishes both tasks of GUT symmetry breaking and Higgs
doublet-triplet splitting \cite{Barr:1997hq}. Let there be a single
adjoint field, $A$, and {\it two} pairs of spinors, $C +
\overline{C}$ and $C' + \overline{C}'$. The complete Higgs
superpotential is assumed to have the form

\begin{equation}
W = W_A + W_C + W_{ACC'} + (T_1 A T_2 + S T_2^2).
\end{equation}

\noindent The precise forms of $W_A$ and $W_C$ do not matter, as
long as $W_A$ gives $\langle A \rangle$ the Dimopoulos-Wilczek form,
and $W_C$ makes the VEVs of $C$ and $\overline{C}$ point in the
$SU(5)$-singlet direction. For specificity we will take $W_A =
\frac{1}{4 M} {\rm tr} A^4 + \frac{1}{2} P_A ({\rm tr} A^2 + M_A^2)
+ f(P_A)$, where $P_A$ is a singlet, $f$ is an arbitrary polynomial,
and $M \sim M_G$. (It would be possible, also, to have simply $m \
{\rm Tr} A^2$, instead of the two terms containing $P_A$. However,
explicit mass terms for adjoint fields may be difficult to obtain in
string theory.) We take $W_C = X(\overline{C} C - P_C^2)$, where $X$
and $P_C$ are singlets, and $\langle P_C \rangle \sim M_G$.

The crucial term that couples the spinor and adjoint sectors
together has the form

\begin{equation}
W_{ACC'} = \overline{C}' \left( \left( \frac{P}{M_P} \right) A + Z
\right) C + \overline{C} \left( \left( \frac{\overline{P}}{M_P}
\right) A + \overline{Z} \right) C',
\end{equation}

\noindent where $Z$, $\overline{Z}$, $P$, and $\overline{P}$ are
singlets. $\langle P \rangle$ and $\langle \overline{P} \rangle$ are
assumed to be of order $M_G$. The critical point is that the VEVs of
the primed spinor fields will vanish, and therefore the terms in Eq.
(3) will not make a destabilizing contribution to $- F_A^* =
\partial W/\partial A$. This is the essence of the mechanism.

$W$ contains several singlets ($P_C$, $P$, $\overline{P}$, and $S$)
that are supposed to acquire VEVs of order $M_G$, but which are left
undetermined at tree-level by the terms so far written down. These
VEVs may arise radiatively when SUSY breaks, or may be fixed at tree
level by additional terms in $W$.

In $SU(5)$ the construction which gives natural Higgs
doublet-triplet splitting requires the $SU(5)$ representations ${\bf
75, \ 50, \ \overline{50}}$ and a superpotential of the form
\cite{missingpartner,Altarelli:2000fu}  \be W \supset 75^3 + M 75^2
+ 5_H \ 75 \ 50 + \bar 5_H \ 75 \ \overline{50} + 50 \ \overline{50}
\ X . \ee  The $50, \overline{50}$ contain Higgs triplets but no
Higgs doublets.  Thus when the $75$ obtains an $SU(5)$ breaking VEV,
the color triplets obtain mass but the Higgs doublets remain
massless.

\section{Orbifold GUTs}

\subsection{GUTs on a Circle}

As the first example of an orbifold GUT consider a pure $SO(3)$
gauge theory in 5 dimensions \cite{Dermisek:2002ri}.  The gauge
field is \be A_M \equiv A^a_M \ T^a,  \; a=1,2,3 ; \; M,N =
\{0,1,2,3,5\} .\ee The gauge field strength is given by \be F_{MN}
\equiv F^a_{MN} \ T^a =
\partial_M A_N - \partial_N A_M + i [A_M, A_N]  \ee where $T^a$ are
$SO(3)$ generators.  The Lagrangian is \be {\cal L}_5 = - \frac{1}{4
g_5^2 k} \ Tr (F_{MN} F^{MN}) \ee and we have $Tr (T^a \ T^b) \equiv
k \delta^{ab}$.  The inverse gauge coupling squared has mass
dimensions one.

Let us first compactify the theory on ${\cal M}_4 \times S^1$ with
coordinates  $\{ x_\mu, y \}$ and $y = [0, 2 \pi R)$. The theory is
invariant under the local gauge transformation \be A_M(x_\mu, y)
\rightarrow U \ A_M(x_\mu, y) \ U^\dagger - i U \ \partial_M \
U^\dagger, \;\; U = \exp( i \theta^a(x_\mu, y) \ T^a). \ee

Consider the possibility $\partial_5 A_\mu \equiv 0$.   We have \be
F_{\mu 5} = \partial_\mu A_5 + i [A_\mu, A_5] \equiv  D_\mu \ A_5 .
\ee  We can then define \be \tilde \Phi \equiv A_5 \ \frac{\sqrt{2
\pi R}}{g_5} \equiv A_5/g \ee where $g_5 \equiv \sqrt{2 \pi R} \; g$
and $g$ is the dimensionless 4d gauge coupling.  The 5d Lagrangian
reduces to the Lagrangian for a 4d $SO(3)$ gauge theory with
massless scalar matter in the adjoint representation, i.e. \be {\cal
L}_5 = \frac{1}{2 \pi R} [ -\frac{1}{4 g^2 k} \ Tr(F_{\mu \nu}
F^{\mu \nu}) + \frac{1}{2 k} \ Tr(D_\mu \tilde \Phi \ D^\mu \tilde
\Phi) ] . \ee

In general we have the mode expansion \be A_M(x_\mu, y) = \sum_n [
a^n_M \cos n \frac{y}{R} + b^n_M \sin n \frac{y}{R} ] \ee where only
the cosine modes with $n = 0$ have zero mass.   Otherwise the 5d
Laplacian $\partial_M \partial^M = \partial_\mu \partial^\mu +
\partial_y \partial^y$ leads to Kaluza-Klein [KK] modes with
effective 4d mass \be m_n^2 = \frac{n^2}{R^2} . \ee

\subsection{Fermions in 5d}

The Dirac algebra in 5d is given in terms of the 4 $\times$ 4 gamma
matrices  $\gamma_M,  \; M = 0,1,2,3,5$ satisfying $\{ \gamma_M,
\gamma_N \} = 2 g_{MN}$.  A four component massless Dirac spinor
$\Psi(x_\mu, y)$ satisfies the Dirac equation \be i \gamma_M
\partial^M \Psi = 0 = i (\gamma_\mu \partial^\mu - \gamma_5
\partial_y) \Psi \ee  with $\gamma_5 = i \left( \begin{array}{cc} -1 & 0 \\ 0 & 1 \end{array}
\right)$. In 4d the four component Dirac spinor decomposes into two
Weyl spinors with \be \Psi = \left( \begin{array}{c} \psi_1 \\  i
\sigma_2 \psi_2^*
\end{array} \right) = \left( \begin{array}{c} \psi_L \\  \psi_R
\end{array} \right)\ee where $\psi_{1,2}$ are two left-handed Weyl
spinors.   In general, we obtain the normal mode expansion for the
fifth direction given by \be  \psi_{L,R} =  \sum ( a_n(x) \cos n
\frac{y}{R} + b_n(x) \sin n \frac{y}{R} ). \ee If we couple this 5d
fermion to a local gauge theory, the theory is necessarily
vector-like; coupling identically to both $\psi_{L,R}$.

 We can obtain a chiral theory in 4d with the
following parity operation \be {\cal P} : \Psi(x_\mu, y) \rightarrow
\Psi(x_\mu, -y) = P \Psi(x_\mu, y) \ee  with $P =
i \gamma_5$.   We then have \bea  \Psi_L \sim \cos n \frac{y}{R} \nn \\
\Psi_R \sim \sin n \frac{y}{R}. \eea

\subsection{GUTs on an Orbi-Circle}

Let us briefly review the geometric picture of orbifold GUT models
compactified on an orbi-circle ${\rm S}^1/{\mathbb Z}_2$. The circle
$S^1 \equiv \mathbb{R}^1/{\cal T}$ where ${\cal T}$ is the action of
translations by $2 \pi R$.  All fields $\Phi$ are thus periodic
functions of $y$ (up to a finite gauge transformation), i.e. \be
{\cal T} : \Phi(x_\mu, y) \rightarrow \Phi(x_\mu, y + 2 \pi R) = T \
\Phi(x_\mu, y)  \ee where $T \in SO(3)$ satisfies $T^2 = 1$.  This
corresponds to the translation ${\cal T}$ being realized
non-trivially by a degree-2 Wilson line (i.e., background gauge
field - $\langle A_5 \rangle \neq 0$ with $T \equiv \exp (i \oint
\langle A_5 \rangle dy)$). Hence the space group of ${\rm
S}^1/\mz_2$ is composed of two actions, a translation, ${\cal
T}:\,y\rightarrow y+2\pi R$, and a space reversal, ${\cal
P}:\,y\rightarrow -y$. There are two (conjugacy) classes of fixed
points, $y=(2n)\pi R$ and $(2n+1)\pi R$, where $n\in\mz$.

\begin{figure}\begin{center}
\includegraphics[height=.05\textheight]{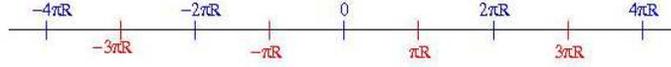} \end{center}
\caption{\label{fig:epsart} The real line modded out by the space
group of translations, ${\cal T}$, and a $\mathbb{Z}_2$ parity,
${\cal P}$.}
\end{figure}

The space group multiplication rules imply ${\cal T}{\cal P}{\cal
T}={\cal P}$, so we can replace the translation by a composite
$\mz_2$ action ${\cal P}'={\cal P}{\cal T}: y\rightarrow -y+2\pi R$.
The orbicircle ${\rm S}^1/{\mathbb Z}_2$ is equivalent to an
${\mathbb R}/({\mathbb Z}_2\times{\mathbb Z}'_2)$ orbifold, whose
fundamental domain is the interval $[0,\,\pi R]$, and the two ends
$y=0$ and $y=\pi R$ are fixed points of the ${\mathbb Z}_2$ and
${\mathbb Z}_2'$ actions respectively.

A generic 5d field $\Phi$  has the following transformation
properties under the ${\mathbb Z}_2$ and ${\mathbb Z}'_2$
orbifoldings (the 4d space-time coordinates are suppressed),
\be {\cal P}:\,\Phi(y)\rightarrow\Phi(-y)=P\Phi(y)\,,\qquad {\cal
P}':\,\Phi(y)\rightarrow \Phi(-y+2{\pi R})=P'\Phi(y)\,, \ee
where $P,\,P' \equiv P T =\pm$ are \textit{orbifold parities} acting
on the field $\Phi$ in the appropriate group representation. Where
it is assumed that $[P, T] = 0$. The four combinations of orbifold
parities give four types of states, with wavefunctions \bea
\zeta_m(++) \sim\cos(my/R), \nn \\
\zeta_m(+-) \sim\cos[(2m+1)y/2R], \nn \\
\zeta_m(-+) \sim\sin[(2m+1)y/2R], \nn \\
\zeta_m(--) \sim\sin[(m+1)y/R], \eea where $m\in\mathbb Z$. The
corresponding KK towers have mass
\bea M_{\rm KK}=\left\{
\begin{array}{ll}
m/R  &\,\,{\rm for}\,(PP')=(++)\,,\\
(2m+1)/2R &\,\,{\rm for}\,(PP')=(+-)\,\,{\rm and}\,\,(-+)\,,\\
(m+1)/R &\,\, {\rm for}\,(PP')=(--)\,.\label{kkmass}
\end{array}\right.
\eea
Note that only the $\Phi_{++}$ field possesses a massless zero mode.

For example, consider the Wilson line $T =  \exp (i \pi T^3) = {\rm
diag} (-1, -1, 1)$. Let $A_\mu(y) \; (A_5(y))$ have parities $P = +
(-)$, respectively.  Then only $A^3_\mu$ has orbifold parity $(++)$
and $A^3_5$ has orbifold parity $(--)$.  Note, $A^3_5(-y) = -
A^3_5(y) + \frac{1}{R}$. Define the fields \be W^\pm =
\frac{1}{\sqrt{2}} ( A^1 \mp i A^2) \ee with $T^\pm =
\frac{1}{\sqrt{2}} ( T^1 \pm i T^2)$ and $[T^3, T^\pm] = \pm T^\pm$.
Then $W^\pm_\mu \; [W^\pm_5]$ have orbifold parity $(+ -) \; [(-
+)]$, respectively. Thus the $SO(3)$ gauge group is broken to $SO(2)
\approx U(1)$ in 4d.  The local gauge parameters preserve the $(P,
T)$ parity/holonomy, i.e. \bea
\theta^3(x_\mu, y) = \theta^3_m(x_\mu) \zeta_m(++) \nn \\
\theta^{1,2}(x_\mu, y) = \theta^{1,2}_m(x_\mu) \zeta_m(+-). \eea
 Therefore $SO(3)$ is {\em not} the symmetry at $y = \pi R$.

\subsection{A Supersymmetric $SU(5)$ orbifold GUT}\label{sec:so10}

Consider the 5d orbifold GUT model of ref.~\cite{Hall:2001pg}. The
model has an $SU(5)$  symmetry broken by orbifold parities to the SM
gauge group in 4d. The compactification scale $M_c = R^{-1}$ is
assumed to be much less than the cutoff scale.

The gauge field is a 5d vector multiplet ${\cal
V}=(A_M,\lambda,\lambda',\sigma)$, where $A_M,\,\sigma$ (and their
fermionic partners $\lambda, \ \lambda'$) are in the adjoint
representation (${\bf 24}$) of $SU(5)$ . This multiplet consists of
one 4d ${\mathcal N}= 1$ supersymmetric vector multiplet
$V=(A_\mu,\lambda)$ and one 4d chiral multiplet
$\Sigma=((\sigma+iA_5)/\sqrt{2},\lambda')$. We also add two 5d
hypermultiplets containing the Higgs doublets, ${\cal H}=(H_5,
{H_5}^c)$, $\overline{\cal H}=(\bar H_{\bar 5}, {\bar H_{\bar
5}}^c)$. The 5d gravitino $\Psi_M=(\psi^1_M,\psi_M^2)$ decomposes
into two 4d gravitini $\psi_\mu^1$, $\psi_\mu^2$ and two dilatini
$\psi_5^1$, $\psi_5^2$. To be consistent with the 5d supersymmetry
transformations one can assign positive parities to
$\psi_\mu^1+\psi_\mu^2$, $\psi_5^1-\psi_5^2$ and negative parities
to $\psi_\mu^1-\psi_\mu^2$, $\psi_5^1+\psi_5^2$; this assignment
partially breaks ${\mathcal N}=2$ to ${\mathcal N}=1$ in 4d.

The orbifold parities for various states in the vector and hyper
multiplets are chosen as follows \cite{Hall:2001pg} (where we have
decomposed all the fields into SM irreducible representations and
under $SU(5)$ we have taken $P = (+++++), \ P^\prime = (---++)$)
\bea
\begin{array}{llllll}
\hline
{\rm States} & P & P' &{\rm States} & P & P'\\
 \hline
V({\bf 8,1,0}) & + & + & \Sigma({\bf 8,1,0}) & - & -\\
V({\bf 1,3,0})  & + & + & \Sigma({\bf 1,3,0}) & - & -\\
V({\bf 1,1,0})  & + & + & \Sigma({\bf 1,1,0}) & - & -\\
V({\bf \bar 3,2,5/3})  & + & - & \Sigma({\bf 3,2,-5/3}) & - & + \\
V({\bf 3,2,-5/3})  & + & - & \Sigma({\bf \bar 3,2,5/3}) & - & + \\
T({\bf 3,1,-2/3})  & + & - & T^c({\bf \bar 3,1,2/3}) & - & + \\
H({\bf 1,2,+1})  & + & + & H^c({\bf \bar 1,2,-1}) & - & - \\
\bar T({\bf \bar 3,1,+2/3})  & + & - & \bar T^c({\bf 3,1,-2/3}) & - & + \\
\bar H({\bf 1,2,-1})  & + & + & \bar H^c({\bf 1,2,+1}) & - & - \\
\hline
\end{array}
\eea
We see the fields supported at the orbifold fixed points $y=0$ and
$\pi R$ have parities $P=+$ and $P'=+$ respectively. They form
complete representations under the $SU(5)$ and SM groups; the
corresponding fixed points are called $SU(5)$ and SM ``branes.'' In
a 4d effective theory one would integrate out all the massive
states, leaving only massless modes of the $P=P'=+$ states. With the
above choices of orbifold parities, the SM gauge fields and the $H$
and $\bar H$ chiral multiplet are the only surviving states in 4d.
We thus have an ${\mathcal N} = 1$ SUSY in 4d.   In addition, the $T
+ \bar T$ and $T^c + \bar T^c$ color-triplet states are projected
out, solving the doublet-triplet splitting problem that plagues
conventional 4d GUTs.

\subsection{Gauge Coupling Unification}

We follow the field theoretical analysis in
ref.~\cite{Dienes:1998vg} (see also
\cite{Contino:2001si,Ghilencea:2002ff}). It has been shown there the
correction to a generic gauge coupling due to a tower of KK states
with masses $M_{\rm KK}=m/R$ is
\be
\alpha^{-1}(\Lambda)=\alpha^{-1}(\mu_0)+ \frac{b}{4\pi}
\int_{r\Lambda^{-2}}^{r\mu_0^{-2}}\frac{{\rm d}t}{t}\,
\theta_3\left(\frac{{\rm i}t}{\pi R^2}\right)\,,
\ee
where the integration is over the Schwinger parameter $t$, $\mu_0$
and $\Lambda$ are the IR and UV cut-offs, and $r=\pi/4$ is a
numerical factor. $\theta_3$ is the Jacobi theta function,
$\theta_3(t)=\sum_{m=-\infty}^\infty {\rm e}^{{\rm i}\pi m^2 t}$,
representing the summation over KK states.

For our $S^1/\mathbb{Z}_2$ orbifold there is one modification in the
calculation.   There are four sets of KK towers, with mass $M_{\rm
KK}=m/R$ (for $P=P'=+$), $(m+1)/R$ (for $P=P'=-$) and $(m+1/2)/R$
(for $P=+$, $P'=-$ and $P=-$, $P'=+$), where $m\geq 0$. The
summations over KK states give respectively
$\frac{1}{2}\left(\theta_3({\rm i}t/\pi R^2)-1\right)$ for the first
two cases and $\frac{1}{2}\theta_2({\rm i}t/\pi R^2)$ for the last
two (where $\theta_2(t)= \sum_{m=-\infty}^\infty {\rm e}^{{\rm i}\pi
(m+1/2)^2 t}$), and we have separated out the zero modes in the
$P=P'=+$ case.

Tracing the renormalization group evolution from low energy scales,
we are first in the realm of the MSSM, and the beta function
coefficients are ${\bf b}^{MSSM}=(- \frac{33}{5},-1,3)$. The next
energy threshold is the compactification scale $M_c$. From this
scale to the cut-off scale, $M_*$, we have the four sets of KK
states.

Collecting these facts, and using $\theta_2({\rm i}t/\pi R^2)\simeq
\theta_3({\rm i}t/\pi R^2)\simeq\sqrt{\frac{\pi}{t}} R$ for
$t/R^2\ll 1$, we find the RG equations,
\bea \label{running_fiveD} \alpha_i^{-1}(M_Z) = & \alpha_{*}^{-1} -
\frac{b_i^{MSSM}}{2\pi} \log \frac{M_*}{M_Z} + \frac{1}{4\pi}
\left(b_i^{++} + b_i^{--}\right) \log \frac{M_*}{M_c}  & \nn
\\ & -
\frac{b^{\mathcal G}}{2 \pi}\left(\frac{M_*}{M_c} - 1\right) +
\delta_i^2 + \delta_i^l & \eea
for $i=1,2,3$, where $\alpha_*^{-1} = \frac{8 \pi^2 R}{g_5^2}$ and
we have taken the cut-off scales, $\mu_0=M_c = \frac{1}{R}$ and
$\Lambda=M_*$.  (Note, this 5d orbifold GUT is a non-renormalizable
theory with a cut-off.   In string theory, the cut-off will be
replaced by the physical string scale, $M_{\rm STRING}$.)
$b^{\mathcal G} = \sum_{P=\pm, P'=\pm} b^{\mathcal G}_{PP'}$, so in
fact it is the beta function coefficient of the orbifold GUT gauge
group, ${\mathcal G} = SU(5)$. The beta function coefficients in the
last two terms have an ${\mathcal N}=2$ nature, since the massive KK
states enjoy a larger supersymmetry.  In general we have
$b^{\mathcal G} = 2 C_2(\mathcal G) - 2 N_{hyper} T_R$.  The first
term (in Eqn. \ref{running_fiveD}) on the right is the 5d gauge
coupling defined at the cut-off scale, the second term accounts for
the one loop RG running in the MSSM from the weak scale to the
cut-off,  the third and fourth terms take into account the KK modes
in loops above the compactification scale and the last two terms
account for the corrections due to two loop RG running and weak
scale threshold corrections.

It should be clear that there is a simple correspondence to the 4d
analysis. We have \bea \alpha_G^{-1} \;\; (4d) & \leftrightarrow &
\alpha_*^{-1} - \frac{b^{\mathcal G}}{2 \pi}\left( \frac{M_*}{M_c} -
1\right) \;\; (5d) \nn \\ \label{eq:deltash} \delta_i^h \;\; (4d) &
\leftrightarrow & \frac{1}{4\pi} \left(b_i^{++} + b_i^{--}\right)
\log \frac{M_*}{M_c} - \frac{b^{MSSM}_i}{2 \pi} \log \frac{M_*}{M_G}
\;\; (5d) . \eea Thus in 5d the GUT scale threshold corrections
determine the ratio $M_*/M_c$ (note the second term in Eqn.
\ref{eq:deltash} does not contribute to $\delta_s^h$). For $SU(5)$
we have ${\bf b^{++} + b^{--}} = ( - 6/5, 2, 6)$ and given
$\delta_s^h$ (Eqn. \ref{eq:deltah}) we have \be \delta_s^h =
\frac{12}{28 \pi} \log \frac{M_*}{M_c} \approx + 0.94 \ee or \be
\frac{M_*}{M_c} \approx 10^3. \ee

\begin{figure*} \begin{center}
\scalebox{0.85}{\includegraphics{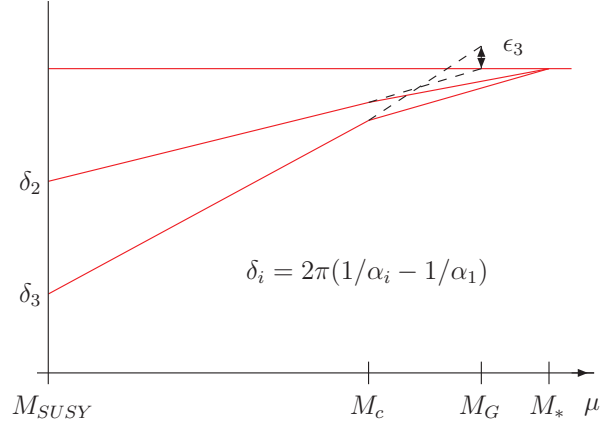}} \caption{\label{fig:gcu}
The differences $\delta_i = 2 \pi( 1/\alpha_i - 1/\alpha_1)$ are
plotted as a function of energy scale $\mu$.  The threshold
correction $\epsilon_3$ defined in the 4d GUT scale is used to fix
the threshold correction in the 5d orbifold GUT. } \end{center}
\end{figure*}

If the GUT scale is defined at the point where $\alpha_1 =
\alpha_2$, then we have $\delta_1^h = \delta_2^h$ or $\log
\frac{M_*}{M_G} \approx 2$.  In 5d orbifold GUTs, nothing in
particular happens at the 4d GUT scale.  However, since the gauge
bosons affecting the dimension 6 operators for proton decay obtain
their mass at the compactification scale,  it is important to
realize that the compactification scale is typically lower than the
4d GUT scale and the cut-off is higher (see Figure \ref{fig:gcu}).

\subsection{Quarks and Leptons in 5d Orbifold GUTs}

Quarks and lepton fields can be put on either of the orbifold
``branes" or in the 5d bulk.   If they are placed on the $SU(5)$
``brane" at $y = 0$, then they come in complete $SU(5)$ multiplets.
As a consequence a coupling of the type \be W \supset \int d^2\theta
\int  dy \  \delta(y) \ \bar H \ 10 \ \bar 5 \ee will lead to bottom
- tau Yukawa unification.   This relation is good for the third
generation and so it suggests that the third family should reside on
the $SU(5)$ brane.   Since this relation does not work for the first
two families, they might be placed in the bulk or on the SM brane at
$y = \pi \ R$.  Without further discussion of quark and lepton
masses (see
\cite{Hall:2002ci,Kim:2004vk,Alciati:2005ur,Alciati:2006sw} for
complete $SU(5)$ or $SO(10)$ orbifold GUT models), let us consider
proton decay in orbifold GUTs.

\subsection{Proton Decay}

\subsubsection{Dimension 6 Operators}

The interactions contributing to proton decay are those between the
so-called $X$ gauge bosons $A_\mu^{(+-)} \in V(+-)$ (where
${A_{\mu}^{(+-)}}^{a i}(x_\mu,y)$ is the five dimensional gauge
boson with quantum numbers $(\bar 3,2,+5/3)$ under SU(3)$\times$
SU(2) $\times$ U(1), $a$ and $i$ are color and SU(2) indices
respectively) and the ${\mathcal N}=1$ chiral multiplets on the
$SU(5)$ brane at $y=0$. Assuming all quarks and leptons reside on
this brane we obtain the $\Delta B \neq 0$ interactions given by \be
{\cal S}_{\Delta B \neq 0} = - \frac{g_5}{\sqrt{2}} \int d^4x
{A_{\mu}^{(+-)}}^{a i}(x_\mu,0) J_{a i}^{\mu}(x_\mu)+ h.c.~~~. \ee
The currents $J_{a i}^{\mu}$ are given by: \bea J_{a i}^{\mu}&=&
\epsilon_{a b c}\,\epsilon_{i j} (\bar u)^*_b \,\bar \sigma^\mu \,
q^{c j} +  q_{a i}^* \, \bar \sigma^\mu\, \bar e - \tilde l_{i}^* \, \bar \sigma^\mu\, (\bar d)_a \nn\\
&=&  (\bar u)^* \, \bar \sigma^\mu\, q + q^* \, \bar \sigma^\mu\,
\bar e - \tilde l^* \, \bar \sigma^\mu\, \bar d  ~~~, \eea

Upon integrating out the $X$ gauge bosons we obtain the effective
lagrangian for proton decay  \be {\cal L} = -\frac{g_G^2}{2 M_X^2}
\sum_{i, j} \left[({q^*_i} \bar \sigma^\mu \bar u_i) \; (\tilde
l^*_j \bar \sigma_\mu \bar d_j) \ + \ (q^*_i \bar \sigma^\mu \bar
e_i) \; (q^*_j \bar \sigma_\mu \bar u_j) \right]~~~, \label{lagpd0}
\ee where all fermions are weak interaction eigenstates and
$i,j,k=1,2,3$ are family indices. The dimensionless quantity \be
g_G\equiv g_5 \frac{1}{\sqrt{2 \pi R}} \ee is the four-dimensional
gauge coupling of the gauge bosons zero modes. The combination \be
M_X=\frac{M_c}{\pi} ~~~, \ee proportional to the compactification
scale \be M_c\equiv \frac{1}{R}~~~, \ee is an effective gauge vector
boson mass arising from the sum over all the Kaluza-Klein levels:
\be \sum_{n=0}^\infty\frac{4}{(2 n+1)^2 M_c^2}=\frac{1}{M_X^2}~~~.
\ee Before one can evaluate the proton decay rate one must first
rotate the quark and lepton fields to a mass eigenstate basis.  This
will bring in both left- and right-handed quark and lepton mixing
angles. However, since the compactification scale is typically lower
than the 4d GUT scale, it is clear that proton decay via dimension 6
operators is likely to be enhanced.

\subsubsection{Dimension 5 Operators}

The dimension 5 operators for proton decay result from integrating
out color triplet Higgs fermions.   However in this simplest $SU(5)$
5d model the color triplet mass is of the form
\cite{ArkaniHamed:2001tb} \be W \supset \int d^2 \theta \ dy \
(T(-+)^c \
\partial_y \ T(+-) + \bar T(-+)^c \ \partial_y \ \bar T(+-)) \ee where a sum over massive
KK modes is understood. Since only $T, \ \bar T$ couple directly to
quarks and leptons,  {\em no dimension 5 operators are obtained when
integrating out the color triplet Higgs fermions}.

\subsubsection{Dimension 4 baryon and lepton violating operators}

If the theory is constructed with an R parity or family reflection
symmetry, then no such operators will be generated.

\section{String Theory}

As mentioned in the introduction,  there are several limiting forms
of string theory, known as Type I, Type II A \& B,  $E_8 \otimes
E_8$ \& $SO(32)$ (or more precisely $Spin(32)/\mathbb{Z}_2$)
heterotic and F and M theories.  The first five are perturbative
string theories defined in terms of quantum superstrings propagating
in 9 + 1 space-time dimensions.\footnote{A superstring theory
denotes a string theory which has two dimensional worldsheet
supersymmetry. Such theories can lead to spacetime supersymmetry,
depending on the vacuum.}  Type I is a theory of open and closed
strings with the open strings coupled to gauge fields taking values
in $Spin(32)/\mathbb{Z}_2$. Type II is a theory of closed strings
with Abelian gauge symmetry in IIA and no gauge symmetry in IIB.
Heterotic strings are a combination of superstrings for right-moving
excitations along the string and bosonic strings for left-moving
excitations.  The right-movers travel in a target space of 10
space-time dimensions, while the left-movers travel in 26 space-time
dimensions.   However 16 spatial dimensions are compactified on a
$E_8 \otimes E_8$ or $Spin(32)/\mathbb{Z}_2$ lattice.  This
compactification leads to the two possible gauge groups of the
heterotic string.

Type I and II strings also include non-perturbative excitations
called $D_p$-branes.  $D_p$-branes are hyper-surfaces in $p$ spatial
dimensions on which open strings can end.   N $D_p$-branes, piled on
top of each other, support massless $U(n)$ gauge excitations.  And
when $D_p$ branes intersect, massless chiral matter appears.  The
low energy effective field theories for the perturbative strings
corresponds to N=1 supergravity in 10 space-time dimensions. In
addition, Type I and II low energy theories include massless gauge
matter living on $D$-branes and chiral matter living at the
intersection of two $D$-branes. M and F theory are only defined in
terms of non-perturbative effective field theories in 10 + 2 and 10
+ 1 space-time dimensions, respectively. F theory is essentially a
Type IIB string with the extra internal two dimensions corresponding
to the complex Type IIB string coupling. $D_p$-branes in the Type
IIB picture occur at the position of singularities in the string
coupling in the F theory description. M theory has no fundamental
string limit.   It is defined as N = 1 supergravity in 11 dimensions
which reduces to N = 8 supergravity in 4 dimensions. This is the
maximal supergravity theory!  $D_0$-branes of Type IIB form a tower
of excitations which is interpreted as the 11th dimension.  In M
theory,  gauge fields live on 3 dimensional singular sub-manifolds
and chiral matter live on zero dimensional singular points in the
internal 7 dimensions. There is also a purely quantum mechanical
formulation of M theory in terms of a quantum theory of matrices,
the so-called M(atrix) theory. However this formulation is even more
difficult to deal with.

All of these limiting forms are related by so-called S,  T or U
dualities
\cite{Schwarz:1996qw,Antoniadis:1997nz,Dijkgraaf:1997ip,Morrison:1999zi}.
The dilaton S is a measure of the string coupling, thus $S
\rightarrow \frac{1}{S}$ corresponds to strong - weak coupling
duality.  Which means to say, one limiting string theory calculated
in the weak coupling limit is dual to a different limiting string
theory, calculated in the strong coupling limit. Examples of S dual
theories are Type I - $SO(32)$ heterotic, Type IIB - Type IIB, Type
IIA - $E_8 \otimes E_8$ heterotic (via the intermediate step of M
theory).  The modulus T is a measure of the volume associated with
the extra 6 internal dimensions of a perturbative string.  Thus T
duality is the equality of two different limiting theories with one
compactified on a space of large volume and the other compactified
on a space of small volume. Finally, if one theory compactified on a
space of large (or small) volume is equivalent to another theory at
strong (or weak) coupling, they are called U dual. If the two
theories are the same, then the two theories are said to be
self-dual. T duality, unlike S or U duality, can be understood
perturbatively.  Type IIA - Type IIB and $E_8 \otimes E_8$ heterotic
- $SO(32)$ heterotic are T dual. Combining S and T dualities, one
can show that Type IIB - $E_8 \otimes E_8$ heterotic are dual.   In
addition, in this way one can also obtain a non-perturbative
definition of Type IIB,i.e.  F theory.

String model building refers to choosing a particular limiting
string theory and compactifying the extra (6 for Type I, II, and
heterotic,  7 for M theory and 8 for F theory) dimensions. If the
compactification manifold is flat, then the string spectrum and
interactions can be calculated perturbatively (with the addition of
open strings attached to $D_p$-brane states in Type I and II
theories).   The $D_p$-branes are massive and will distort the flat
background metric of the internal dimensions.   In order to include
the back reaction on the metric, one studies the effective low
energy supergravity limit of the string.   In this case one
compactifies on an internal manifold which preserves at least one
supersymmetry in 4 dimensions.   This requires compactifying on a
Calabi-Yau 3-fold for Type I, II and heterotic strings,  a $G_2$
manifold for M theory and a Calabi-Yau 4-fold for F theory.   In
order to complete the picture of string model building, it is
necessary to also mention string theories defined solely in 4
dimensions in terms of tensor products of two dimensional conformal
field theories (another term for strings).  Such theories go under
the name of free fermionic heterotic strings
\cite{Kawai:1986va,Narain:1985jj,Antoniadis:1986rn} or Gepner models
\cite{Gepner:1987qi}.  Note, the free fermionic construction can be
put into one-to-one correspondence with orbifold constructions
compactified, at the first step, on toroidal manifolds with radii
determined as rational ratios of the string scale.  For arbitrary
radii, as allowed in bosonic constructions, one requires additional
Thirring interactions included in the fermionic construction. Hence,
the free fermionic construction describes isolated points in the
bosonic moduli space.

Now consider $E_8 \otimes E_8$ heterotic - F theory duality.  It
turns out that F theory compactified on $K_3$ is dual to the  $E_8
\otimes E_8$ heterotic string compactified on a two dimensional
torus, $T^2$.  In addition  F theory on a $CY_3$ is dual to the $E_8
\otimes E_8$ heterotic string compactified on $K_3$.   Finally,
there is a duality between F theory on $CY_4$ and the  $E_8 \otimes
E_8$ heterotic string on $CY_3$.

The string landscape refers to all possible supersymmetry breaking
solutions of the string in 4 space-time dimensions.  This, however,
is a particularly ill-defined concept, since in string theory the
coordinates of space-time are themselves dynamical variables and
hence the space-time background geometry is quantum mechanical.  In
this sense it may be said that in string theory space-time itself is
an emergent concept.  Moreover, since energy and momentum are
derived quantities in Lagrangian systems with space and time
translation invariance,  one must first define the Lagrangian for an
effective field theory describing the modes of a string.   String
field theory exists describing all the string excitations.  But this
theory is quite difficult to handle.   Thus in most situations, when
one speaks of the string vacua or the string landscape one is most
likely referring to the ground state for the massless degrees of
freedom in terms of the supergravity limit of the string. As
discussed in the introduction, the string landscape is immense.  It
has been estimated that the landscape of SUSY breaking string vacua
numbers of order $10^{500}$ \cite{Bousso:2000xa}. For a recent
animated history of the subject, see \cite{Schellekens:2008kg}.

\subsection{Random searches for the MSSM in the String Landscape \label{sec:random}}

The literature is replete with searches over the string landscape.
Among these have been random searches in the string landscape
looking for features common to the MSSM.  In particular, vacua with
N = 1 supersymmetry [SUSY],  the Standard Model gauge group and
three families of quarks and leptons.   These random searches have
for the most part shown that the MSSM is an extremely rare point in
the string landscape. For example, searches in Type II intersecting
D-brane models~\cite{Gmeiner:2005vz} have found nothing looking like
the MSSM in $10^9$ tries.  Searches in Gepner orientifolds have been
a bit more successful finding one MSSM-like model for every 10,000
tries~\cite{Anastasopoulos:2006da}. Even searches in the heterotic
string, using the free fermionic construction,  have shown that the
MSSM is a very rare point in the string
landscape~\cite{Dienes:2007ms}.  The bottom line: if you want to
find the MSSM, then a random search is not the way to go.   In fact,
MSSM-like models have been found in Type II D-brane vacua, BUT by
directed searches~\cite{Blumenhagen:2005mu} AND not random ones. For
a recent discussion of random searches in the string landscape, see
\cite{GatoRivera:2009yt}.

\subsection{String model building}

In recent years, some progress has been made in finding MSSM-like
theories starting from different points in the string
landscape~\cite{Cleaver:1998saa,Braun:2005ux,Cvetic:2005bn,Buchmuller:2005jr,Verlinde:2005jr,Buchmuller:2006ik,Lebedev:2006kn,Kim:2007mt,Chen:2007px,Lebedev:2007hv,Lebedev:2008un,Beasley:2008dc,Donagi:2008ca,Beasley:2008kw,Donagi:2008kj,Blumenhagen:2008zz,Conlon:2008wa,Bourjaily:2009vf,Hayashi:2009ge,Huh:2009nh,Marsano:2009gv},
i.e. free fermionic, orbifold or smooth Calabi-Yau constructions of
the heterotic string, intersecting (and local) D-brane constructions
in type II string,  and M or F theory constructions. Much of this
progress has benefited from the requirement of an intermediate grand
unified gauge symmetry which naturally delivers the standard model
particle spectrum.  I will discuss a few recent results.

String model building is typically a two stage process.  At the
first stage one searches for an MSSM-like theory with ${\cal N} = 1$
supersymmetry.  One is interested in the Yukawa and gauge coupling
constants which, at this stage, are functions of moduli, i.e. SM
singlet fields with no potential.   These moduli also determine the
masses of any vector-like exotic fields which may exist in this
vacuum.   These moduli must be stabilized!  However, stabilization
requires lifting the flat directions and this requires spontaneously
breaking the ${\cal N} = 1$ supersymmetry.   Supersymmetry breaking
is also required for phenomenological purposes.   Supersymmetric
partners of ordinary particles have not been discovered.  Hence they
must be heavier than their SM namesakes.  We discuss supersymmetry
breaking and moduli stabilization in Section
\ref{sec:moduli_stabilization}.

\subsection{Heterotic constructions on smooth manifolds}

\begin{figure} \begin{center}
  \includegraphics[height=3cm]{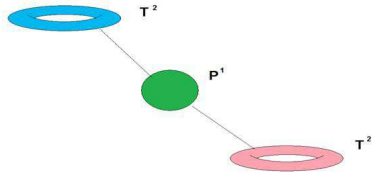}
  \caption{Calabi-Yau 3-fold defined in terms of a double elliptic fibration. The line connecting
  the tori to the 2-sphere represents the fibration.}
  \label{fig:CY3} \end{center}
\end{figure}

Bouchard et al.~\cite{Bouchard:2005ag}  have obtained an $SU(5)$ GUT
model on a CY$_3$ with the following properties.  They have three
families of quarks and leptons, and one or two pairs of Higgs
doublets. They accomplish GUT symmetry breaking and Higgs
doublet-triplet splitting via a Wilson line in the weak hypercharge
direction.  The CY$_3$ is defined by a double elliptic fibration,
i.e. two tori whose radii change as the tori move over the surface
of a sphere (see Fig. \ref{fig:CY3}).

In addition, they obtain a non-trivial up Yukawa matrix given
by~\cite{Bouchard:2006dn}
\begin{equation}
\lambda_u = \left( \begin{array}{ccc} a & b & c \\ b & d & e \\ c & e & 0
\end{array} \right) .
\end{equation}
The parameters $a, \cdots , e$ are functions of the moduli. The down
and charged lepton Yukawa matrices are however zero and would
require non-perturbative effects to change this.

Braun et al.~\cite{Braun:2005ux} obtain an $SU(5)\otimes U(1)_{B-L}$
GUT model on a CY$_3$.  GUT symmetry breaking and Higgs
doublet-triplet splitting is accomplished via a Wilson line in the
weak hypercharge direction. The low energy theory contains three
families of quarks and leptons, one pair of Higgs doublets and the
Standard Model gauge symmetry plus the additional $U(1)_{B-L}$. The
latter forbids R parity violating operators and is only
spontaneously broken near the weak scale via right-handed sneutrino
VEVs.  Hence in this theory the tau neutrino mixes with neutralinos
and is Majorana.  However, in the simplest scenario, the electron
and muon neutrinos are pure Dirac.  Moreover they are light due to
the assumption of very small Yukawa couplings. The phenomenology of
this model has been discussed in Ref. \cite{Ambroso:2010pe}.

\subsection{Heterotic String Orbifolds and Orbifold GUTs \label{sec:heterotic}}

Early work on orbifold constructions of the heterotic string was
started over 20 years
ago~\cite{Dixon:1985jw,Dixon:1986jc,Ibanez:1986tp,Ibanez:1987sn,Font:1989aj}.
The first complete MSSM model from the $E_8 \otimes E_8$ heterotic
string was obtained using the free fermionic construction by Faraggi
et al. \cite{Faraggi:1989ka,Cleaver:1998saa}.  The authors impose an
intermediate $SO(10)$ SUSY GUT.

Progress has been made
recently~\cite{Kobayashi:2004ud,Kobayashi:2004ya,Buchmuller:2005jr,Buchmuller:2006ik,Lebedev:2006kn,Lebedev:2006tr,Lebedev:2007hv,Kim:2006hv,Kim:2006zw,Kim:2007mt}.
In a ``mini-landscape" search of the $E(8) \times E(8)$ heterotic
landscape~\cite{Lebedev:2007hv} 223 models with 3 families, Higgs
doublets and ONLY vector-like exotics were found out of a total of
order 30,000 models or approximately 1 in 100 models searched looked
like the MSSM!  We called this a ``fertile patch" in the heterotic
landscape. Let me describe this focussed search in more detail.

\subsubsection{Phenomenological guidelines} We use the following
guidelines when searching for ``realistic" string
models~\cite{Lebedev:2006kn,Lebedev:2007hv}.   We want to:
\begin{enumerate}
\item Preserve gauge coupling unification; \item  Keep low energy SUSY as solution to the gauge hierarchy problem,
i.e. why is $M_Z << M_G$; \item Put quarks and leptons in {\bf
16} of SO(10); \item Put Higgs in {\bf 10}, thus quarks and
leptons are distinguished from Higgs by their SO(10) quantum
numbers; \item Preserve GUT relations for 3rd family Yukawa
couplings; \item  Use the fact that GUTs accommodate a
``Natural" See-Saw scale ${\cal O}(M_G)$;  \item Use intuition
derived from Orbifold GUT constructions,
\cite{Kobayashi:2004ud,Kobayashi:2004ya} and \item Use local
GUTs to enforce family structure
\cite{Forste:2004ie,Buchmuller:2005jr,Buchmuller:2006ik}.
\end{enumerate}
It is the last two guidelines which are novel and characterize our
approach.

\subsubsection{$E_8 \times E_8$ heterotic string compactified on
$\mz_3 \times \mz_2$ 6D orbifold}

There are many reviews and books on string theory.  I cannot go into
great detail here, so I will confine my discussion to some basic
points.  We start with the 10d heterotic string theory, consisting
of a 26d left-moving bosonic string and a 10d right-moving
superstring. Modular invariance requires the momenta of the internal
left-moving bosonic degrees of freedom (16 of them) to lie in a 16d
Euclidean even self-dual lattice, we choose to be the
$\lgp{E}{8}\times\lgp{E}{8}$ root lattice.  Note, for an orthonormal
basis, the $\lgp{E}{8}$ root lattice consists of the following
vectors, $(n_1,n_2,\cdots,n_8)$ and
$(n_1+\frac{1}{2},n_2+\frac{1}{2},\cdots,n_8+\frac{1}{2})$, where
$n_1, n_2,\cdots n_8$ are integers and $\sum_{i=1}^8 n_i=0\,{\rm
mod}\,2$.

\subsubsection{Heterotic string compactified on $T^6/\mz_6$}

\begin{figure*}  \begin{center}
\scalebox{0.5}{\includegraphics{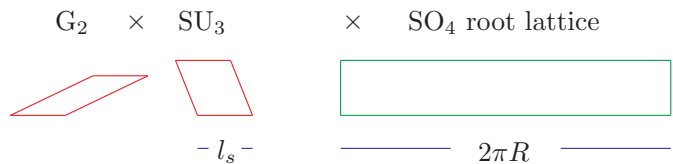}} \caption{$G_2 \oplus SU(3)
\oplus SO(4)$ lattice. Note, we have taken 5 directions with string
scale length $\ell_s$ and one with length $2 \pi R \gg \ell_s$. This
will enable the analogy of an effective 5d orbifold field theory.
\label{fig:untwisted} }  \end{center}
\end{figure*}

We first compactify the theory on a 6d torus defined by the space
group action of translations on $\mathbb{R}^6$ giving the torus
$T^6$ in terms of a factorizable Lie algebra lattice $G_2\oplus
SU(3)\oplus SO(4)$ (see Fig. \ref{fig:untwisted}). Then we mod out
by the $\mz_6$ action on the three complex compactified coordinates
given by $Z^i\rightarrow e^{2\pi i {\bf r}_i\cdot{\bf v}_6}Z^i$,
$i=1,2,3$, where ${\bf v}_6=\frac{1}{6}(1,2,-3)$ is the twist
vector, and ${\bf r}_1=(1,0,0,0)$, ${\bf r}_2=(0,1,0,0)$, ${\bf
r}_3=(0,0,1,0)$. [Together with ${\bf r}_4=(0,0,0,1)$, they form the
set of positive weights of the ${\bf 8}_v$ representation of the
$SO(8)$, the little group in 10d. $\pm{\bf r}_4$ represent the two
uncompactified dimensions in the light-cone gauge. Their space-time
fermionic partners have weights ${\bf
r}=(\pm\frac{1}{2},\pm\frac{1}{2},\pm\frac{1}{2},\pm\frac{1}{2})$
with even numbers of positive signs; they are in the ${\bf 8}_s$
representation of $SO(8)$. In this notation, the fourth component of
${\bf v}_6$ is zero.\label{fn1}]

The $\mz_6$ orbifold is equivalent to a $\mz_2\times\mz_3$ orbifold,
where the two twist vectors are ${\bf v}_2=3{\bf
v}_6=\frac{1}{2}(1,0,-1)$ and ${\bf v}_3=2{\bf
v}_6=\frac{1}{3}(1,-1,0)$. The $\mz_2$ and $\mz_3$ sub-orbifold
twists have the $SU(3)$ and $SO(4)$ planes as their fixed torii. In
Abelian symmetric orbifolds, gauge embeddings of the point group
elements and lattice translations are realized by shifts of the
momentum vectors, ${\bf P}$, in the $E_8\times E_8$ root
lattice.~\cite{Ibanez:1987sn,Bailin:1987xm,Ibanez:1987pj,Font:1989aj,Katsuki:1989bf},
{\it i.e.}, ${\bf P}\rightarrow {\bf P}+k{\bf V}+l{\bf W}$, where
$k, l$ are some integers, and ${\bf V}$ and ${\bf W}$ are known as
the gauge twists and Wilson lines \cite{Ibanez:1986tp}. These
embeddings are subject to modular invariance requirements
\cite{Dixon:1985jw,Dixon:1986jc,Vafa:1986wx}. [Note, the $E_8$ root
lattice is given by the set of states ${\bf P} = \{n_1, n_2, \cdots,
n_8 \}, \ \{n_1 +\frac{1}{2}, n_2 +\frac{1}{2}, \cdots, n_8
+\frac{1}{2} \}$ satisfying $n_i \in \mz, \ \sum_{i = 1}^8 n_i = 2
\mz$.]  The Wilson lines are also required to be consistent with the
action of the point group. In the $\mz_6$ model, there are at most
three consistent Wilson lines \cite{kobayashi}, one of degree 3
(${\bf W}_3$), along the $SU(3)$ lattice, and two of degree 2 (${\bf
W}_2,\,{\bf W}_2'$), along the $SO(4)$ lattice.

\begin{figure*}  \begin{center}
\scalebox{0.5}{\includegraphics{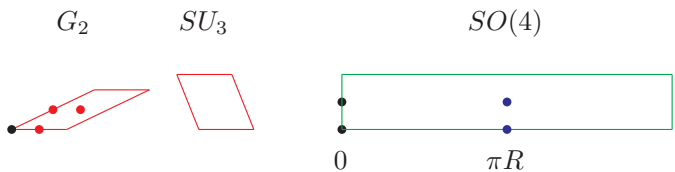}} \caption{$G_2 \oplus SU(3)
\oplus SO(4)$ lattice with $\mz_2$ fixed points. The $T_{3}$ twisted
sector states sit at these fixed points.  The fixed point at the
origin and the symmetric linear combination of the red (grey) fixed
points in the $G_2$ torus have $\gamma =1$. \label{fig:T3} }
\end{center}
\end{figure*}

\begin{figure*} \begin{center}
\scalebox{0.5}{
\includegraphics{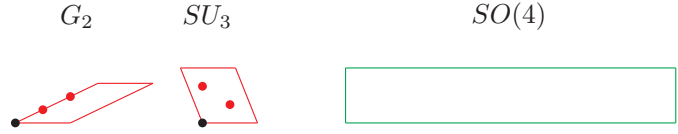}}
\caption{\label{fig:T2} $G_2 \oplus SU(3) \oplus SO(4)$ lattice with
$\mz_3$ fixed points for the $T_2$ twisted sector.  The fixed point
at the origin and the symmetric linear combination of the red (grey)
fixed points in the $G_2$ torus have $\gamma =1$.}  \end{center}
\end{figure*}

\begin{figure*} \begin{center}
\scalebox{0.5}{
\includegraphics{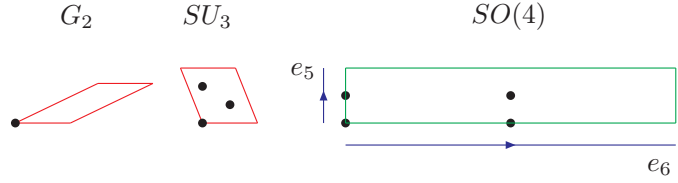}}
\caption{\label{fig:T1} $G_2 \oplus SU(3) \oplus SO(4)$ lattice with
$\mz_6$ fixed points. The $T_{1}$ twisted sector states sit at these
fixed points.}  \end{center}
\end{figure*}

The $\mz_6$ model has three untwisted sectors ($U_i,\,i=1,2,3$) and
five twisted sectors ($T_i,\,i=1,2,\cdots,5$). (The $T_k$ and
$T_{6-k}$ sectors are CPT conjugates of each other.) The twisted
sectors split further into sub-sectors when discrete Wilson lines
are present. In the $SU(3)$ and $SO(4)$ directions, we can label
these sub-sectors by their winding numbers, $n_3=0,1,2$ and
$n_2,\,n'_2=0,1$, respectively. In the $G_2$ direction, where both
the $\mz_2$ and $\mz_3$ sub-orbifold twists act, the situation is
more complicated.  There are four $\mz_2$ fixed points in the $G_2$
plane. Not all of them are invariant under the $\mz_3$ twist, in
fact three of them are transformed into each other. Thus for the
$T_3$ twisted-sector states one needs to find linear combinations of
these fixed-point states such that they have definite eigenvalues,
$\gamma=1$ (with multiplicity 2), $e^{i2\pi/3}$, or $e^{i4\pi/3}$,
under the orbifold twist \cite{DFMS,kobayashi} (see
Fig.~\ref{fig:T3}). Similarly, for the $T_{2,4}$ twisted-sector
states, $\gamma=1$ (with multiplicity 2) and $-1$ (the fixed points
of the $T_{2,4}$ twisted sectors in the $G_2$ torus are shown in
Fig.~\ref{fig:T2}). The $T_{1}$ twisted-sector states have only one
fixed point in the $G_2$ plane, thus $\gamma=1$ (see
Fig.~\ref{fig:T1}). The eigenvalues $\gamma$ provide another piece
of information to differentiate twisted sub-sectors.


Massless states in 4d string models consist of those momentum
vectors ${\bf P}$ and ${\bf r}$ ($\bf r$ are in the $SO(8)$ weight
lattice) which satisfy the following mass-shell equations
\cite{Dixon:1985jw,Dixon:1986jc,Ibanez:1987sn,Bailin:1987xm,Ibanez:1987pj,Font:1989aj,Katsuki:1989bf},
\bea 
&&\frac{\alpha'}{2}m_{R}^2=N^k_R+\frac{1}{2}\left|{\bf r}+k{\bf v}\right|^2+a^k_R=0\,,\label{masscond1}\\
&&\frac{\alpha'}{2}m_L^2=N_L^k+\frac{1}{2}\left|{\bf P}+k{\bf
X}\right|^2+a_L^k=0\,, \label{masscond2} 
\eea 
where $\alpha'$ is the Regge slope, $N^k_R$ and $N^k_L$ are
(fractional) numbers of the right- and left-moving (bosonic)
oscillators, ${\bf X}={\bf V}+n_3{\bf W}_3+n_2{\bf W}_2+n_2'{\bf
W}_2'$, and $a^k_R$, $a_L^k$ are the normal ordering constants, 
\bea 
a^k_R
&=&-\frac{1}{2}+\frac{1}{2}\sum_{i=1}^3|{\widehat{kv_i}}|\left(1-|{\widehat{kv_i}}|\right)\,,\nn\\
a_L^k
&=&-1+\frac{1}{2}\sum_{i=1}^3|{\widehat{kv_i}}|\left(1-|{\widehat{kv_i}}|\right)\,,
\eea 
with $\widehat{kv_i}={\rm mod}(kv_i, 1)$.

These states are subject to a generalized Gliozzi-Scherk-Olive (GSO)
projection ${\cal
P}=\frac{1}{6}\sum_{\ell=0}^{5}\Delta^\ell$~\cite{Ibanez:1987sn,Bailin:1987xm,Ibanez:1987pj,Font:1989aj,Katsuki:1989bf}.
For the simple case of the $k$-th twisted sector ($k=0$ for the
untwisted sectors) with no Wilson lines ($n_3 = n_2 = n^\prime_2 =
0$) we have
\be
\Delta= \gamma\phi  \exp\left\{i\pi \biggl[(2{\bf P}+k{\bf X})
\cdot{\bf X} -(2{\bf r}+k{\bf v})\cdot {\bf
v}\biggr]\right\},\label{GSO}
\ee
where $\phi$ are phases from bosonic oscillators.   However, in the
$\mz_6$ model, the GSO projector must be modified for the
untwisted-sector and $T_{2,4}$, $T_3$ twisted-sector states in the
presence of Wilson lines \cite{Kobayashi:2004ya}. The Wilson lines
split each twisted sector into sub-sectors and there must be
additional projections with respect to these sub-sectors. This
modification in the projector gives the following projection
conditions,
\be
{\bf P}\cdot{\bf V}-{\bf r}_i\cdot{\bf v}=\mz\,\,\,\,(i=1,2,3),\quad
{\bf P}\cdot{\bf W}_3,\,\,{\bf P}\cdot{\bf W}_2,\,\,{\bf P}\cdot{\bf
W}_2'=\mz, \label{eq1}
\ee
for the untwisted-sector states, and
\be
T_{2,4}:\,{\bf P}\cdot{\bf W}_2,\,\,{\bf P}\cdot{\bf W}_2'={\mathbb
Z}\,,\qquad T_3:\,{\bf P}\cdot{\bf W}_3={\mathbb Z}\,,\label{eq2}
\ee
for the $T_{2,3,4}$ sector states (since twists of these sectors
have fixed torii). There is no additional condition for the $T_1$
sector states.

\subsubsection{An orbifold GUT -- heterotic string dictionary}

We first implement the $\mz_3$ sub-orbifold twist, which acts only
on the $G_2$ and $SU(3)$ lattices. The resulting model is a 6d gauge
theory with ${\cal N}=2$ hypermultiplet matter, from the untwisted
and $T_{2,4}$ twisted sectors. This 6d theory is our starting point
to reproduce the orbifold GUT models.  The next step is to implement
the $\mz_2$ sub-orbifold twist. The geometry of the extra dimensions
closely resembles that of 6d orbifold GUTs. The $SO(4)$ lattice has
four $\mz_2$ fixed points at $0$, $\pi R$, $\pi R^\prime$ and $\pi(R
+ R^\prime)$, where $R$ and $R'$ are on the $e_5$ and $e_6$ axes,
respectively, of the lattice (see Figs.~\ref{fig:T3} and
\ref{fig:T1}). When one varies the modulus parameter of the $SO(4)$
lattice such that the length of one axis ($R$) is much larger than
the other ($R'$) and the string length scale ($\ell_s$), the lattice
effectively becomes the $S^1/\mz_2$ orbi-circle in the 5d orbifold
GUT, and the two fixed points at $0$ and $\pi R$ have degree-2
degeneracies. Furthermore, one may identify the states in the
intermediate $\mz_3$ model, {\it i.e.} those of the untwisted and
$T_{2,4}$ twisted sectors, as bulk states in the orbifold GUT.

Space-time supersymmetry and GUT breaking in string models work
exactly as in the orbifold GUT models.  First consider supersymmetry
breaking. In the field theory, there are two gravitini in 4d, coming
from the 5d (or 6d) gravitino. Only one linear combination is
consistent with the space reversal, $y\rightarrow -y$; this breaks
the ${\cal N}=2$ supersymmetry to that of ${\cal N}=1$. In string
theory, the space-time supersymmetry currents are represented by
those half-integral $SO(8)$ momenta.  [Together with ${\bf
r}_4=(0,0,0,1)$, they form the set of positive weights of the ${\bf
8}_v$ representation of the $SO(8)$, the little group in 10d.
$\pm{\bf r}_4$ represent the two uncompactified dimensions in the
light-cone gauge. Their space-time fermionic partners have weights
${\bf
r}=(\pm\frac{1}{2},\pm\frac{1}{2},\pm\frac{1}{2},\pm\frac{1}{2})$
with even numbers of positive signs; they are in the ${\bf 8}_s$
representation of $SO(8)$. In this notation, the fourth component of
${\bf v}_6$ is zero.]  The $\mz_3$ and $\mz_2$ projections remove
all but two of them, ${\bf r}=\pm\frac{1}{2}(1,1,1,1)$; this gives
${\cal N}=1$ supersymmetry in 4d.

Now consider GUT symmetry breaking. As usual, the $\mz_2$ orbifold
twist and the translational symmetry of the $SO(4)$ lattice are
realized in the gauge degrees of freedom by degree-2 gauge twists
and Wilson lines respectively. To mimic the 5d orbifold GUT example,
we impose only one degree-2 Wilson line, ${\bf W}_2$, along the long
direction of the $SO(4)$ lattice, ${\bf R}$.  [Wilson lines can be
used to reduce the number of chiral families. In all our models, we
find it is sufficient to get three-generation models with two Wilson
lines, one of degree 2 and one of degree 3. Note, however, that with
two Wilson lines in the $SO(4)$ torus we can break $SO(10)$ directly
to $SU(3) \times SU(2) \times U(1)_Y \times U(1)_X$ (see for
example, Ref.~\cite{6dOGUT}).]  The gauge embeddings generally break
the 5d/6d (bulk) gauge group further down to its subgroups, and the
symmetry breaking works exactly as in the orbifold GUT models. This
can clearly be seen from the following string theoretical
realizations of the orbifold parities
\be P=p\,e^{2\pi i\,[{\bf P}\cdot{\bf V}_2-{\bf r}\cdot{\bf
v}_2]}\,,\quad P'=p\,e^{2\pi i\,[{\bf P}\cdot({\bf V}_2+{\bf
W}_2)-{\bf r}\cdot{\bf v}_2]}\,,\label{PP'}
\ee
where ${\bf V}_2=3{\bf V}_6$, and $p=\gamma\phi$ can be identified
with intrinsic parities in the field theory language. [For gauge and
untwisted-sector states, $p$ are trivial. For non-oscillator states
in the $T_{2,4}$ twisted sectors, $p=\gamma$ are the eigenvalues of
the $G_2$-plane fixed points under the ${\mathbb Z}_2$ twist. Note
that $p=+$ and $-$ states have multiplicities $2$ and $1$
respectively since the corresponding numbers of fixed points in the
$G_2$ plane are $2$ and $1$.] Since $2({\bf P}\cdot{\bf V}_2-{\bf
r}\cdot{\bf v}_2),\,2{\bf P}\cdot{\bf W}_2=\mz$, by properties of
the $E_8\times E_8$ and $SO(8)$ lattices, thus $P^2=P'^2=1$, and
Eq.~(\ref{PP'}) provides a representation of the orbifold parities.
From the string theory point of view, $P=P'=+$ are nothing but the
projection conditions, $\Delta=1$, for the untwisted and $T_{2,4}$
twisted-sector states (see Eqs.~(\ref{GSO}), (\ref{eq1}) and
(\ref{eq2})).

To reaffirm this identification, we compare the masses of KK
excitations derived from string theory with that of orbifold GUTs.
The coordinates of the $SO(4)$ lattice are untwisted under the
$\mz_3$ action, so their mode expansions are the same as that of
toroidal coordinates. Concentrating on the ${\bf R}$ direction, the
bosonic coordinate is
$X_{L,R}=x^{}_{L,R}+p^{}_{L,R}(\tau\pm\sigma)+{\rm oscillator\,
terms}$, with $p^{}_L$, $p^{}_R$ given by
\bea
p^{}_L= & \frac{m}{2R}+\left(1-\frac{1}{4}|{\bf W}_2|^2\right)
\frac{n_2R}{\ell_s^2}+\frac{{\bf P}\cdot{\bf W}_2}{2R} & \,,\qquad
\nn
\\
p^{}_R= & p^{}_L-\frac{2n_2R}{\ell_s^2}\,,\label{PLR} &
\eea
where $m$ ($n_2$) are KK levels (winding numbers). The ${\mathbb
Z}_2$ action maps $m$ to $-m$, $n_2$ to $-n_2$ and ${\bf W}_2$ to
$-{\bf W}_2$, so physical states must contain linear combinations,
$|m,n_2\rangle\pm|-m,-n_2\rangle$; the eigenvalues $\pm 1$
correspond to the first ${\mathbb Z}_2$ parity, $P$, of orbifold GUT
models. The second orbifold parity, $P'$, induces a non-trivial
degree-2 Wilson line; it shifts the KK level by $m\rightarrow m+{\bf
P}\cdot{\bf W}_2$. Since $2{\bf W}_2$ is a vector of the (integral)
$E_8\times E_8$ lattice, the shift must be an integer or
half-integer. When $R\gg R'\sim\ell_s$, the winding modes and the KK
modes in the smaller dimension of $SO(4)$ decouple.  Eq.~(\ref{PLR})
then gives four types of KK excitations, reproducing the field
theoretical mass formula in Eq.~(\ref{kkmass}).

\subsubsection{MSSM with R parity  \label{sec:minilandscape}}

In this section we discuss just one ``benchmark" model (Model 1)
obtained via a ``mini-landscape" search
\cite{Lebedev:2006kn,Lebedev:2007hv,Lebedev:2008un} of the $E_8
\times E_8$ heterotic string compactified on the $\mz_6$ orbifold
\cite{Lebedev:2007hv}.  [For earlier work on MSSM models from
$\mz_6$ orbifolds of the heterotic string, see
\cite{Buchmuller:2005jr,Buchmuller:2006ik}. For reviews, see
\cite{Nilles:2008gq,Raby:2009dm}]

The model is defined by the shifts and Wilson lines
\begin{eqnarray}
 V & = &
 \left( \frac{1}{3},-\frac{1}{2},-\frac{1}{2},0,0,0,0,0\right)\,
   \left(\frac{1}{2},-\frac{1}{6},-\frac{1}{2},-\frac{1}{2},-\frac{1}{2},-\frac{1}{2},-\frac{1}{2},\frac{1}{2}\right)\;, \\
 W_2 & = &
 \left( 0,-\frac{1}{2},-\frac{1}{2},-\frac{1}{2},\frac{1}{2},0,0,0\right)\,
   \left(
   4,-3,-\frac{7}{2},-4,-3,-\frac{7}{2},-\frac{9}{2},\frac{7}{2}\right)\;, \\
 W_3 & = &
 \left(-\frac{1}{2},-\frac{1}{2},\frac{1}{6},\frac{1}{6},\frac{1}{6},\frac{1}{6},\frac{1}{6},\frac{1}{6}\right)\,
   \left( \frac{1}{3},0,0,\frac{2}{3},0,\frac{5}{3},-2,0\right)\;.
\end{eqnarray}
A possible second order 2 Wilson line is set to zero.

The shift $V$ is defined to satisfy two criteria. \begin{itemize}
\item The first criterion is the existence of a local $SO(10)$ GUT at the $T_1$ fixed points at
$x_6 =0$ in the $SO(4)$ torus (Fig. \ref{fig:T1}). \be P \cdot V =
\mathbb{Z};  \;  P \in SO(10) \;\; {\rm momentum \; lattice} .\ee
Since the $T_1$ twisted sector has no invariant torus and only one
Wilson line along the $x_6$ direction,  all states located at these
two fixed points must come in complete $SO(10)$ multiplets. [For
more discussion on local GUTs, see
\cite{Forste:2004ie,Buchmuller:2005jr}]
\item  The second criterion is that two massless spinor representations of $SO(10)$ are
located at the $x_6 = 0$ fixed points.
\end{itemize}
Hence, the two complete families on the local $SO(10)$ GUT fixed
points gives us an excellent starting point to find the MSSM.  The
Higgs doublets and third family of quarks and leptons must then come
from elsewhere.

Let us now discuss the effective 5d orbifold GUT
\cite{Dundee:2008ts}.  Consider the orbifold $(T^2)^3/\mz_3$ plus
the Wilson line $W_3$ in the $SU_3$ torus.  The $\mz_3$ twist does
not act on the $SO_4$ torus, see Fig. \ref{fig:T2}. As a consequence
of embedding the $\mz_3$ twist as a shift in the $E_8 \times E_8$
group lattice and taking into account the $W_3$ Wilson line, the
first $E_8$ is broken to $SU(6)$.  This gives the effective 5d
orbifold gauge multiplet contained in the ${\mathcal N}=1$ vector
field $V$.  In addition we find the massless states $\Sigma = {\bf
35}$, and ${\bf 20} + {\bf 20}^c$ and 18 (${\bf 6} + {\bf 6}^c$) in
the 6d untwisted sector and $T_2, \ T_4$ twisted sectors. Together
these form a complete ${\mathcal N}=2$ gauge multiplet ($V +
\Sigma$) (see Eqns. \ref{eq:V6trans} and \ref{eq:phi6trans}) and a
{\bf 20} + 18 ({\bf 6}) dimensional hypermultiplets.\footnote{Note,
in four dimensions massless chiral adjoints cannot be obtained at
level 1 Kac-Moody algebras.  Clearly this is not true in 5d.} In
fact the massless states in this sector can all be viewed as ``bulk"
states moving around in a large 5d space-time.

Now consider the $\mz_2$ twist and the Wilson line $W_2$ along the
$x_6$ axis in the $SO_4$ torus.   The action of the $\mz_2$ twist
breaks the gauge group to $SU(5)$, while $W_2$ breaks $SU(5)$
further to the SM gauge group $SU(3)_C \times SU(2)_L \times U(1)_Y$
(the SM is represented by the red states in Eqn. \ref{eq:V6trans}).

Let us focus on those MSSM states located in the bulk.  From two of
the pairs of ${\mathcal N}=1$ chiral multiplets $\bf{6} + \bf{6}^c$,
which decompose as \bea \nonumber 2 \times (\bf 6 + \bf 6^c) \supset
& & \\ \nn  \left[(1,\bf 2)_{1,1}^{--}+(\bf
3,1)_{-2/3,1/3}^{-+}\right]+ \left[(1,\bf
2)_{-1,-1}^{++}+(\overline{\bf 3},1)_{2/3,-1/3}^{--}\right] & &
\\
 +\left[(1,\bf 2)_{1,1}^{-+}+(\bf 3,1)_{-2/3,1/3}^{--}\right]+
\left[(1,\bf 2)_{-1,-1}^{+-}+(\overline{\bf
3},1)_{2/3,-1/3}^{++}\right], & & \eea we obtain the third family
$b^c$ and lepton doublet, $l$. The rest of the third family comes
from the $\mathbf{10} + \mathbf{10}^c$ of $SU(5)$ contained in the
$\mathbf{20} + \mathbf{20}^c$ of $SU(6)$, in the untwisted sector.

Now consider the Higgs bosons.  The bulk gauge symmetry is $SU(6)$.
Under $SU(5) \times U(1)$, the adjoint decomposes as
\begin{equation}
   \mathbf{35} \rightarrow \mathbf{24}_0 + \mathbf{5}_{+1} + \mathbf{5}^c_{-1} + 1_0.
\end{equation}
The Higgs doublets are represented by the blue states in Eqn.
\ref{eq:phi6trans}.  Thus the MSSM Higgs sector emerges from the
breaking of the $SU(6)$ adjoint by the orbifold and the model
satisfies the property of ``gauge-Higgs unification."

To summarize, in models with gauge-Higgs unification, the Higgs
multiplets come from the 5d \textit{vector} multiplet ($V, \Sigma$),
both in the adjoint representation of $SU(6)$.   $V$ is the 4d gauge
multiplet and the 4d chiral multiplet $\Sigma$ contains the Higgs
doublets. These states transform as follows under the orbifold
parities $(P\,\,P')$:
\begin{equation}
  V:\: \left( \begin{array}{ccc|cc|c}
   {\color{red} (+ +)} & {\color{red}  (+ +)} & {\color{red}  (+ +)}  & (+ -) & (+ -) & (- +) \\
   {\color{red}  (+ +)} & {\color{red} (+ +) } & {\color{red} (+ +)} & (+ -) & (+ -) & (- +) \\
    {\color{red} (+ +)} & {\color{red} (+ +)} & {\color{red} (+ +)} & (+ -) & (+ -) & (- +) \\ \hline
    (+ -) & (+ -) & (+ -) & {\color{red} (+ +)} & {\color{red} (+ +)} & (- -) \\
    (+ -) & (+ -) & (+ -) & {\color{red} (+ +)} & {\color{red} (+ +)} & (- -) \\ \hline
    (- +) & (- +) & (- +) & (- -) & (- -) & (+ +)
  \end{array} \right)
\label{eq:V6trans}
\end{equation}
\begin{equation}
  \Phi:\: \left( \begin{array}{ccc|cc|c}
    (- -) & (- -) & (- -) & (- +) & (- +) & (+ -) \\
    (- -) & (- -) & (- -) & (- +) & (- +) & (+ -) \\
    (- -) & (- -) & (- -) & (- +) & (- +) & (+ -) \\ \hline
    (- +) & (- +) & (- +) & (- -) & (- -) & {\color{blue} (+ +)} \\
    (- +) & (- +) & (- +) & (- -) & (- -) & {\color{blue} (+ +)} \\ \hline
    (+ -) & (+ -) & (+ -) & {\color{blue}(+ +)} & {\color{blue}(+ +)} & (- -)
  \end{array} \right).
\label{eq:phi6trans}
\end{equation}
Hence,  we have obtained doublet-triplet splitting via orbifolding.

\subsubsection{$D_4$ Family Symmetry}

\begin{figure*} \begin{center}
 \scalebox{0.5} {
  \includegraphics{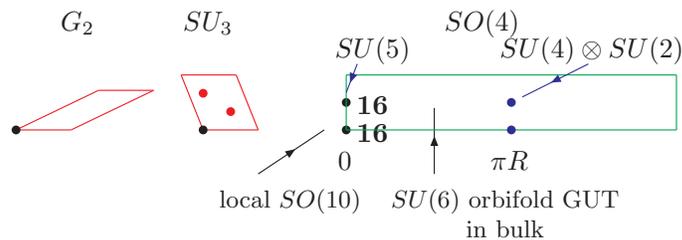}}
  \caption{The two families in the $T_1$ twisted sector.
  \label{fig:2fam}} \end{center}
\end{figure*}

Consider the $\mz_2$ fixed points.   We have four fixed points,
separated into an $SU(5)$ and SM invariant pair by the $W_2$ Wilson
line (see Fig. \ref{fig:2fam}).   We find two complete families, one
on each of the $SO(10)$ fixed points and a small set of vector-like
exotics (with fractional electric charge) on the other fixed points.
Since $W_2$ is in the direction orthogonal to the two families, we
find a non-trivial $D_4$ family symmetry. This will affect a
possible hierarchy of fermion masses. We will discuss the family
symmetry and the exotics in more detail next.

The discrete group $D_4$ is a non-abelian discrete subgroup of
$SU_2$ of order 8.  It is generated by the set of $2 \times 2$ Pauli
matrices
\begin{equation}
D_4 =  \{ \pm 1, \pm \sigma_1, \pm \sigma_3,  \mp i \sigma_2 \}.
\end{equation}
In our case, the action of the transformation  $\sigma_1 = \left(
\begin{array}{cc} 0 & 1 \\ 1 & 0 \end{array} \right)$ takes $F_1
\leftrightarrow F_2$,  while the action of $\sigma_3 = \left(
\begin{array}{cc} 1 &0 \\ 0 & -1
\end{array} \right)$ takes  $F_2 \rightarrow - F_2$.   These are symmetries of the string.
The first is a symmetry which interchanges the two light families
located at opposite sides of one cycle of the orbifolded $SO(4)$
torus.   This symmetry corresponds to the geometric translation half
way around the cycle which is not broken by the Wilson line lying
along the orthogonal cycle.   The second is a result of so-called
space-group selection rules which require an even number of states
at each of these two fixed points. As a result, the theory is
invariant under the action of multiplying each state located at,
say, the lower fixed point by minus one.

Under $D_4$ the three families of quarks and leptons transform as a
doublet, ($F_1, \ F_2$), and a singlet, $F_3$.  As a consequence of
$D_4$ (and additional U(1) symmetries), only the third family can
have a tree level Yukawa coupling to the Higgs (which is also a
$D_4$ singlet).   All others Yukawa couplings can only be obtained
once the family symmetries are broken.  Thus the string theory
includes a natural Froggatt-Nielsen mechanism~\cite{Froggatt:1978nt}
for generating a hierarchy of fermion masses.
In summary:
\begin{itemize}
\item Since the top quarks and the Higgs are derived from the $SU(6)$ chiral adjoint and {\bf 20} hypermultiplet in the 5D
bulk, they have a tree level Yukawa interaction given by
\begin{equation}
W \supset \frac{g_5}{\sqrt{\pi R}} \int_0^{\pi R} dy {\bf {20^c}} \ \Sigma \
{\bf 20} = g_G \ Q_3 \ H_u \ U^c_3
\end{equation}
where $g_5$ ($g_G$) is the 5d (4d) $SU(6)$ gauge coupling
constant evaluated at the string scale.  [For a detailed
analysis of gauge-Yukawa unification in this context, see
\cite{Hosteins:2009xk}.]

\item The first two families reside at the $\mz_2$ fixed points, resulting in a $D_4$ family symmetry.
Hence family symmetry breaking may be used to generate a
hierarchy of fermion masses. [For a discussion of $D_4$ family
symmetry and phenomenology, see Ref. \cite{Ko:2007dz}. For a
general discussion of discrete non-Abelian family symmetries
from orbifold compactifications of the heterotic string, see
\cite{Kobayashi:2006wq}.]
\end{itemize}

\subsubsection{More details of ``Benchmark" Model 1
\cite{Lebedev:2007hv}}

In order to analyze the theory further one must construct the
superpotential and Kahler potential.  We have assumed the zeroth
order perturbative Kahler potential in our analysis.   The
superpotential on the other hand can be obtained as a power series
in chiral superfields.   Each monomial in this series is a
holomorphic product of chiral superfields to some given order, $n$,
in the fields.   We have considered terms up to $n = 8$. The
monomials in the superpotential are strongly constrained by string
selection rules and gauge invariance. In our analysis we allow all
monomials consistent with these selection rules. Although the
coefficients of each term can in principle be determined by
calculating the vacuum expectation value of the product of the
appropriate vertex operators; such a calculation is prohibitive due
to the fact that there are hundreds of terms in the polynomial at
each order $n$. Thus we assume that any term that is permitted by
the selection rules appears in the superpotential with order one
coefficient.

Let us now consider the spectrum, exotics, R parity, Yukawa
couplings, and neutrino masses.   In Table \ref{tab:states} we list
the states of the model.   In addition to the three families of
quarks and leptons and one pair of Higgs doublets, we have
vector-like exotics (states which can obtain mass without breaking
any SM symmetry) and SM singlets.   The SM singlets enter the
superpotential in several important ways.   They can give mass to
the vector-like exotics via effective mass terms of the form \be E
E^c \tilde S^n  \ee  where  $E, E^c$ ($\tilde S$) represent the
vector-like exotics and SM singlets respectively.   We have checked
that all vector-like exotics and unwanted U(1) gauge bosons obtain
mass at supersymmetric points in moduli space with $F = D = 0$;
leaving only the MSSM states at low energy.   The SM singlets also
generate effective Yukawa matrices for quarks and leptons, including
neutrinos.  The charged fermion Yukawa matrices are
\begin{eqnarray}
Y_u&=& \left(
\begin{array}{ccc}
 \widetilde{s}^5 & \widetilde{s}^5 & \widetilde{s}^5 \\
 \widetilde{s}^5 & \widetilde{s}^5 & \widetilde{s}^5 \\
 \widetilde{s}^6 & \widetilde{s}^6 & 1
\end{array}
\right)\;,\quad Y_d~=~ \left(
\begin{array}{ccc}
 \widetilde{s}^5 & \widetilde{s}^5 & 0  \\
 \widetilde{s}^5 & \widetilde{s}^5 & 0  \\
 \widetilde{s}^6 & \widetilde{s}^6 & 0
\end{array}
\right)\;,\quad Y_e~=~ \left(
\begin{array}{ccc}
 \widetilde{s}^5 & \widetilde{s}^5 & \widetilde{s}^6 \\
 \widetilde{s}^5 & \widetilde{s}^5 & \widetilde{s}^6 \\
 \widetilde{s}^6 & \widetilde{s}^6 & 0
\end{array}
\right)\;.\nonumber\\
& &
\end{eqnarray}
where $\tilde s^n$ represents a polynomial in SM singlets beginning
at order $n$ in the product of fields. And we have shown that the
three left-handed neutrinos get small mass due to a non-trivial
See-Saw mechanism involving the 16 right-handed neutrinos and their
13 conjugates. All in all, this ``benchmark" model looks very much
like the MSSM!

In addition, one of the most important constraints in this
construction is the existence of an exact low energy R parity. In
this model we identified a generalized \BmL (see Table
\ref{tab:states}) which is standard for the SM states and
vector-like on the vector-like exotics.  This $\BmL$ naturally
distinguishes the Higgs and lepton doublets. Moreover we found SM
singlet states \be \tilde S = \{ h_i, \ \chi_i, \ s_i^0 \} \ee which
can get vacuum expectation values preserving a matter parity
$\mathbb{Z}^{\cal M}_2$ subgroup of $U(1)_{\BmL}$.  It is this set
of SM singlets which give vector-like exotics mass and effective
Yukawa matrices for quarks and leptons. The $\chi$ fields
spontaneously break \BmL leaving over a discrete ${\mathbb Z}_2$
matter parity under which all quarks and leptons are odd and Higgs
doublets are even. This symmetry enforces an exact R-parity
forbidding the baryon or lepton number violating operators, $ \bar U
\bar D \bar D, \  Q L \bar D, \ L L \bar E, \ L H_u$.

Finally the mu term vanishes in the supersymmetric limit.  This is a
consequence of the fact that the coefficient of the $H_u H_d$ term
in the superpotential has vacuum quantum numbers.  Thus any product
of SM singlets which can appear in the pure singlet superpotential
can appear as an effective mu term.   In fact both the mu term and
the singlet superpotential vanish to order 6 in the product of
fields.   Hence in the supersymmetric vacuum the VEV of the
superpotential and the mu term both vanish. As a consequence, when
supergravity is considered, the supersymmetric vacuum is consistent
with flat Minkowski space. [For a recent discussion on the vanishing
of the $\mu$ term and its connection to approximate $R$ symmetries,
see \cite{Brummer:2010fr}.]

Note,   Yukawa couplings,  gauge couplings and  vector-like exotic
masses are functions  of moduli  (along SUSY flat directions). Some
of these moduli are blow up modes for some,  BUT NOT ALL,  of the
orbifold fixed points.   In fact,  two fixed points are NOT blown
up!

\begin{table}[h]
\centerline{
\begin{tabular}{|c|l|l|c|c|l|l|}
\hline
\# & irrep & label & & \# & irrep & label\\
\hline
 3 &
$\left(\boldsymbol{3},\boldsymbol{2};\boldsymbol{1},\boldsymbol{1}\right)_{(1/3,1/3)}$
 & $q_i$
 & &
 3 &
$\left(\overline{\boldsymbol{3}},\boldsymbol{1};\boldsymbol{1},\boldsymbol{1}\right)_{(-4/3,-1/3)}$
 & $\bar u_i$
 \\
 3 &
$\left(\boldsymbol{1},\boldsymbol{1};\boldsymbol{1},\boldsymbol{1}\right)_{(2,1)}$
 & $\bar e_i$
 & &
 8 &
$\left(\boldsymbol{1},\boldsymbol{2};\boldsymbol{1},\boldsymbol{1}\right)_{(0,*)}$
 & $m_i$
 \\
 4 &
$\left(\overline{\boldsymbol{3}},\boldsymbol{1};\boldsymbol{1},\boldsymbol{1}\right)_{(2/3,-1/3)}$
 & $\bar d_i$
 & &
 1 &
$\left(\boldsymbol{3},\boldsymbol{1};\boldsymbol{1},\boldsymbol{1}\right)_{(-2/3,1/3)}$
 & $d_i$
 \\
 4 &
$\left(\boldsymbol{1},\boldsymbol{2};\boldsymbol{1},\boldsymbol{1}\right)_{(-1,-1)}$
 & $\ell_i$
 & &
 1 &
$\left(\boldsymbol{1},\boldsymbol{2};\boldsymbol{1},\boldsymbol{1}\right)_{(1,1)}$
 & $\bar \ell_i$
 \\
 1 &
$\left(\boldsymbol{1},\boldsymbol{2};\boldsymbol{1},\boldsymbol{1}\right)_{(-1,0)}$
 & $\phi_i$
 & &
 1 &
$\left(\boldsymbol{1},\boldsymbol{2};\boldsymbol{1},\boldsymbol{1}\right)_{(1,0)}$
 & $\bar \phi_i$
 \\
 6 &
$\left(\overline{\boldsymbol{3}},\boldsymbol{1};\boldsymbol{1},\boldsymbol{1}\right)_{(2/3,2/3)}$
 & $\bar \delta_i$
 & &
 6 &
$\left(\boldsymbol{3},\boldsymbol{1};\boldsymbol{1},\boldsymbol{1}\right)_{(-2/3,-2/3)}$
 & $\delta_i$
 \\
 14 &
$\left(\boldsymbol{1},\boldsymbol{1};\boldsymbol{1},\boldsymbol{1}\right)_{(1,*)}$
 & $s^+_i$
 & &
 14 &
$\left(\boldsymbol{1},\boldsymbol{1};\boldsymbol{1},\boldsymbol{1}\right)_{(-1,*)}$
 & $s^-_i$
 \\
 16 &
$\left(\boldsymbol{1},\boldsymbol{1};\boldsymbol{1},\boldsymbol{1}\right)_{(0,1)}$
 & $\bar n_i$
 & &
 13 &
$\left(\boldsymbol{1},\boldsymbol{1};\boldsymbol{1},\boldsymbol{1}\right)_{(0,-1)}$
 & $n_i$
 \\
 5 &
$\left(\boldsymbol{1},\boldsymbol{1};\boldsymbol{1},\boldsymbol{2}\right)_{(0,1)}$
 & $\bar \eta_i$
 & &
 5 &
$\left(\boldsymbol{1},\boldsymbol{1};\boldsymbol{1},\boldsymbol{2}\right)_{(0,-1)}$
 & $\eta_i$
 \\
 10 &
$\left(\boldsymbol{1},\boldsymbol{1};\boldsymbol{1},\boldsymbol{2}\right)_{(0,0)}$
 & $h_i$
 & &
 2 &
$\left(\boldsymbol{1},\boldsymbol{2};\boldsymbol{1},\boldsymbol{2}\right)_{(0,0)}$
 & $y_i$
 \\
 6 &
$\left(\boldsymbol{1},\boldsymbol{1};\boldsymbol{4},\boldsymbol{1}\right)_{(0,*)}$
 & $f_i$
 & &
 6 &
$\left(\boldsymbol{1},\boldsymbol{1};\overline{\boldsymbol{4}},\boldsymbol{1}\right)_{(0,*)}$
 & $\bar f_i$
 \\
 2 &
$\left(\boldsymbol{1},\boldsymbol{1};\boldsymbol{4},\boldsymbol{1}\right)_{(-1,-1)}$
 & $f_i^-$
 & &
 2 &
$\left(\boldsymbol{1},\boldsymbol{1};\overline{\boldsymbol{4}},\boldsymbol{1}\right)_{(1,1)}$
 & ${\bar f_i}^+$
 \\
 4 &
$\left(\boldsymbol{1},\boldsymbol{1};\boldsymbol{1},\boldsymbol{1}\right)_{(0,\pm2)}$
 & $\chi_i$
 & &
 32 &
$\left(\boldsymbol{1},\boldsymbol{1};\boldsymbol{1},\boldsymbol{1}\right)_{(0,0)}$
 & $s^0_i$
 \\
 2 &
$\left(\overline{\boldsymbol{3}},\boldsymbol{1};\boldsymbol{1},\boldsymbol{1}\right)_{(-1/3,2/3)}$
 & $\bar v_i$
 & &
 2 &
$\left(\boldsymbol{3},\boldsymbol{1};\boldsymbol{1},\boldsymbol{1}\right)_{(1/3,-2/3)}$
 & $v_i$
 \\
\hline
\end{tabular}
} \caption{Spectrum. The quantum numbers under
$\SU3\times\SU2\times[\SU4\times\SU2']$ are shown in boldface;
hypercharge and \BmL\ charge appear as subscripts.  Note that the
states $s_i^\pm$, $f_i$, $\bar f_i$ and $m_i$ have different $B-L$
charges for different $i$, which we do not explicitly list
\cite{Lebedev:2007hv}.} \label{tab:states}
\end{table}

\subsubsection{\label{sec:protondecay} Gauge Coupling Unification and Proton Decay}

\begin{figure}[ht!]
        \centering
        \includegraphics[scale = 0.3]{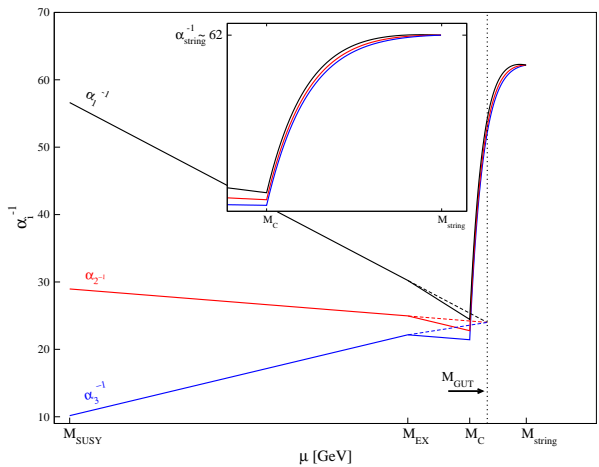}
        \caption{An example of the type of gauge coupling evolution we see in these models,
        versus the typical behavior in the MSSM.  The ``tail'' is due to the power law running
        of the couplings when towers of Kaluza-Klein modes are involved.  Unification in this
        model occurs at $M_{\rm STRING} \simeq 5.5\times 10^{17}$ GeV, with a compactification scale
        of $M_{c} \simeq 8.2 \times 10^{15}$ GeV, and an exotic mass scale of $M_{\rm EX} \simeq 8.2 \times 10^{13}$ GeV.}
        \label{fig:beta_functions}
\end{figure}

We have checked whether the SM gauge couplings unify at the string
scale in the class of models similar to Model 1 above
\cite{Dundee:2008ts}.  All of the 15 MSSM-like models of Ref.
\cite{Lebedev:2007hv} have 3 families of quarks and leptons and one
or more pairs of Higgs doublets. They all admit an $SU(6)$ orbifold
GUT with gauge-Higgs unification and the third family in the bulk.
They differ, however, in other bulk and brane exotic states.   We
show that the KK modes of the model, including only those of the
third family and the gauge sector, are {\em not} consistent with
gauge coupling unification at the string scale.  Nevertheless, we
show that it is possible to obtain unification if one adjusts the
spectrum of vector-like exotics below the compactification scale. As
an example, see Fig. \ref{fig:beta_functions}.  Note, the
compactification scale is less than the 4d GUT scale and some
exotics have mass two orders of magnitude less than $M_{c}$, while
all others are taken to have mass at $M_{\rm STRING}$.   In
addition,  the value of the GUT coupling at the string scale,
$\alpha_G(M_{STRING}) \equiv \alpha_{string}$, satisfies the weakly
coupled heterotic string relation \be \label{het_string_BC} G_N =
\frac{1}{8}\,\alpha_{string}\,\alpha^\prime \ee  or  \be
\alpha_{string}^{-1} = \frac{1}{8}\, (\frac{M_{Pl}}{M_{\rm
STRING}})^2 . \ee

%
%
\begin{figure}[ht!]
        \centering
        \includegraphics[scale = 0.20]{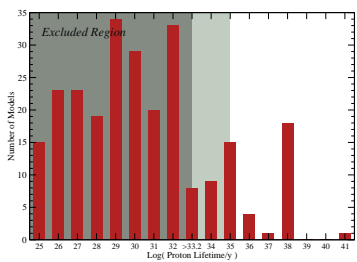}
        \caption{Histogram of solutions with $M_{\rm STRING} > M_{c} \gtrsim M_{\rm EX}$, showing the
        models which are excluded by Super-K bounds (darker green) and those which are
        potentially accessible in a next generation proton decay experiment (lighter green).
        Of 252 total solutions, 48 are not experimentally ruled out by the current
        experimental bound, and most of the remaining parameter space can be eliminated
        in the next generation of proposed proton decay searches.}
        \label{fig:histogram}
\end{figure}

In Fig. \ref{fig:histogram} we plot the distribution of solutions
with different choices of light exotics.   On the same plot we give
the proton lifetime due to dimension 6 operators.  Recall in these
models the two light families are located on the $SU(5)$ branes,
thus the proton decay rate is only suppressed by $M_{c}^{-2}$. Note,
90\% of the models are already excluded by the Super-Kamiokande
bounds on the proton lifetime.  The remaining models may be tested
at a next generation megaton water \v{c}erenkov detector.

Note that dimension 5 operators are generated when integrating out
color triplet exotics.  These operators may be suppressed by fine
tuning.   But this is not a satisfactory solution.   Recently,
however, it was shown that a discrete $\mathbb{Z}_4$ R symmetry can
be used to naturally suppress both the $\mu$ term and dimension 5
proton decay operators.  These terms are forbidden at the
perturbative level, but can be generated through non-perturbative
interactions. Moreover string models with such symmetries have been
constructed \cite{Lee:2010gv}.

\subsection{F  theory /  Type IIB}

\begin{figure} \begin{center}
  \includegraphics[height=2.5cm]{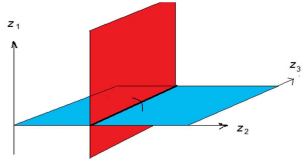}
  \caption{The figure represents 3 complex planes labeled by $z_i, i = 1,2,3$. The four dimensional
  blue surface is the gauge brane and the matter brane is red.  Open strings at the intersection give
  chiral matter in bi-fundamental representations.}
  \label{fig:branes} \end{center}
\end{figure}

\begin{figure} \begin{center}
  \includegraphics[height=2.5cm]{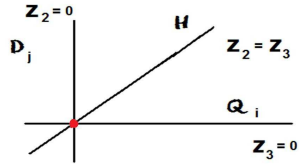}
  \caption{Yukawa couplings are generated at the intersection of two quark branes with a Higgs brane.}
  \label{fig:fermionmasses}  \end{center}
\end{figure}

We now change directions and discuss some recent progress in F
theory model
building~\cite{Beasley:2008dc,Beasley:2008kw,Donagi:2008ca,Donagi:2008kj,Blumenhagen:2008zz,Chen:2009me,Marsano:2009ym}.
An $SU(5)$ GUT is obtained on a D7 ``gauge" brane $S \times
R^{3,1}$.  D7 ``matter" branes on $S' \times R^{3,1}$ also exist
with chiral matter in 6D on $\Sigma \times R^{3,1}$ at the
intersection of the gauge and matter branes (Fig. \ref{fig:branes}).
Yukawa couplings enter at the triple intersections $\Sigma_1 \bigcap
\Sigma_2 \bigcap \Sigma_3$ of matter sub-manifolds (Fig.
\ref{fig:fermionmasses}).

$SU(5)$ is broken to the SM gauge group with non-vanishing
hypercharge flux $\langle F_Y \rangle$.  Note, this is not possible
in the Heterotic string!  This is because of the term in the
Lagrangian $\int d^{10}x ( dB + A_Y \bigwedge \langle F_Y \rangle
)^2$ which leads to a massive hypercharge gauge boson and
consequently a massive photon. In addition, $\langle F_Y \rangle$ on
the Higgs brane leads to doublet-triplet splitting.  Finally, spinor
representations of $SO(10)$ are possible in F theory; although they
are not possible in the perturbative type IIB string.

It has been argued that it is difficult to find multiply connected
manifolds, with non-zero $\Pi(1)$ and thus breaking GUTs with
hypercharge flux is novel advantage for F theory model building.  On
the other hand, it has been shown
\cite{Donagi:2008kj,Blumenhagen:2008aw} that breaking GUTs with
hypercharge field strengths induces significant threshold
corrections to gauge coupling unification.   This is a real problem
that must be dealt with in any GUT model obtained in this way.

It was also demonstrated that a fermion flavor hierarchy is
natural~\cite{Heckman:2008qa,Randall:2009dw,Cecotti:2009zf}, due to
flux in the $z_2-z_3$ surface breaking geometric flavor symmetry,
with Yukawa matrices of the form
\begin{equation} \lambda \sim  \left(
\begin{array}{ccc} \epsilon^8 & \epsilon^6 & \epsilon^4 \\
\epsilon^6 & \epsilon^4 & \epsilon^2 \\ \epsilon^4 & \epsilon^2 & 1
\end{array} \right)
\end{equation}

Finally, gravity decouples (i.e. $M_{Pl} \rightarrow \infty$) with a
non-compact $z_1$ direction.   These are so-called ``local"
constructions.  A bit of progress has also been made in ``global"
compact constructions~\cite{Blumenhagen:2008zz,Marsano:2009ym}.

\section{Heterotic - F theory duals}

\begin{figure}  \begin{center}
  \includegraphics[height=2.5cm]{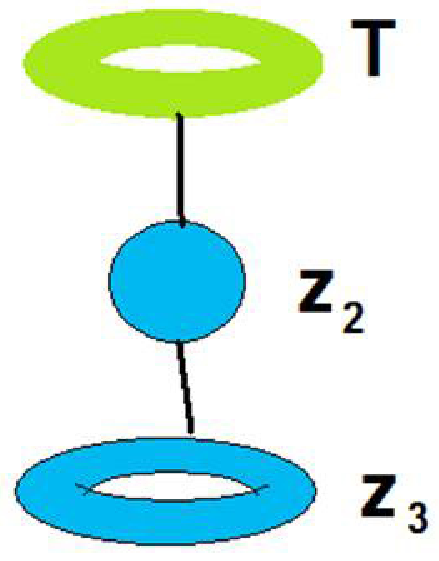}
  \includegraphics[height=2.5cm]{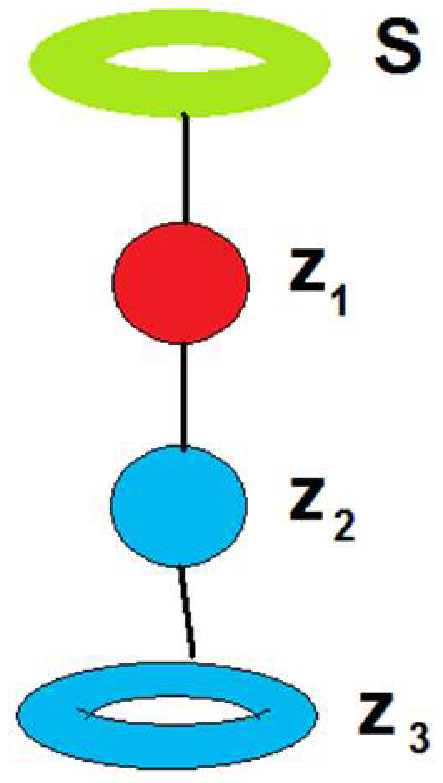}
  \caption{The heterotic string (left) is compactified on the product of two tori fibered over a common 2-sphere (defining a
  Calabi-Yau three-fold), while F theory
  (right) is defined in terms of a torus (whose complex structure defines the coupling strength of a Type IIB string) fibered
  over the product of two 2-spheres and the additional torus fibered over a common 2-sphere (defining a Calabi-Yau four-fold).}
  \label{fig:heteroticvsFtheory}  \end{center}
\end{figure}

F theory defined on a CY$_4$ is dual to the heterotic string defined
on a CY$_3$ (Fig. \ref{fig:heteroticvsFtheory}).  We are now
attempting to construct the F theory dual to our MSSM-like
models~\cite{bobkov}.  The motivation is three-fold.
\begin{enumerate} \item Three family MSSM-like
models have been found using an $SO(5) \times SO(5) \times
SO(4)$ torus mod $\mathbb{Z}_4 \times \mathbb{Z}_2$~\cite{RVW}
and an $SO(4)^3$ torus mod $\mathbb{Z}_2 \times \mathbb{Z}_2$
\cite{Blaszczyk:2009in}. This suggests a larger class of
MSSM-like models. We hope to find a more general description of
MSSM-like models, i.e. all models in the same universality
class. \item It may also provide a general understanding of
moduli space, since from the orbifold view point we must first
construct the superpotential before it is possible to identify
the moduli, and
\item it may help us understand moduli stabilization and SUSY
breaking. \end{enumerate}

Why should we expect F theory duals of heterotic orbifold models to
exist. First uplift our $E(8) \times E(8)$ orbifold models onto a
smooth Calabi-Yau manifold. Recall that after the first ${\mathbb
Z}_3$ orbifold plus Wilson line $W_3$ we find a 6D $SU(6)$ orbifold
GUT compactified on $(T_2)^2/{\mathbb Z}_3 \times T_2$.  The
complete massless spectrum in this case (including the hidden
sector) is given in Table \ref{tab:6dspectrum}~\cite{Dundee:2008ts}.
This spectrum satisfies the gravity anomaly constraint $N_H - N_V^6
+ 29 N_T = 273$, where ($N_H = 320, N_V^6 = 76, N_T = 1$) are the
number of (hyper-, vector,tensor) multiplets.

\begin{table}[h!]
\centering \caption{The full (six dimensional) spectrum of the
``benchmark" model with gauge group $SU(6) \times \left[SO(8) \times
SU(3) \right]'$. Note that $\mathbf{8}_{v+c+s} \equiv \mathbf{8}_{v}
+ \mathbf{8}_{c} + \mathbf{8}_{s}$.   In addition, the states are
written in the language of $D=6,  \ N=1$ supersymmetry.}
\label{tab:6dspectrum} \vspace{5mm}
\begin{footnotesize}
\begin{tabular}{c|c|c}
\hline
Multiplet Type&Representation&Number\\
\hline
\hline
tensor&singlet&1\\
\hline
vector&$(\mathbf{35},1,1)\oplus(1,\mathbf{28},1)$&35 + 28\\
&$\oplus(1,1,\mathbf{8})\oplus5\times(1,1,1)$&8 + 5\\
\hline
hyper&$(\mathbf{20},1,1)\oplus(1,\mathbf{8}_{v+c+s},1)\oplus 4\times(1,1,1)$&20+24+4 \\
&$\oplus9\times\left\{(\mbsix,1,1)\oplus(\mbsixb,1,1)\right\}$&108\\
&$\oplus9\times\left\{(1,1,\mbthree)\oplus(1,1,\mbthreeb)\right\}$&54\\
&$\oplus3\times(1,\mathbf{8}_{v+c+s},1)$&72\\
&$\oplus 36 \times (1,1,1)$&36\\
&SUGRA singlets&2\\
\hline
\end{tabular}
\end{footnotesize}
\end{table}

\begin{figure} \begin{center}
  \includegraphics[height=2.5cm]{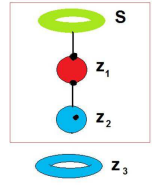}
  \caption{At the 6D level we have a Calabi-Yau three-fold times a torus.  The gauge branes are located at the points $z_1 = 0$ and
  $\infty$.   The matter branes are located schematically at the solid point in $Z_2$.}
  \label{fig:Ftheory2}  \end{center}
\end{figure}

Moreover, using the results of Bershadsky et
al.~\cite{Bershadsky:1996nh} we show that the $E(8) \times E(8)$
heterotic string compactified on a smooth $K_3 \times T^2$, with
instantons imbedded into $K_3$, is equivalent to the orbifolded
theories.   For example, with 12 instantons imbedded into an $SU(3)
\times SU(2)$ subgroup of the first $E(8)$ leaves an $SU(6)$ 6D GUT
with the massless hypermultiplets $(20 + cc) +  18 (6 + c.c.)$. Then
imbedding 12 instantons into an $E(6)$ subgroup of the second $E(8)$
plus additional higgsing leaves an unbroken $SO(8)$ gauge symmetry
with the massless hypermultiplets $4 (8v+ 8s+ 8c + c.c.)$. This is
identical to the massless spectrum of the orbifold GUT, IF we
neglect the additional $SU(3) \times U(1)^5$ symmetry which is
broken when going to the smooth limit.  In fact, it is expected that
the blow up modes necessary to smooth out the orbifold singularities
carry charges under some of the orbifold gauge symmetries;
spontaneously breaking these symmetries. Therefore $K_3 \times T_2$
with instantons is the smooth limit of $T_4/{\mathbb Z}_3 \times
T_2$ orbifold plus Wilson line.  In addition, it was shown that F
theory compactified on a Calabi-Yau 3-fold [defined in terms of a
torus $T_2$ fibered over the space $F_n$ $\times T_2$] is dual to an
$E(8) \times E(8)$ heterotic string compactified on $K_3 \times T_2$
with instantons~\cite{Bershadsky:1996nh} (see Fig.
\ref{fig:Ftheory2}).

Pictorially we see that the $SU(6) [SO(8)]$ gauge branes are
localized at the upper [lower] points on the $z_1$ 2-sphere (Fig.
\ref{fig:Ftheory2}).  These 7 branes wrap the four dimensional
surface $S = (z_2,z_3)$.  The matter 7 branes intersect the gauge
branes at points in $z_2$ and wrap the four dimensional surface $S'
= (z_1, z_3)$.   The intersection of the matter and gauge branes is
along the two dimensional surface $\Sigma = (z_3)$.

\begin{figure}  \begin{center}
  \includegraphics[height=2.5cm]{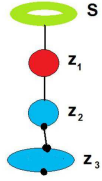}
  \caption{At the 4D level we fiber the last torus over the 2-sphere
  and retain two orbifold fixed points.  These two
  fixed points are where the two light families are conjectured to be located.
  In addition, as long as the fixed points
  remain, the Wilson line wrapping the last torus is stable.}
  \label{fig:Ftheory3}  \end{center}
\end{figure}

We now need to break the 6D $SU(6)$ GUT to $SU(5)$ and then to the
Standard Model.  At the same time we must break the N=1 SUSY in 6D
to N=1 in 4D.   This is accomplished by acting with the ${\mathbb
Z}_2$ orbifold on the torus and the 2-sphere.   A $U(1)$ flux in
$SU(6)$ on the gauge and matter branes breaks $SU(6)$ to $SU(5)$.
The breaking to the Standard Model requires a Wilson line on the
torus. However, we now encounter a possible obstruction to finding
the F theory dual of our Heterotic orbifold model. We need to keep
two orbifold fixed points (Fig. \ref{fig:Ftheory3}) -
\begin{enumerate} \item otherwise hypercharge gets mass~\cite{Nibbelink:2009sp},
\item and the Wilson line shrinks to a point, since $T^2/{\mathbb Z}_2$ is
topologically equivalent to a 2-sphere;
\item  and blow up modes on the heterotic side leave two orbifold fixed
points.
\end{enumerate}

In addition, on the heterotic side the two light families are
located at 4D orbifold fixed points.   We expect that on the F
theory side they will be located on $D_3$ branes fixed at the two
remaining 4D fixed points.  [These problems may be absent in the
model of Ref. \cite{Blaszczyk:2009in} in which the GUT breaking
Wilson line wraps a non-contractible cycle, even when all orbifold
fixed points are blown up.]

\section{Supersymmetry breaking and Moduli stabilization \label{sec:moduli_stabilization}}

Up until now we have focused on MSSM-like models obtained in the
supersymmetric limit.   The massless spectrum of these theories
contain three families of quarks and leptons, one or more pairs of
Higgs doublets and the Standard Model gauge sector $SU(3) \otimes
SU(2) \otimes U(1)_Y$.   However these theories typically include
many Standard Model singlet fields.  In the effective field theory
limit, these singlet fields have flat potentials.  All such fields
are called moduli.  Some of these moduli are geometric, describing
the volume or shape of the internal 6 dimensions.  While in orbifold
models there are also blow-up modes which parametrize the smoothing
out of the orbifold fixed points.  The gauge and Yukawa couplings of
the Standard Model are functions of these moduli.  In addition there
are, in many cases, extra gauge interactions. Typically a
non-Abelian hidden sector with new hidden sector quarks and
anti-quarks carrying only the hidden sector gauge charge. There may
be extra $U(1)$ gauge interactions and also new states which are
vector-like under the Standard Model gauge symmetry.  Blow-up modes
typically have charge under some of the additional gauge symmetries.
This is because these symmetries are only realized in the orbifold
limit.

Supersymmetry breaking is studied in the effective supergravity
field theory describing the massless states of the particular string
model.  If we can manage to break supersymmetry in this field theory
limit, then moduli stabilization will generically manage itself.
Radiative corrections to scalar potentials in non-supersymmetric
theories will, in most cases, introduce curvature into tree level
flat directions. Supersymmetry prevents this, however once
supersymmetry is broken then there will be few if any flat
directions left.   But how to break supersymmetry?  It is easy to
show that if the tree level theory is supersymmetric, then
supersymmetry cannot be broken in any finite order of perturbation
theory.   Hence in order to break supersymmetry, one needs
non-perturbative effects and non-perturbative effects come in at
least two ways,  either via non-local tunneling in field space or
via strong interaction dynamics, such as gaugino condensates.

There is one additional constraint in any successful SUSY breaking
solution.  The vacuum energy must be nearly zero in Planck units,
i.e.  $\langle T_{\mu \nu} \rangle = M_{Pl}^2 \Lambda g_{\mu \nu}$
where the observed cosmological constant has a value  $\Lambda \sim
{\cal O} (10^{-120} \ M_{Pl}^2)$, corresponding to a vacuum energy
density of order ($10^{-3}$eV)$^4$.   The scalar potential in
supergravity has the form
\begin{equation}  V = e^\fk (\fk_{I \bar J}^{-1} (D_I \fw)(D_J \fw)^* - 3
|\fw|^2) + \sum_a D_a^2   \end{equation}  where $\fk$ is the
K\"{a}hler potential, $\fw$ is the superpotential, $D_a$ is the D
term associated to the gauge group $a$ and $D_I =
\frac{\partial}{\partial \Phi_I} + \fk_I$ is the K\"{a}hler
covariant derivative with respect to the chiral field $\Phi_I$.
Note,  $\fk_I = \frac{\partial \fk}{\partial \Phi_I}, \; \fk_{I \bar
J} = \frac{\partial^2 \fk}{\partial \Phi_I \partial \Phi_J^*}$.
Supersymmetry breaking requires that $D_I \fw \neq 0$ and/or $D_a
\neq 0$.  In the supersymmetric limit,  the vacuum energy is less
than or equal to zero, where $V < 0$ if $\langle \fw \rangle \neq
0$. Typically, even when supersymmetry is broken, one requires some
up-lifting of the vacuum energy in order to obtain a small positive
cosmological constant.   The study of supersymmetry breaking via
non-perturbative effects goes back to the work of Witten and others
\cite{Witten:1981nf}.  Supersymmetry is broken in a hidden sector
and then transmitted to the visible SM sector via SM gauge
interactions.  The messengers of supersymmetry breaking couple
directly to the hidden sector and they are also charged under SM
gauge interactions.  This mechanism for supersymmetry breaking is
known as gauge mediated supersymmetry breaking or GMSB.  No self
consistent supersymmetry breaking model of this kind existed until
the work of Affleck, Dine and Seiberg and others
\cite{Affleck:1983rr}. These authors showed that instanton effects
can generate non-perturbative contributions to the superpotential.
The non-perturbative terms, along with perturbative contributions,
can sometimes lead to supersymmetry breaking. Simple gauge mediated
SUSY breaking models were found in Ref.~\cite{Dine:1993yw}. However
the most general understanding of non-perturbative effects in
supersymmetric gauge theories came with the work of Seiberg
\cite{Seiberg:1994pq}.   All of the aforementioned progress was in
the context of global SUSY.   It was however shown in local
supersymmetry, i.e. supergravity, by Nilles and others
\cite{Nilles:1982ik} that gaugino condensates lead to SUSY breaking.
This SUSY breaking is then transmitted to the SM sector via
supergravity, with the low energy SUSY breaking scale set by the
gravitino mass,  $m_{3/2}$.  In recent years, more mechanisms for
supersymmetry breaking have been discovered, including anomaly
mediated SUSY breaking and gaugino mediated SUSY breaking.  In
summary, non-perturbative mechanisms for supersymmetry breaking have
a long history.   Some, or all, of these mechanisms may be applied
in a string theory context.   I will only discuss some of the more
recent literature on SUSY breaking in string models.

Moduli stabilization and supersymmetry breaking in Type II string
models have been considered in \cite{Giddings:2001yu,Kachru:2003aw}.
Most of the geometric moduli in these theories can be stabilized via
gauge fluxes on internal manifolds.   In the context of Type II
string models, KKLT \cite{Kachru:2003aw} use fluxes to stabilize
most geometric moduli and SUSY breaking comes from non-perturbative
contributions to the superpotential? The superpotential has the form
\be \fw = w_0 + C e^{-a T} \ee where the constant $w_0$ is a
function of integer valued flux contributions which stabilize
internal volumes in $CY_3$.  Note, $w_0$ sets the scale of
supersymmetry breaking and for $m_{3/2} \sim 10$ TeV one needs $w_0
\sim 10^{-13}$(in Planck units) \cite{Choi:2004sx}.  $T$ is the
volume modulus of a 4 cycle wrapped by a $D_7$-brane. The
non-perturbative contribution to the superpotential can arise either
from an instanton configuration or a gaugino condensate. There is a
problem, however, with this simple set-up, i.e. the vacuum energy in
this simplest case is negative. Thus one also needs to introduce an
up-lifting sector. KKLT add an anti-$D_3$-brane which adds a
positive contribution to the vacuum energy, $V_D = \frac{D}{(Re
T)^3}$, and the constant $D$ can be fine-tuned to obtain a small
positive cosmological constant.

Soft SUSY breaking contributions to squark, slepton and gaugino
masses have been calculated in Refs. \cite{Choi:2004sx}.   They find
that matter VEVs dominate over moduli VEVs leading to a scenario for
SUSY breaking termed,  mirage mediation.   In this case, scalar
masses are of order $m_{3/2}/\langle a T \rangle \sim 10^2 - 10^3$
GeV (with $\langle a T \rangle \sim \ln(M_{Pl}/m_{3/2}) \sim {\cal
O}  (8 \pi^2$)), while gaugino masses, similarly suppressed, obtain
significant contributions from both moduli and anomaly mediation. As
a result, gaugino masses measured at the LHC may appear to unify at
a mirage scale which is intermediate between a TeV and GUT scales.

In F theory constructions
\cite{Heckman:2008es,Marsano:2008jq,Heckman:2008qt,Heckman:2010xz}
the authors have used the vector-like exotics under SM gauge
symmetries as mediators of supersymmetry breaking to the MSSM
states. This is known as gauge mediated SUSY breaking and typically
results in a gravitino LSP and a long-lived stau or bino NLSP.   The
idea is to generate a Polonyi potential to induce SUSY breaking. The
models contain two $D_7$-branes wrapping 4-cycles and intersecting
along a 2d curve $\Sigma$.  With the appropriate cohomology, there
is one massless chiral multiplet, $X$, at the intersection.  Then
including a $D_3$-instanton wrapping one of the 4-cycles, a Polonyi
term of the form $W \supset F_X \ X$ is generated with $F_X$
suppressed as compared to the compactification scale of the 4-cycle.
The phenomenology is somewhat model dependent. In Ref.
\cite{Marsano:2008jq}, the authors obtain a gravitino mass $m_{3/2}
\sim \frac{F_X}{M_{Pl}} \sim 1$ GeV, with a messenger mass $M \sim
M_{GUT}$.   In Ref. \cite{Heckman:2010xz}, on the other hand, the
authors obtain a gravitino mass $m_{3/2} \sim \frac{F_X}{M_{Pl}}
\sim 10 - 100$ MeV, with a messenger mass $M \sim 10^{12}$ GeV. It
should be noted that earlier and similar SUSY breaking
constructions, in the context of Type II strings can be found in
Refs. \cite{Aharony:2007db,Aparicio:2008wh} or Type I in Refs.
\cite{Cvetic:2007qj}.

Stabilization of moduli and SUSY breaking within the context of the
heterotic string has a long history.  The racetrack mechanism has
been discussed the most \cite{Krasnikov:1987jj}.\footnote{The
racetrack mechanism has also been discussed in the context of F
theory \cite{Kaplunovsky:1997cy} and more recently in terms of M
theory \cite{Acharya:2006ia}.}  In this scenario, two gaugino
condensates generate a non-perturbative superpotential of the form
\begin{equation} \nonumber \label{racetrack}
   \mathcal{W}_{\textsc{np}} = \sum_{a = 1,2} \left[ C_a
   \left[\Lambda_a(S,T)\right]^{3} \right]
\end{equation} where \be \Lambda_a(S,T) = e^{-\frac{8\pi^2}{\beta_a} f_a(S,T)}
\ee is the strong dynamics scale, $f_a \sim S$ (at zeroth order) is
the gauge kinetic function and $\beta_a = 3 N_a$ for $SU(N_a)$ is
the coefficient of the one-loop beta function.

In Section \ref{sec:minilandscape} we discussed ``benchmark model 1"
of the ``mini-landscape" of heterotic orbifold
constructions~\cite{Lebedev:2007hv} which has properties very close
to that of the MSSM. This model has been analyzed in the
supersymmetric limit. It contains an MSSM spectrum with three
families of quarks and leptons, one pair of Higgs doublets and an
exact R parity. In the orbifold limit, it also contains a small
number of vector-like exotics and extra $U(1)$ gauge interactions
felt by standard model particles. This theory also contains a large
number of standard model singlet fields, some of which are moduli,
i.e.\thinspace blow up modes of the orbifold fixed points. It was
shown that all vector-like exotics and additional $U(1)$ gauge
bosons acquire mass at scales of order the string scale at
supersymmetric minima satisfying $ F_I = D_a = 0 $ for all chiral
fields labeled by the index $I$ and all gauge groups labeled by the
index $a$.  In addition, the value of the gauge couplings at the
string scale and the effective Yukawa couplings are determined by
the presumed values of the vacuum expectation values [VEVs] for
moduli including the dilaton, $S$, the bulk volume and complex
structure moduli, $T_i, i = 1,2,3$ and $U$ and the SM singlet fields
containing the blow-up moduli
\cite{Nibbelink:2007pn,Nibbelink:2009sp}. Finally the theory also
contains a hidden sector $SU(4)$ gauge group with QCD-like chiral
matter.

Supersymmetry breaking and moduli stabilization in ``benchmark model
1" has been discussed in Ref. \cite{Dundee:2010sb}.  In general, the
model has a perturbative superpotential satisfying modular
invariance constraints, an anomalous $U(1)_A$ gauge symmetry with a
dynamically generated Fayet-Illiopoulos $D$-term and the hidden
QCD-like non-Abelian gauge sector generating a non-perturbative
superpotential.  The model also has of order 50 chiral singlet
moduli.  The perturbative superpotential is a polynomial in products
of chiral superfields with 100s of terms.  It is not possible at the
present time to analyze this model in complete detail.  Thus in Ref.
\cite{Dundee:2010sb} a simple model with a dilaton, $S$, one volume
modulus, $T$, and three standard model singlets was studied. The
model has only one gaugino condensate, as is the case for the
``benchmark model 1." We obtain a `hybrid KKLT' kind of
superpotential that behaves like a single-condensate for the dilaton
$S$, but as a racetrack for the $T$ and, by extension, also for the
$U$ moduli. An additional matter $F$ term, driven by the cancelation
of an anomalous $U(1)_A$ $D$-term, is the seed for successful
up-lifting. Note, similar analyses in the literature have also used
an anomalous $U(1)_A$ $D$-term in coordination with other
perturbative or non-perturbative terms in the superpotential to
accomplish SUSY breaking and
up-lifting~\cite{Binetruy:1996uv,Gallego:2008sv}.  The discussion in
Ref. \cite{Gallego:2008sv} is closest in spirit to that of
\cite{Dundee:2010sb}.

The structure of the superpotential of the form $\fw \sim w_0 e^{- b
T} + \phi_2 \ e^{- a S - b_2 T}$ gives the crucial progress-
\begin{enumerate}[i.)]
\item  a `hybrid KKLT' kind of superpotential that behaves like a
single-condensate for the dilaton $S$, but as a racetrack for
the $T$ and, by extension, also for the $U$ moduli; and
\item an additional matter $F_{\phi_2}$ term driven by the
cancelation of the anomalous $U(1)_A$ $D$-term seeds SUSY
breaking with successful uplifing.
\end{enumerate}
Note, the constants $b, \ b_2$ can have either sign. For the case
with $b, \ b_2 > 0$ the superpotential for $T$ is racetrack-like.
However for $b, \ b_2 < 0$ the scalar potential for $T$ diverges as
$T$ goes to zero or infinity and compactification is guaranteed
\cite{Font:1990nt}.  Soft SUSY breaking terms for squarks, sleptons
and gauginos were evaluated in this simple model.  Note, the
gravitino mass scale is set by the constant $w_0$ which was shown to
be naturally small in Planck units due to approximate R symmetries
\cite{Kappl:2008ie}.

\subsection{Some thoughts on the string landscape and the anthropic
principle}

The energy balance of the universe includes of order 4\% visible
matter, 23\% dark matter and 73\% dark energy.  The first
corresponds to the matter we are made of, while the second is
postulated to be made of stable particles with mass of order one
TeV; perhaps the lightest supersymmetric particle.   The quantity
known as dark energy is quite frankly a complete conundrum.  It is
most simply described as a cosmological constant or vacuum energy
density.  Why it is so very small has no good explanation (in
natural units ($M_{Pl}$) the cosmological constant is of order
$10^{-120}$). So perhaps the best explanation to date is known as
the anthropic principle. Only if the cosmological constant is small
enough can galaxies form and thus provide the a priori conditions
for humans to be around to observe this universe
\cite{Weinberg:1987dv}.   If there is an ensemble of possible
universes with all possible values of the cosmological constant,
then we might expect a cosmological constant of the observed value.
Indeed this may be the case, and it is good to know that string
theory provides such a large ensemble of possible vacua, i.e. the
string landscape
\cite{Bousso:2000xa,Denef:2007pq,Schellekens:2008kg}. The string
landscape thus provides a possible explanation for why the
cosmological constant is so small.  But what other questions does it
answer?  Can it explain why we have the observed Standard Model
gauge interactions with three families of quarks and leptons and a
Higgs? As discussed earlier in Section \ref{sec:random}, random
searches in the string landscape suggest that the Standard Model is
very rare. This may also suggest that string theory cannot make
predictions for low energy physics.  This would be a shame, but
perhaps we are being too hasty.  One possibility is that some or all
of the above questions are also answered by anthropic arguments.
Attempts in this direction have been taken to try and understand the
weak scale and the Higgs mass \cite{ArkaniHamed:2004fb} and also the
top quark mass \cite{Feldstein:2006ce} or axion cosmology
\cite{Wilczek:2004cr}.

On the other hand, perhaps understanding the value of the
cosmological constant is premature.  After all String Theory, as
such, is a work in progress.  In particular, the most complete
description of string theory is in terms of the quantization of
conformal field theories [CFT] defined on the two dimensional string
world sheets.  The fields themselves span, in the case of
superstrings, a 10 dimensional space-time background.  In fact,
space-time itself is an emergent quantity.  Perturbative
calculations rely on assuming a particular classical background
configuration.  Non-perturbative physics comes in through strong
non-Abelian gauge interactions (fluxes and/or condensates),
instantons (gauge and world sheet) and $D$-branes. Whether the
string CFT is defined on a lattice or in terms of fermions and/or
bosons is an arbitrary choice.  But each choice determines a
different massless sector. The massless sector can be described by
an effective field theory, but how does one distinguish one
arbitrary choice of background from another?  Each effective field
theory has it's own calculable ground state, but how does one
compare one ground state for one background to another with a
different background choice?   Why, in practice, are four space-time
dimensions large, while the other 6 are curled up into an, as yet,
unobservable ``ball"?   Why should we have N=1 SUSY in 4 dimensions?
All of these questions are crucial to an understanding of our
universe and deeply profound, but they have proven to be extremely
difficult nuts to crack.

However, there is much more data in our low energy phenomenology
than just the cosmological constant. Much in this data suggests more
symmetry - the family structure and charge quantization,  the
hierarchy of fermion masses and mixing angles, and the absence of
large flavor violating processes all suggest grand unification with
family symmetries.   Perhaps string theory can be predictive,  IF we
understood the rules for choosing the correct position in the string
landscape.  At the moment, these rules are not understood, so the
best guess relies on statistical analyses.   However,  we have
argued in these pages that finding the Standard Model (or Minimal
Supersymmetric Standard Model) is best achieved by requiring SUSY
GUTs at the first step.  I would like to argue that if we could
better understand the string landscape in these ``fertile patches"
we might then be able to understand the rules needed to choose this
region of the landscape.  Of course, a major caveat in this whole
discussion is the assumption that our low energy universe is
describable by the Minimal Supersymmetric Standard Model.  This
assumption will soon be tested at the Large Hadron Collider located
at CERN near Geneva, Switzerland.

One small hint in this direction is provided by the heterotic
orbifold models discussed in Section \ref{sec:heterotic}.  These
heterotic orbifold models have some amazing properties:
\begin{enumerate}
\item They incorporate local GUTs with two complete families
localized at orbifold fixed points; \item They incorporate a 5d
$SU(6)$ orbifold GUT with gauge-Higgs unification and the third
family in the bulk; \item As a consequence, they have
gauge-Yukawa unification for the top quark (thus explaining why
the top quark is heavy);
\item They incorporate doublet-triplet splitting with a $\mu$
term which is naturally small;   \item They have an exact R
parity. [Moreover recently it was discovered that similar models
can incorporate a $\mathbb{Z}_4^R$ symmetry which allows all
Yukawa interactions and neutrino masses while forbidding the
$\mu$ term and dimension 5 baryon and lepton number violating
operators at the perturbative level \cite{Lee:2010gv}.  The
$\mathbb{Z}_4^R$ symmetry is possible due to the final
$\mathbb{Z}_2$ orbifold.];
\item  As a consequence of the $\mathbb{Z}_2$ orbifold, the
model has a $D_4$ family symmetry which can ameliorate problems
with flavor changing neutral currents while at the same time
accommodating a hierarchy of quark and lepton masses; \item
Approximate R symmetries naturally generate a small constant
contribution to the superpotential, setting the scale for the
gravitino mass once supersymmetry breaking is generated.
\end{enumerate}
Such models are generically related to smooth heterotic models
compactified on $(K_3 \times T^2)/\mathbb{Z}_2$.  Perhaps a clue to
why the Standard Model is special can be found in this class of
heterotic models.

\section{Conclusion}

In this paper, we have discussed an evolution of SUSY GUT model
building. We saw that 4d SUSY GUTs have many virtues.  However there
are some problems which suggest that these model may be difficult to
derive from a more fundamental theory, i.e. string theory.   We then
discussed orbifold GUT field theories which solve two of the most
difficult problems of 4d GUTs, i.e. GUT symmetry breaking and Higgs
doublet-triplet splitting.   We then showed how some orbifold GUTs
can find an ultra-violet completion within the context of heterotic
string theory.

The flood gates are now wide open.  In recent work
\cite{Lebedev:2007hv} we have obtained many models with features
like the MSSM:  SM gauge group with 3 families, and vector-like
exotics which can, in principle, obtain large mass.   The models
have an exact R-parity and non-trivial Yukawa matrices for quarks
and leptons.  In addition, neutrinos obtain mass via the See-Saw
mechanism.   We also showed that gauge coupling unification can be
accommodated.

Of course, this is not the end of the story.  It is just the
beginning.  In order to obtain predictions for the LHC, one must
stabilize the moduli and break supersymmetry.  In fact, these two
conditions are {\em not} independent, since once SUSY is broken, the
moduli will be stabilized.  The scary fact is that the moduli have
to be stabilized at just the right values to be consistent with low
energy phenomenology.  Interesting first steps in the construction
of complete and phenomenologically testable string models have been
taken.

\section*{Acknowledgements}

This work was accomplished with partial support from DOE grant
DOE/ER/01545-888.  I also wish to acknowledge support from CERN,
where most of this article was written.

\end{document}